\documentclass[useAMS,usenatbib]{mn2e}

\usepackage[pdftex,pdfpagemode={UseOutlines},bookmarks,bookmarksopen,colorlinks,linkcolor={blue},citecolor={blue},urlcolor={red}]{hyperref}
\usepackage{float}
\usepackage{graphicx}
\usepackage{caption}
\usepackage{subcaption}
\usepackage{natbib}
\usepackage{rotating,longtable,lscape}

\title[Colour-magnitude diagrams of Exoplanets II]{Colour-magnitude diagrams of transiting Exoplanets - II. A larger sample from photometric distances}
\author[Amaury H.~M.~J. Triaud et al.]
{Amaury H.~M.~J. Triaud$^{1,2}$\thanks{E-mail: triaud@mit.edu}, 
Audrey A. Lanotte$^{3}$, 
Barry Smalley$^{4}$ and
Micha\"el Gillon$^{3}$\\
$^{1}$Kavli Institute for Astrophysics \& Space Research, Massachusetts Institute of Technology, Cambridge, MA 02139, USA\\
$^{2}$Fellow of the Swiss national science foundation\\
$^{3}$Institut d'Astrophysique et de G\'eophysique, Universit\'e de Li\`ege, All\'ee du 6 Ao\^ut 17, Sart Tilman, 4000 Li\`ege 1, Belgium\\
$^{4}$Astrophysics Group, Keele University, Staffordshire, ST5 5BG, UK}
\begin{document}

\date{Accepted ?. Received ?; in original form ?}

\pagerange{\pageref{firstpage}--\pageref{lastpage}} \pubyear{2014}

\maketitle

\label{firstpage}

\begin{abstract}
Colour-magnitude diagrams form a traditional way of { presenting} luminous objects in the Universe and compare them to each others. Here, we estimate the photometric distance of 44 transiting exoplanetary systems. Parallaxes for seven systems confirm our methodology. Combining those measurements with fluxes obtained while planets were
occulted by their host stars, we compose colour-magnitude diagrams in the near and mid-infrared. When possible, planets are plotted alongside
very low-mass stars and field brown dwarfs, who often share similar sizes and equilibrium temperatures. They offer a natural, empirical, comparison sample. We also include directly imaged exoplanets and the expected loci of pure blackbodies. \\
Irradiated planets do not match blackbodies; their emission spectra are not featureless.  For a given luminosity, hot Jupiters' daysides show a larger variety in colour than brown dwarfs do and display an increasing diversity in colour with decreasing intrinsic luminosity.
{ The presence of an extra absorbent within the 4.5 $\mu$m band would reconcile outlying hot Jupiters with ultra-cool dwarfs' atmospheres. Measuring the emission of gas giants cooler than 1\,000 K would disentangle whether planets' atmospheres behave more similarly to brown dwarfs' atmospheres than to blackbodies, whether they are akin to the young directly imaged planets, or if irradiated gas giants form their own sequence.}

\end{abstract}



It is trivial to convert fluxes measured at occultation, or obtained while observing the phase curves of transiting exoplanets 
into absolute magnitudes. One only needs a distance measurement. Two colour-magnitude diagrams for transiting --or occulting-- exoplanets were presented in \citet{Triaud:2014kq} for seven systems
that have {\it Hipparcos} parallaxes \citep{van-Leeuwen:2007qy}. Coincidentally, this happened { approximately a century after the first Herzsprung-Russell diagrams were composed \citep{Hertzsprung:1911lr,Russell:1914fr,Russell:1914kx,Russell:1914nr}}.

Colour-magnitude diagrams offer a means to compare exoplanets with each others, using natural units for observers. In addition, they allow to infer global properties without requiring the need to fit complex atmospherical models through the sparse data points that can only be gathered at this stage. Those inferences can be made by comparing exo-atmospheres to other objects having similar temperatures and sizes; very low-mass stars and field brown dwarfs are a readily available and well-studied sample. Young, directly imaged planets are routinely compared to field brown dwarfs for this very reason (e.g. \citealt{Bonnefoy:2013yu}). { Finally, irradiated and non-irradiated gas giants  can be compared to each others in colour-magnitude space. Those diagrams can offer a tool to pinpoint the processes that lead highly irradiated planets to be bloated (e.g. \citealt{Demory:2011lr}).}

\begin{figure*}  
\begin{center}  
	\begin{subfigure}[b]{0.33\textwidth}
		\caption{$\chi^2_{\rm r} = 0.3 \pm 0.4$}
		\includegraphics[width=\textwidth]{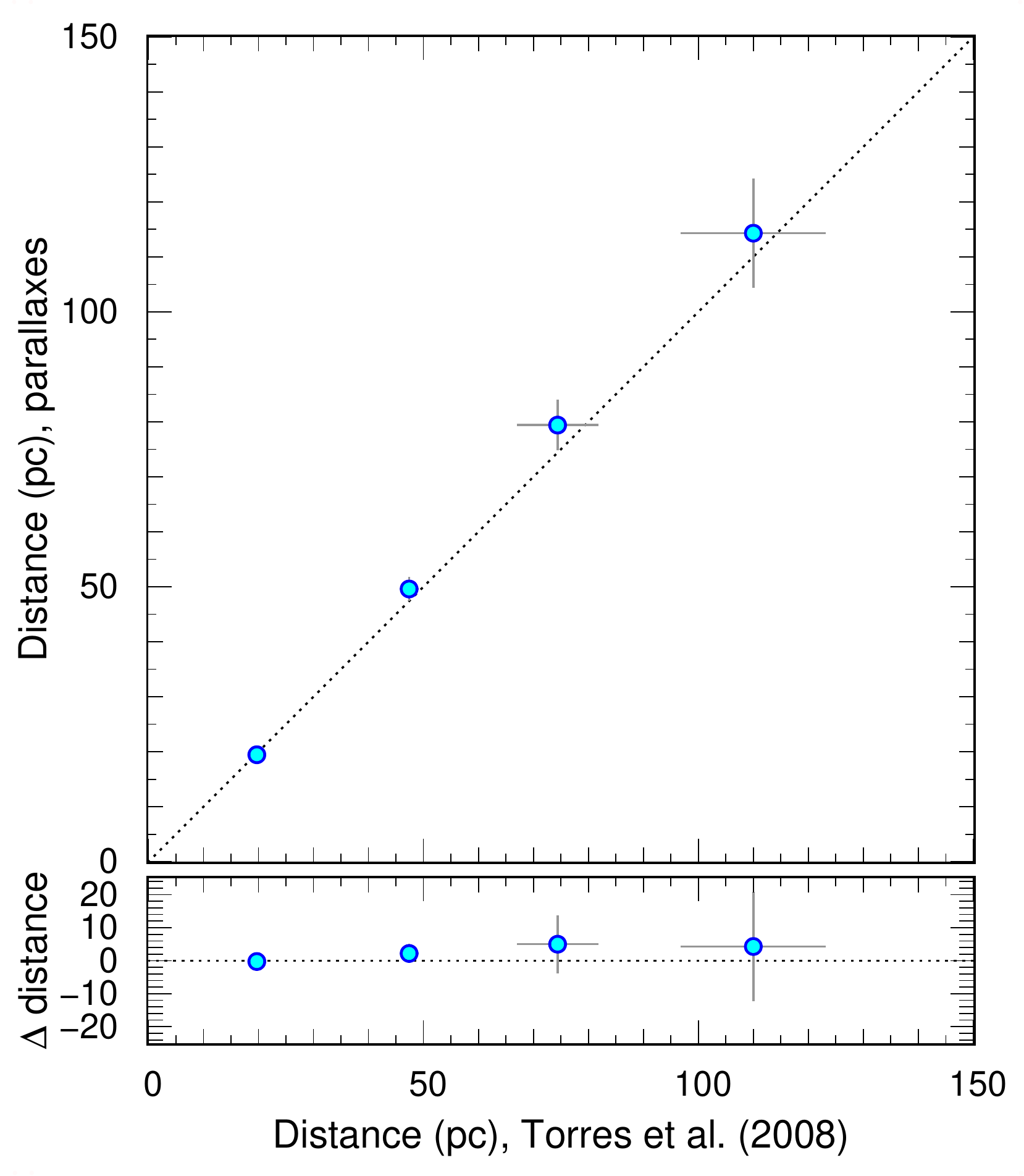}  
		\label{fig:torres_par}  
	\end{subfigure}
	\begin{subfigure}[b]{0.33\textwidth}
		\caption{$\chi^2_{\rm r} = 2.7 \pm 0.8$}
		\includegraphics[width=\textwidth]{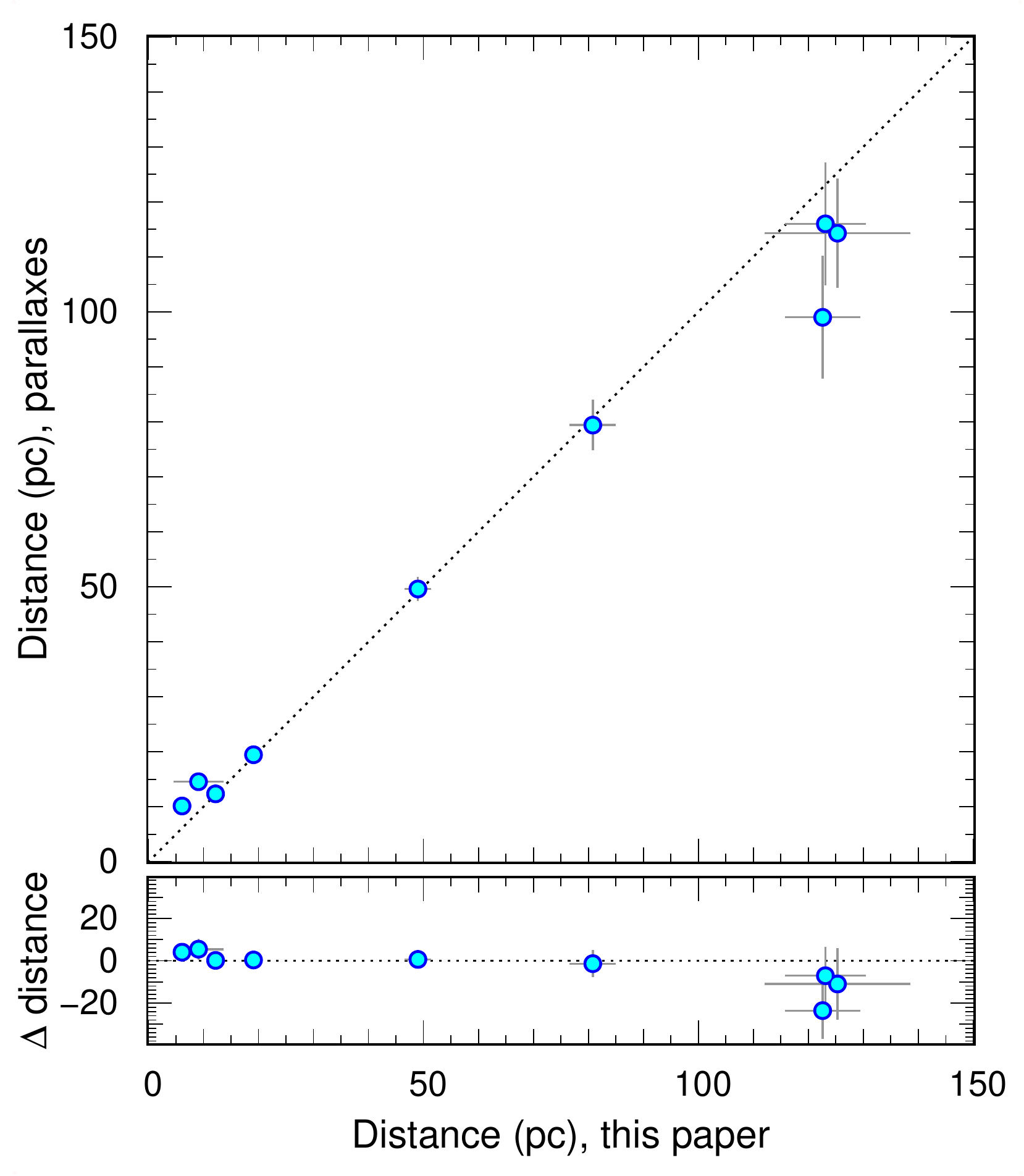}  
		\label{fig:barry_par}  
	\end{subfigure}
	\begin{subfigure}[b]{0.33\textwidth}
		\caption{$\chi^2_{\rm r} = 1.1 \pm 0.4$}
		\includegraphics[width=\textwidth]{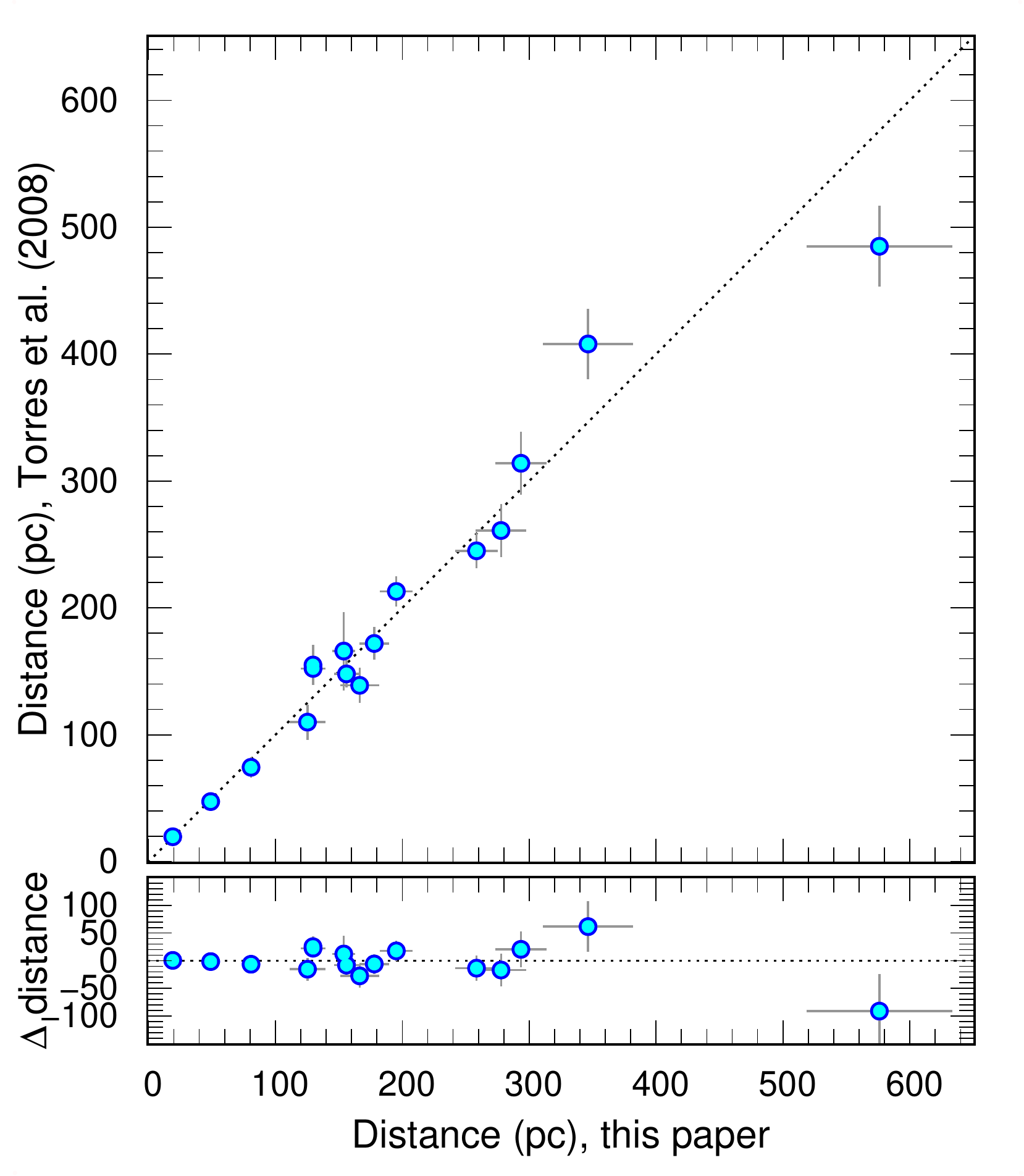}  
		\label{fig:barry_torres}  
	\end{subfigure}
	\caption{Distance measurements compared with one another, from our sample, including the two discrepant stars GJ\,436 and GJ\,1214. Reduced $\chi^2_{\rm r}$ are given. a) parallactic distances from {\it Hipparcos} versus photometric distances from \citet{Torres:2008yq}. b) parallactic distances from {\it Hipparcos} versus photometric distances estimated in this paper. c) photometric distances from  \citet{Torres:2008yq} versus photometric distances estimated in this paper.
}\label{fig:dist_comp}  
\end{center}  
\end{figure*} 

Just as the { construction} of the Herzsprung-Russell diagram led to vast advances in stellar formation and evolution, { the compilation of} colour-magnitude diagrams for transiting exoplanets will likely spur similar developments. Models in colour-space may predict that different planet families have distinct locations or sequences { (dependent on their gravity, their atmospheric structure, their relative abundances...)}. This would provide diagnostics to select suitable targets for further follow-up, in a fashion similar to selecting a particular stellar population, for instance, to remove giant contaminants prior to a survey focusing on G and K dwarfs. { In the case of irradiated gas giants specifically, the lack of cloud cover may cause objects to fall in a specific region in colour-space. Being identifiable, it will help optimise the detection of atmospheric features in transmission. In addition, if planets follow} defined sequences, magnitudes obtained in one band lead to accurate predictions for others bands. It can only encourage observations at wavelengths more difficult to obtain.

In total, 44 systems (43 planets and one brown dwarf) have been observed at occultation and were present in the literature. Rather than waiting for {\it GAIA} (e.g. \citealt{Perryman:2001lr}) to deliver its much awaited parallaxes, this paper will instead use photometric distances. Thanks to their transiting configurations and to the intensive observational efforts that { has been undertaken} both in the confirmation and in the characterisation of these objects, the fundamental stellar parameters are accurately known. This means that reliable distances can be computed such as was done for example by \citet{Torres:2008yq}. 
Hertzsprung-Russell diagrams can be represented as luminosity versus effective temperature. We instead opted for using colours instead of temperatures \citep{Beatty:2014lr}, because magnitudes are closer to direct observables.

The paper is organised in the following way: we first outline our procedure to measure photometric distances (Sec.~\ref{sec:dist}) and then describe how the host stars' apparent magnitudes were determind from the {\it Spitzer} photometry (Sec.~\ref{sec:spitz_mag}). In the following section, different colour-magnitude diagrams are drawn and described in qualitative and quantitative ways. We then discuss our results and conclude.

\begin{figure*}  
\begin{center}  
	\begin{subfigure}[b]{0.33\textwidth}
		\caption{$\chi^2_{\rm r} = 0.7 \pm 0.2$}
		\includegraphics[width=\textwidth]{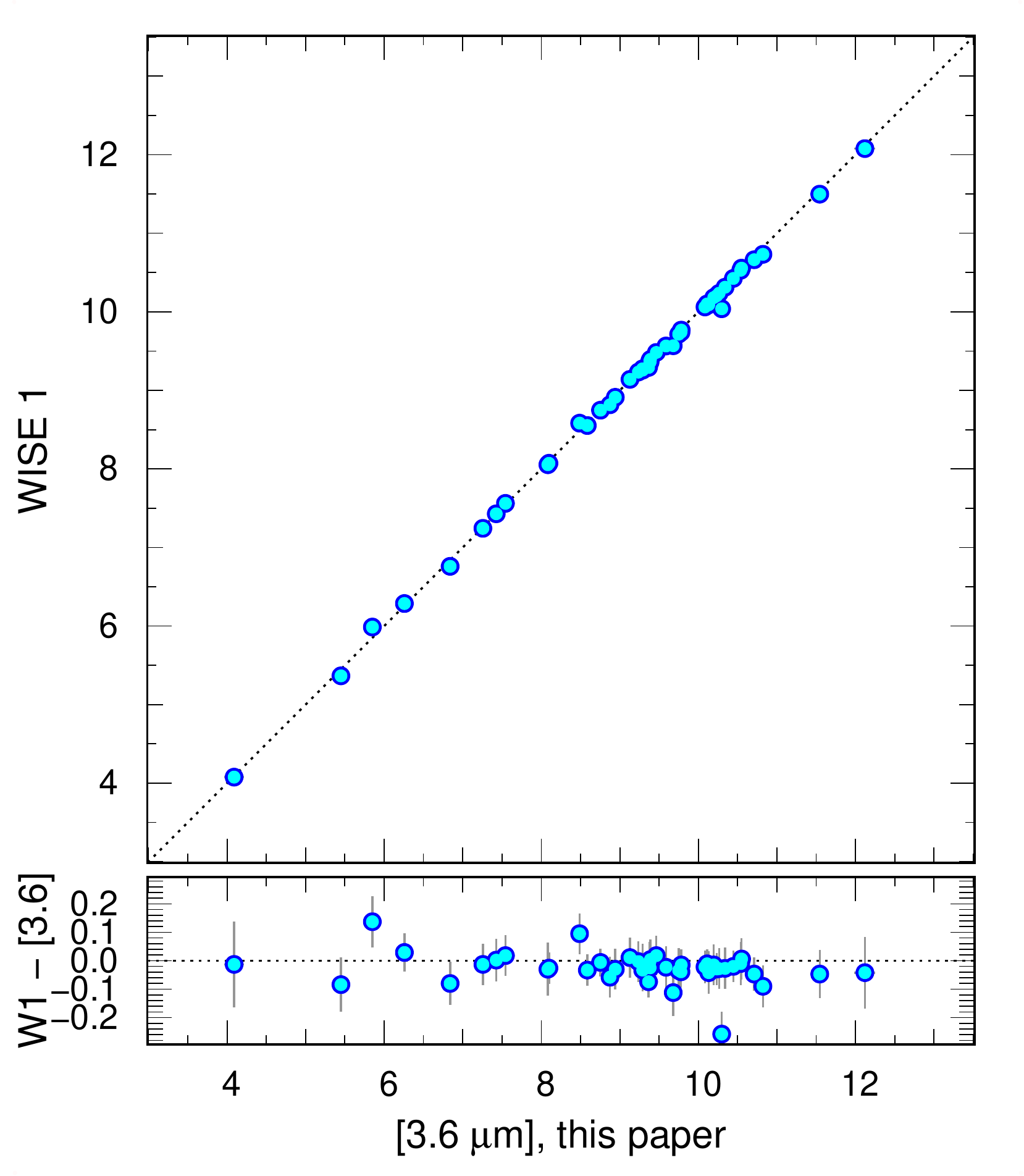}  
		\label{fig:wise_spit3p6}  
	\end{subfigure}
	\begin{subfigure}[b]{0.33\textwidth}
		\caption{$\chi^2_{\rm r} = 1.7 \pm 0.3$}
		\includegraphics[width=\textwidth]{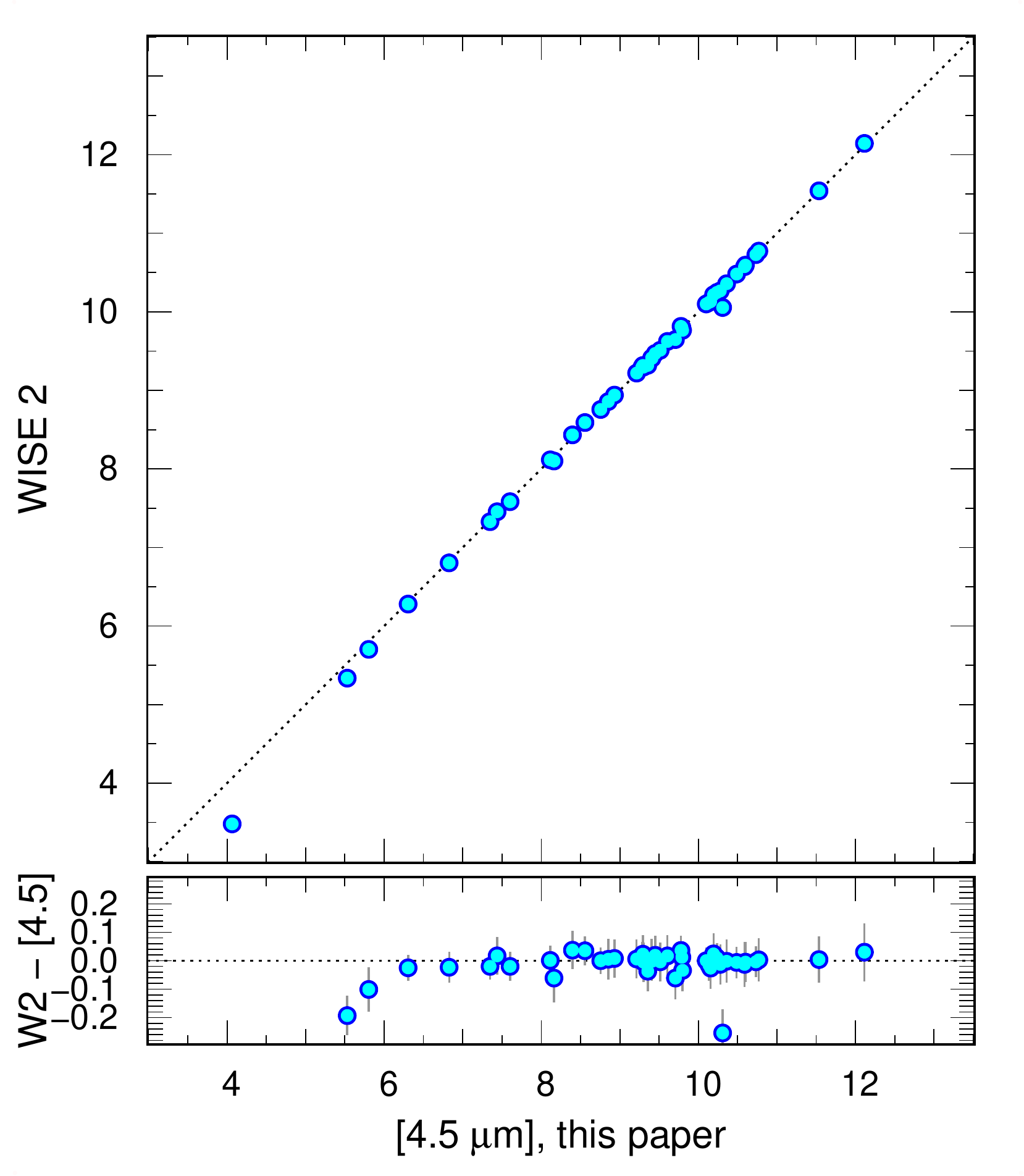}  
		\label{fig:wise_spit4p5}  
	\end{subfigure}
	\caption{Apparent magnitude measurements comparing those obtained by {\it WISE} to those that we estimated, from the {\it Spitzer} images. CoRoT-2A is clearly discrepant in both, because it is blended with CoRoT-2B the {\it WISE} data. Reduced $\chi^2_{\rm r}$ are given. a) on a band centred around 3.6 $\mu m$. b) on a band centred around 4.5 $\mu m$. The discrepant point at $\sim 10.3$ is the CoRoT-2 system. Objects $>6^{\rm th}$ magnitude appear brighter in the WISE 2 band, which may be due to some detector effects. Discrepant points removed, $\chi^2_{\rm r} < 1$.
}\label{fig:wise_spit}  
\end{center}  
\end{figure*} 

\section{The determination of photometric distances}\label{sec:dist}


{ Our distances are derived from catalogued parameters:} we obtained effective temperatures ($T_{\rm eff}$), surface gravities ($\log\,{\rm g}$) and metallicities ([Fe/H]) from TEPCAT\footnote{\href{http://www.astro.keele.ac.uk/jkt/tepcat/}{www.astro.keele.ac.uk/jkt/tepcat/}} \citep{Southworth:2011kx} and used those to compute stellar radii ($R_\star$) thanks to a relation provided in \citet{Torres:2010uq} { (Ch. 8)}. $R_\star$ and $T_{\rm eff}$ directly lead to stellar luminosities ($L_\star$) that were in turn converted into bolometric magnitudes ($M_{\rm bol}$) using the following relation \citep{Cox:2000vn}:
\begin{equation}\label{eq:mbol}
M_{\rm bol} = 4.75 - 2.5 \log L_\star
\end{equation}
Absolute visual magnitudes ($M_{\rm V}$) were estimated thanks to bolometric corrections estimated by \citet{Flower:1996yq}; values are provided in Table~\ref{tab:stars}.

We explored the Tycho2 catalogue \citep{Hog:2000uq} to compile a list of apparent visual magnitude $m_{\rm V}$. Failing to find a number of systems we turned to APASS/UCAC4 \citep{Zacharias:2013fj} and then to TASS \citep{Droege:2006yq}. Distances were obtained from the distance modulus ($m_{\rm V} - M_{\rm V}$). Errors are propagated throughout.

No reddening corrections, $E(B-V)$ were applied since they are not available for most of our sample. We expect most $E(B-V) < 0.1$, leading to offsets $A_{\rm V} < 0.33$ on ($m_{\rm V} - M_{\rm V}$) \citep{Maxted:2011vn}.

The distances we calculated are given in Table~\ref{tab:stars} and are visually represented in Figure~\ref{fig:dist_comp}. Those plots show our results compared to corresponding distances from the revised {\it Hipparcos} catalog \citep{van-Leeuwen:2007qy}. { We also compare our estimates to photometric distances from \citet{Torres:2008yq}, which provides a wider range and greater overlap of systems than {\it Hipparcos}}. Our two most discrepant  distance measurements are on GJ\,436 and GJ\,1214.
This is most probably caused by the late type of both stars{, who, with masses $< 0.6$~M$_\odot$, fall outside the range over which  the Torres relation has been calibrated for.} We thus adopt the most recent distance estimates, from \citet{van-Leeuwen:2007qy} and from \citet{Anglada-Escude:2013lr} respectively.
Removing those two objects, the reduced $\chi^2_{\rm r}$ for Fig.~\ref{fig:barry_par} changes from $2.7 \pm 0.8$ to $0.6\pm0.4$. All comparisons lead to reduced $\chi^2_{\rm r} \sim 1$. Reddening is thus contained within our error bars. 

\section{The determination of {\it Spitzer} apparent magnitudes}\label{sec:spitz_mag}

{ The {\it WISE} satellite  \citep{Cutri:2012fk} has two bandpasses, W1 and W2, that resemble two of {\it Spitzer}'s IRAC channels. \citet{Kirkpatrick:2011jk} showed the colour agreement between both spacecrafts, on field brown dwarfs. Needing to use the IRAC 3 \& 4, for which there is no WISE equivalent, we derived photometry from all {\it Spitzer} channels and compared the [3.6] and [4.5] to W1 and W2, to validate our measurements in the redder channels.}

{ We searched the {\it Spitzer} Heritage Archive\footnote{\href{http://sha.ipac.caltech.edu/applications/Spitzer/SHA/}{http://sha.ipac.caltech.edu/applications/Spitzer/SHA/}} for all frames obtained on the targets with reported occultations in the published literature (Table~\ref{tab:app_pl}). Apparent magnitudes were obtained for each set of observations. Our methods for extracting the photometry are located in appendix~\ref{app:phot}, and here summarised. We perform aperture photometry on the IRAC images calibrated by the standard {\it Spitzer} pipeline according to the {\tt EXOPHOT} {\it pyraf} pipeline following Lanotte et al. (in prep). Stellar flux is corrected for  contribution from visual companions, if relevant. We average those flux and convert them into Vega apparent magnitudes following the methods described by \citet{Reach:2005lr}.} When several observations were made of the same stars, we computed the optimal average of their apparent magnitude in each of { Spitzer's Astronomical Observation Request (AOR)} to produce the values located in table~\ref{tab:stars}.

Our estimates are graphically compared in Fig.~\ref{fig:wise_spit} to corresponding bands employed by the {\it WISE} satellite. Reduced $\chi^2_{\rm r}$ are calculated. They indicate very good agreement between both set of values. Despite good agreement some objects are clearly discrepant. For example CoRoT-2A, that is $\sim 0.3$ mag fainter in our estimation. We suspect this is because {\it WISE} could not distinguish CoRoT-2A from its visual companion, as we have done when deconvoluting. In the 4.5 $\mu$m band, objects brighter than the 6$^{\rm th}$ magnitude are also discrepant. Those removed, $\chi^2_{\rm r}$ drops from 1.7 to 0.4. { All our bright targets remained well within IRAC 2's region of linearity. The discrepancy likely emanates from {\it WISE}}. Our values can therefore be considered as being more accurate. The low $\chi^2_{\rm r}$ we obtain reveals we probably overestimate our error bars. We assume the same of the IRAC 3 \& 4 channels and use our apparent magnitudes to compute the planets'.

\section{Colour-Magnitude diagrams}\label{sec:col_mag}

Planet to star flux ratios, measured at occultation in the J, H and K band as well as observed by {\it Spitzer}'s IRAC 1, 2, 3 and 4 bands were obtained from the literature and transformed into a change in magnitude. Using stellar apparent magnitudes (Table~\ref{tab:stars}), planetary fluxes were thus transformed into apparent magnitudes (Table~\ref{tab:app_pl}). Although only a technicality, this step is interesting in immediately providing an estimate of whether a certain instrument, or mirror-size is sufficient to detect a given planet. This way we realise that 55~Cnc~e, a rocky planet, is a $14^{\rm th}$ magnitude at 4.5~$\mu$m, meaning it can be detectable with a medium-size telescope, which it was \citep{Demory:2012uq}. This is also a practical way to compare transiting planets with directly-detected planets.
Using our computed distance moduli (Table~\ref{tab:stars}) we obtain absolute magnitudes for stars and planets, that are listed, respectively in Tables~\ref{tab:abs}~\&~\ref{tab:planets}. 

The planets' absolute magnitudes are represented by circular, blue symbols arranged as colour-magnitude diagrams in Figs.~\ref{fig:NIR} \& \ref{fig:MidIR}. We will now describe how planets are spread with respect to each other but also to ultra-cool dwarfs. Very low-mass stars and brown dwarfs are represented in the background of the same diagrams as diamonds whose colours move from orange to black as a function of their assigned spectral type (ranging from M5 to Y1).

\subsection{Comparing with ultra-cool dwarfs}

Information comes from comparing a new sample to one already well studied or to a model. Since models for irradiated planets have yet to be computed for colour space, very low-mass stars and field brown dwarfs, who have similar effective temperatures and sizes come as a readily available comparison sample. We can now see if planets follow or depart from the known location of those objects. Our comparison sample was borrowed from \citet{Dupuy:2012lr} who recently compiled a vast list of  ultra-cool dwarf magnitudes and parallaxes. Later in the paper, a comparison will be made to the expected location of blackbodies (Sec.~\ref{sec:bb}) and to the position of directly detected planets (Fig.~\ref{fig:bb_NIR}).

\begin{figure*}  
\begin{center}  
	\begin{subfigure}[b]{0.33\textwidth}
		\includegraphics[width=\textwidth]{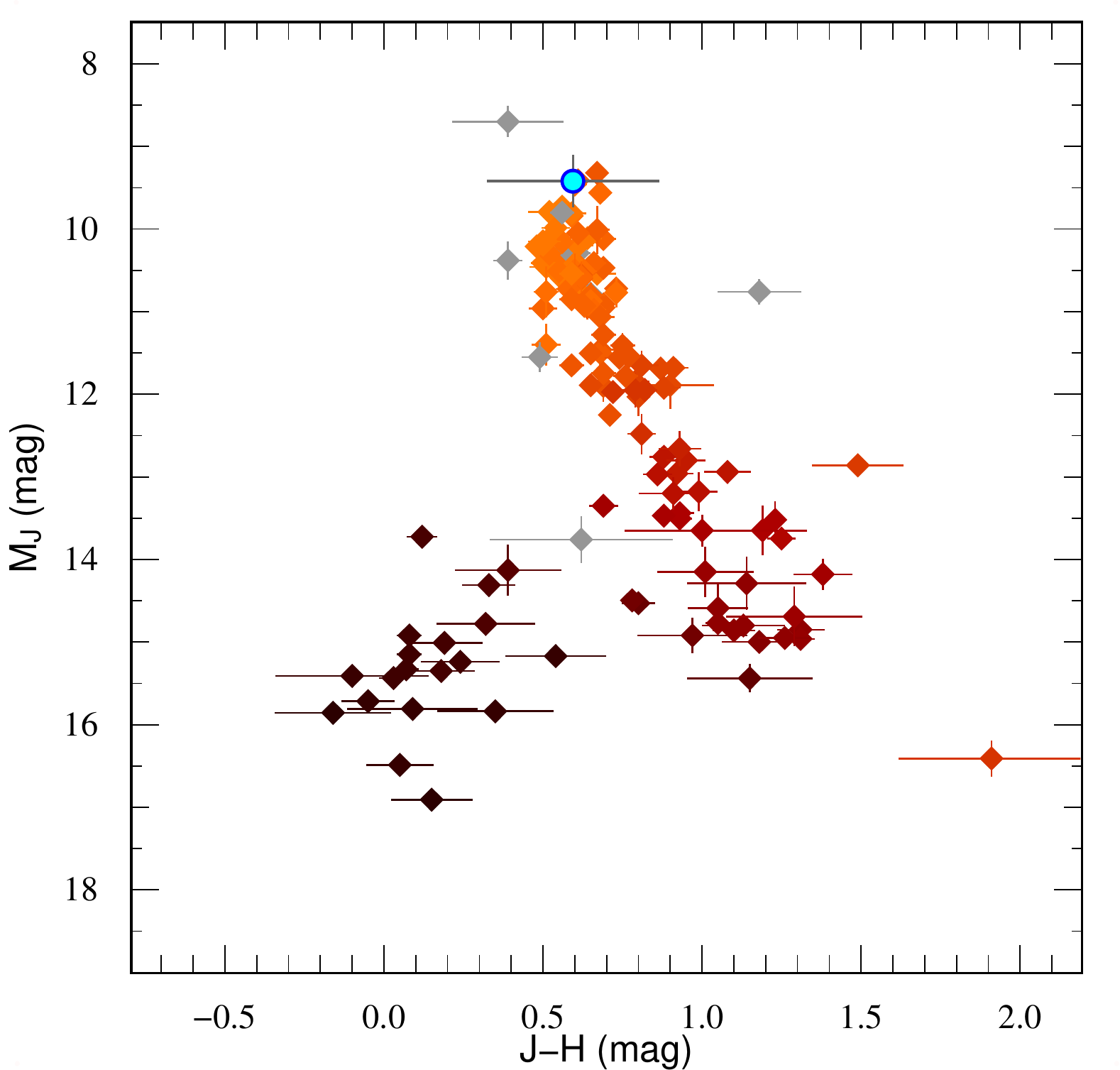}  
		\label{fig:M_J_a}  
	\end{subfigure}
	\begin{subfigure}[b]{0.33\textwidth}
		\includegraphics[width=\textwidth]{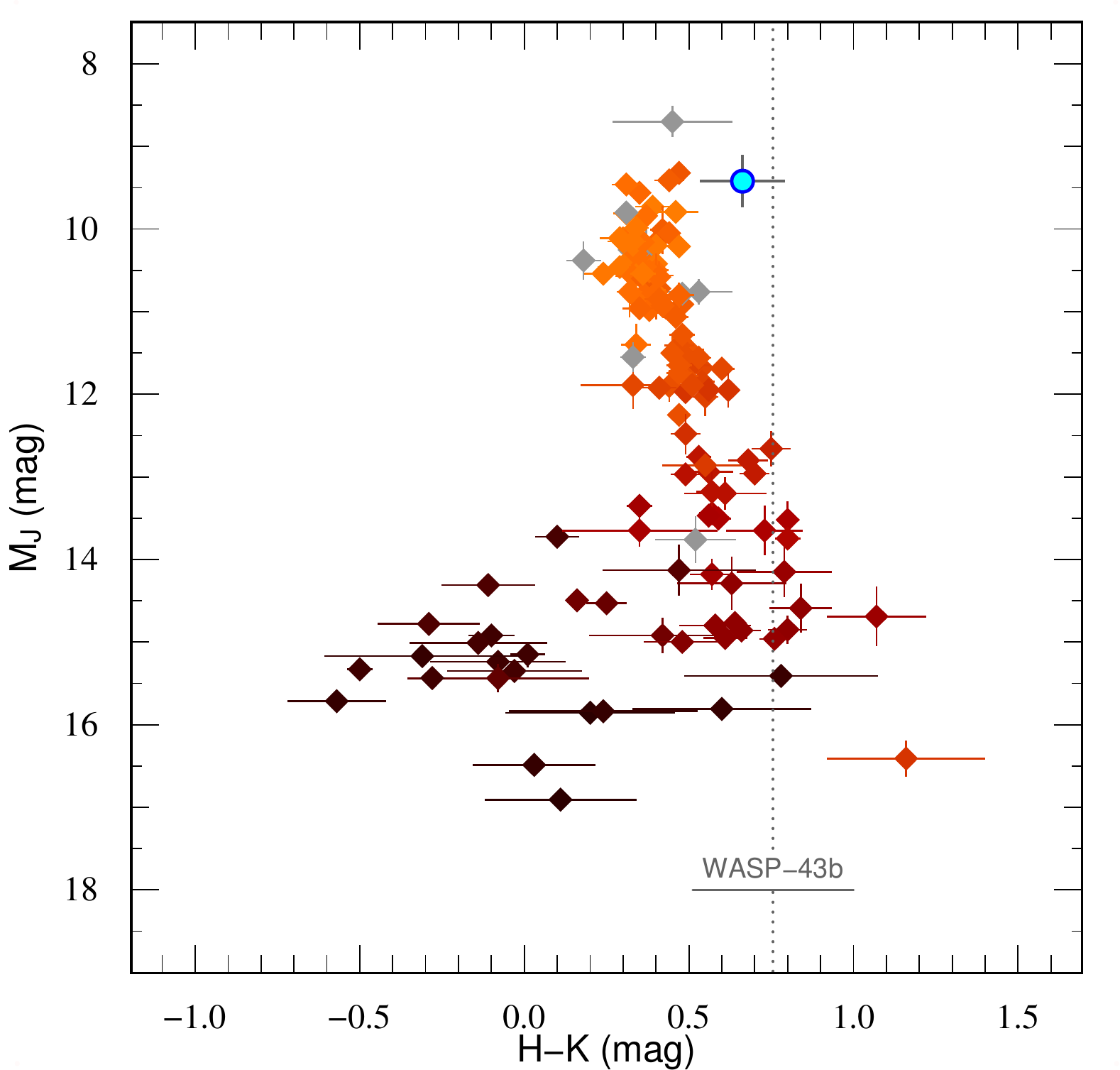}  
		\label{fig:M_J_b}  
	\end{subfigure}
	\begin{subfigure}[b]{0.33\textwidth}
		\includegraphics[width=\textwidth]{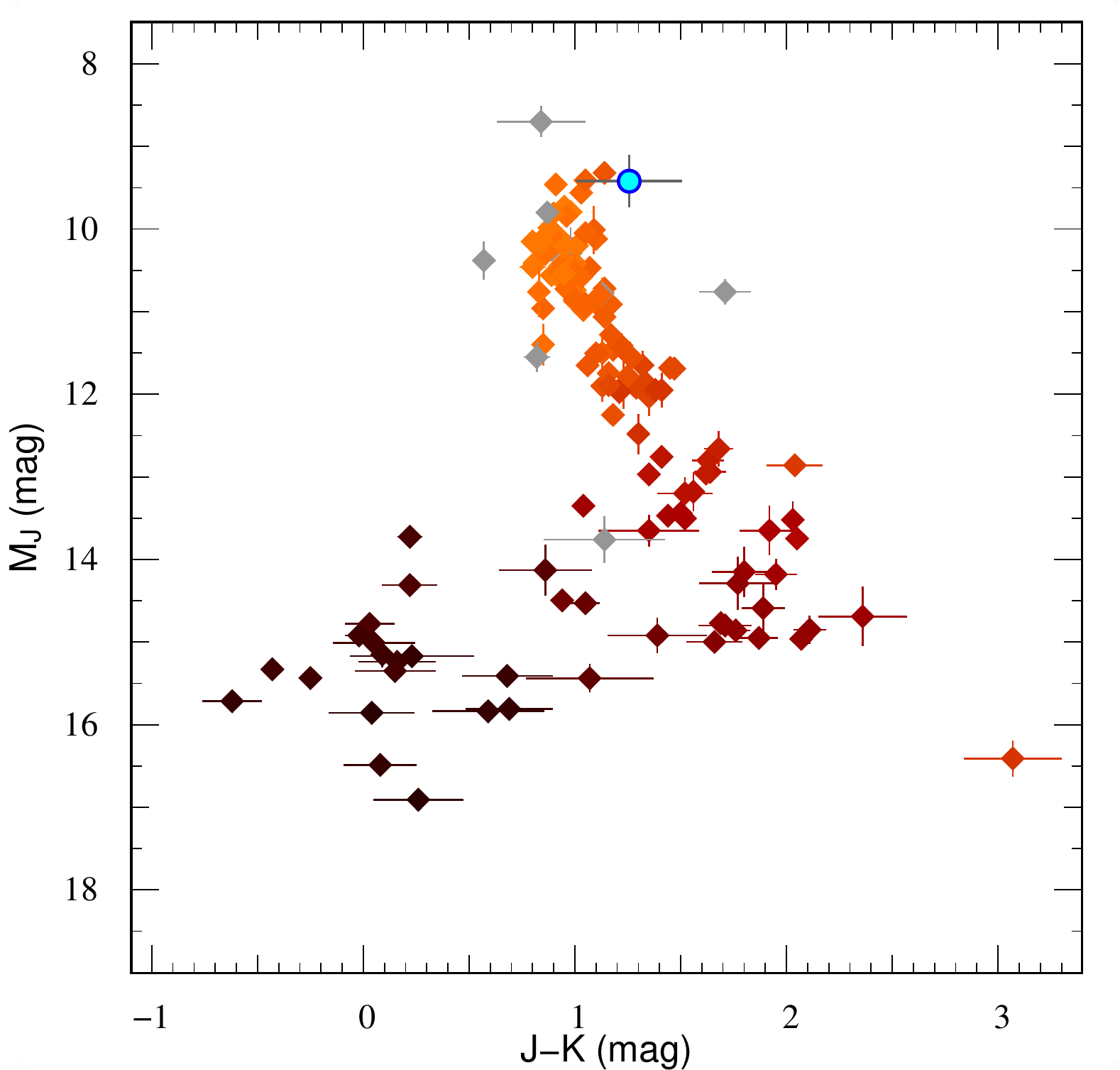}  
		\label{fig:M_J_c}  
	\end{subfigure}

	\begin{subfigure}[b]{0.33\textwidth}
		\includegraphics[width=\textwidth]{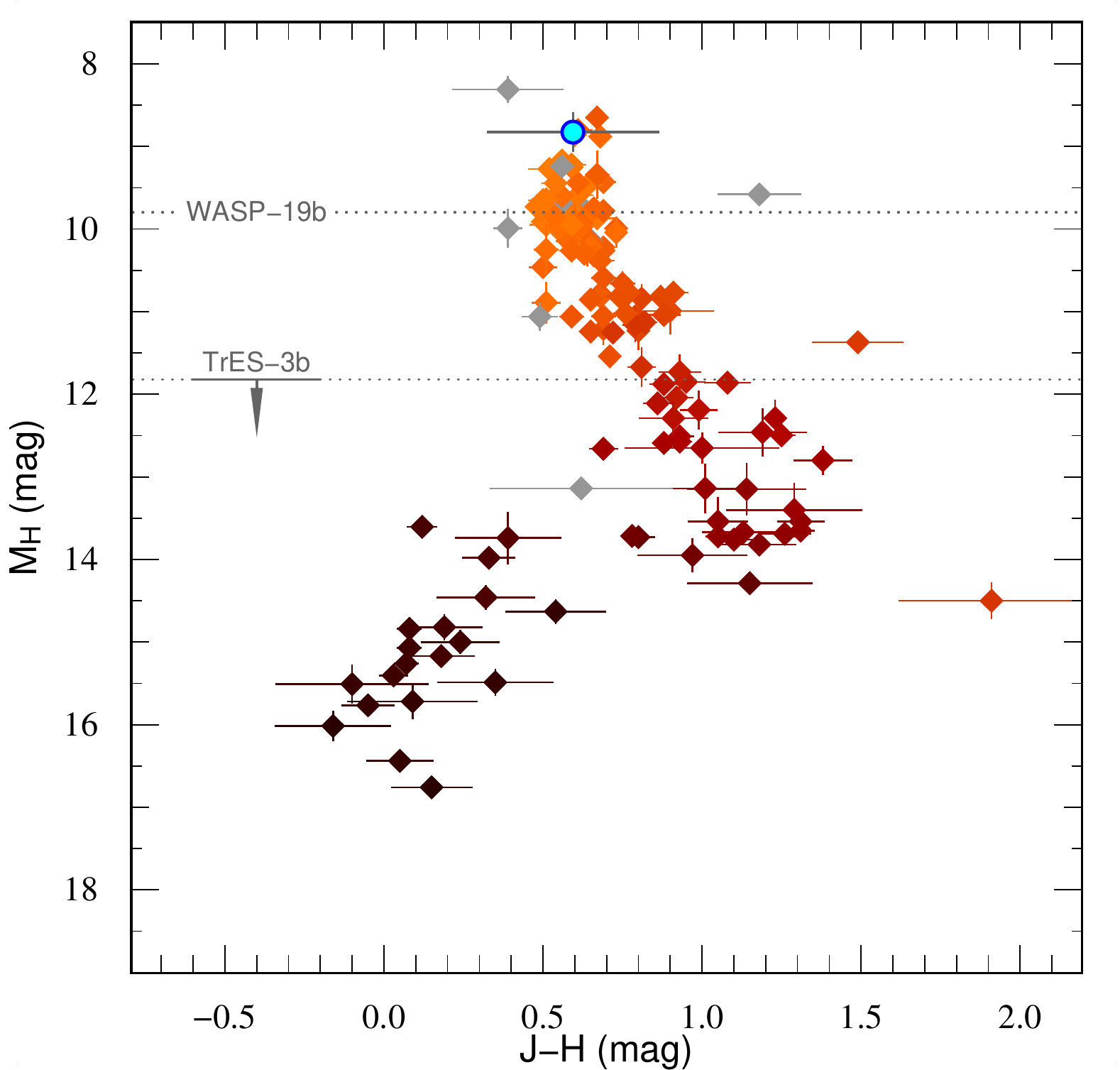}  
		\label{fig:M_H_a}  
	\end{subfigure}
	\begin{subfigure}[b]{0.33\textwidth}
		\includegraphics[width=\textwidth]{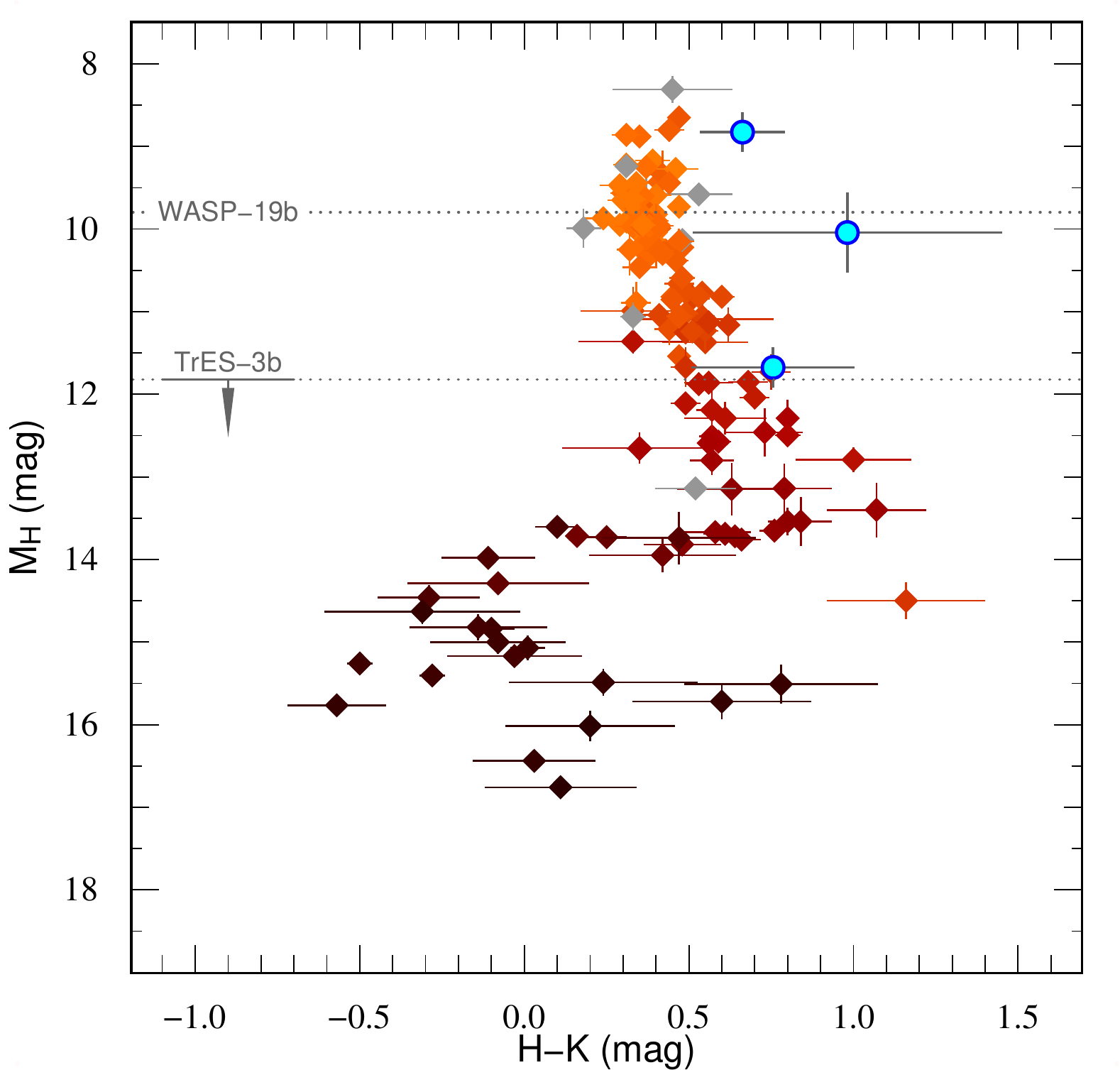}  
		\label{fig:M_H_b}  
	\end{subfigure}
	\begin{subfigure}[b]{0.33\textwidth}
		\includegraphics[width=\textwidth]{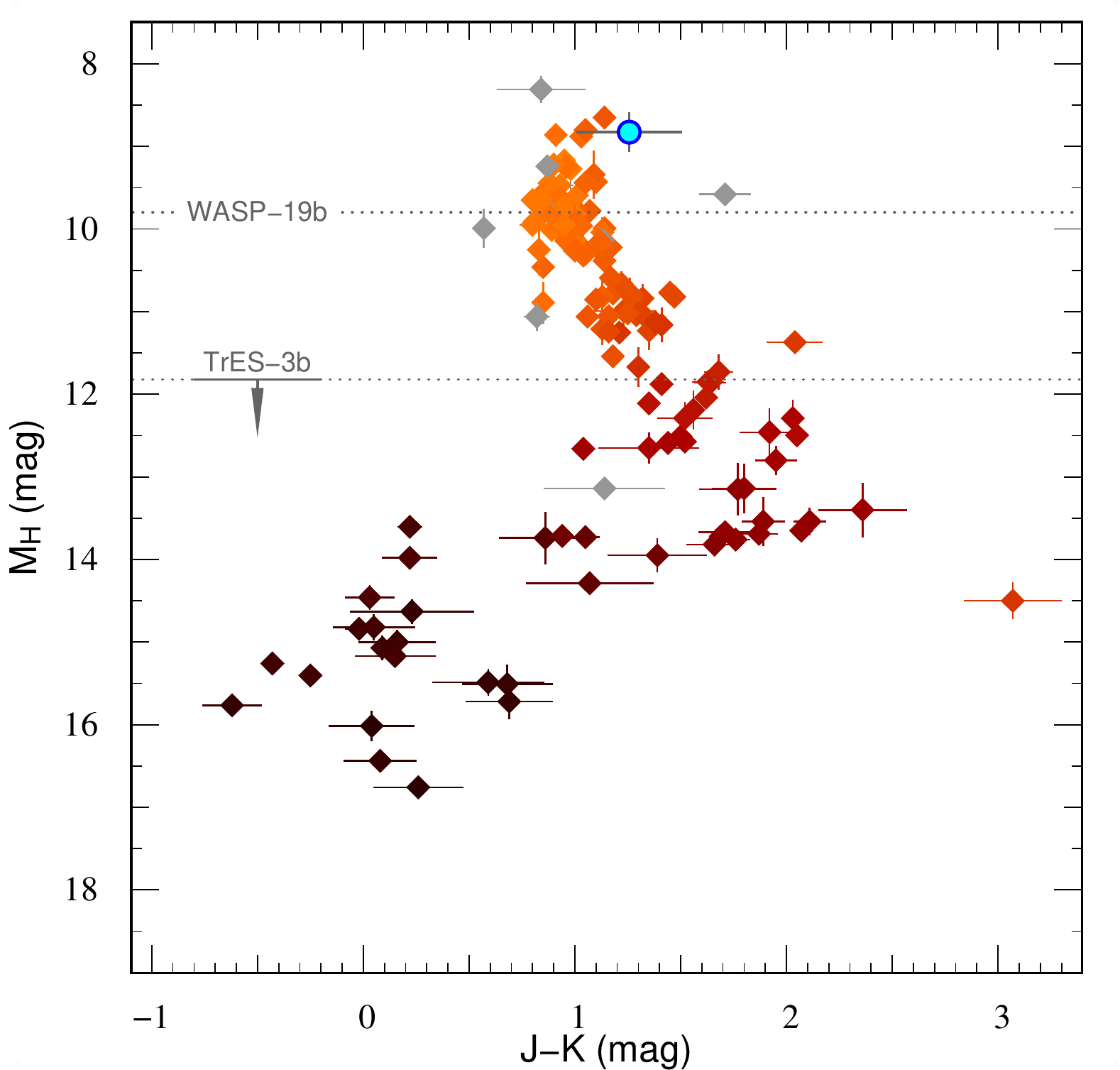}  
		\label{fig:M_H_c}  
	\end{subfigure}

	\begin{subfigure}[b]{0.33\textwidth}
		\includegraphics[width=\textwidth]{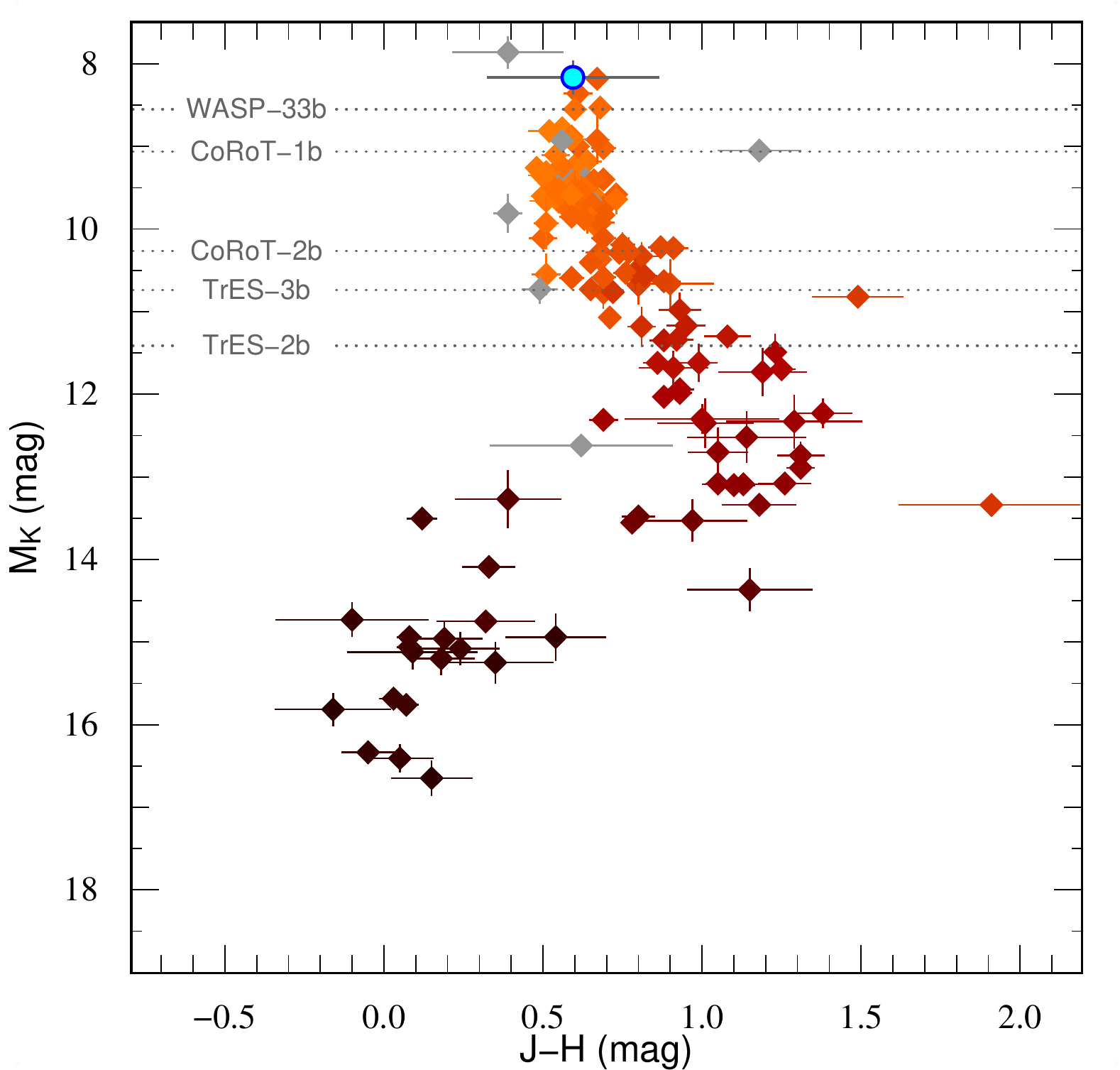}  
		\label{fig:M_K_a}  
	\end{subfigure}
	\begin{subfigure}[b]{0.33\textwidth}
		\includegraphics[width=\textwidth]{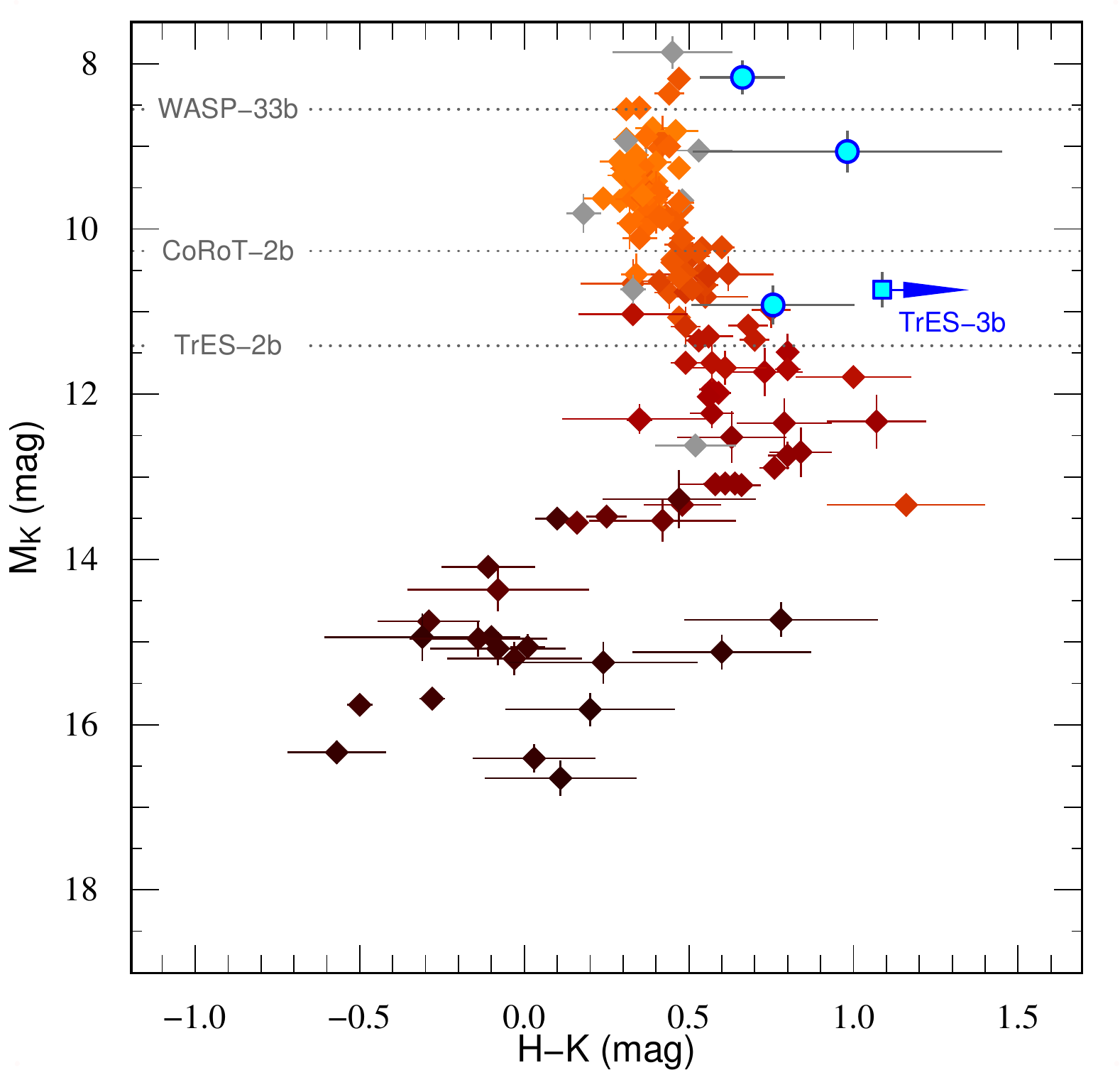}  
		\label{fig:M_K_b}  
	\end{subfigure}
	\begin{subfigure}[b]{0.33\textwidth}
		\includegraphics[width=\textwidth]{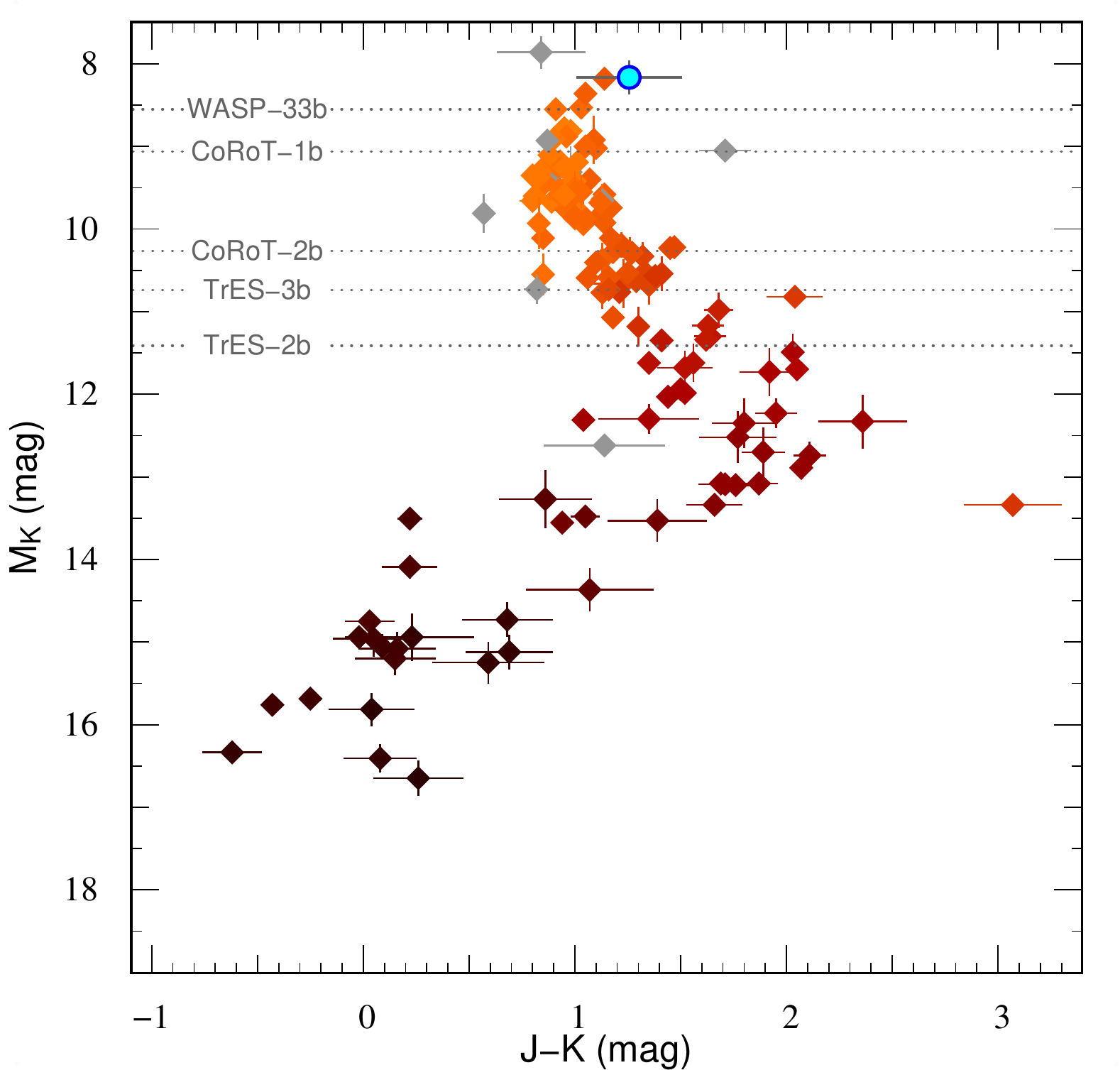}  
		\label{fig:M_K_c}  
	\end{subfigure}
	\begin{subfigure}[b]{0.9\textwidth}
		\includegraphics[width=\textwidth]{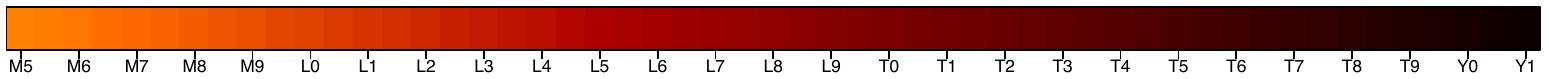}  
		\label{fig:bar}  
	\end{subfigure}

	\caption{Near-infrared colour-magnitude diagrams, using the 2MASS photometric system (i.e., the J, H and K$_{\rm S}$ bands). The blue dots show the dayside emission of transiting planets observed during occultation. Squares and arrows represent upper limits. Lines labelled with the name of a planet show the position of systems { where colour or absolute magnitude is missing} (not all cases are represented, for clarity). The coloured diamonds underlying the plots are brown dwarfs and directly imaged planets, whose magnitudes are listed in \citet{Dupuy:2012lr}. Colours represent the spectral class of the object, spanning from M5 (orange) to Y1 (black). Unclassified objects are in grey.
}\label{fig:NIR}  
\end{center}  
\end{figure*}

Ultra-cool dwarfs comprise very late M dwarfs and brown dwarfs. They span the M, L, T and Y spectral classes. The distinction between the M, L and T spectral classes is described by \citet{Kirkpatrick:2005th}, while the Y class is defined in \citet{Cushing:2011rf}. Covering effective temperatures ranging from roughly 2\,500 to 1\,300~K, the L-dwarf sequence is identified by the disappearance of TiO and VO absorption as those species and others condensate into dust clouds that are thickening with decreasing temperatures, causing an accrued reddening. A rapid blueward change in near-infrared colours for objects with similar effective temperature outlines the transition between spectral classes L7 to T4 (Fig.~\ref{fig:NIR}). This colour variation is interpreted as the disappearance of suspended dust from the photosphere. The process through which these condensates  of atomic and molecular species vanish is the scene of very active research. \citet{Tsuji:2002qf}, \citet{Marley:2002yu} and \citet{Knapp:2004lq} proposed that as the atmosphere cools it reaches a temperature at which dust sedimentation efficiency increases dramatically producing a drain of the cloud decks via a ``sudden downpour". \citet{Ackerman:2001fk} and \citet{Burgasser:2002kx} instead proposed that, very much like what can be observed on Jupiter where clouds are discretised in separate bands, brown dwarfs' silicate clouds could fragment and progressively reveal the deeper, hotter regions of the atmosphere. This scenario produces clear signatures, such as photometric variability caused by inhomogenous structures rotating in and out of view. Those are being detected on an increasing number of { brown dwarfs \citep{Artigau:2009lr,Radigan:2012rt,Heinze:2013lr,Radigan:2014zr}, with some contention which spectral types are more likely to vary and about what causes variability} \citep{Wilson:2014mz}. One could also expect near stochastic modulations like has been noticed on Luhman-16B by \citet{Gillon:2013qv}. Further observations confirmed the presence of patchy clouds on Luhman-16B \citep{Crossfield:2014db}. From spectral type T5 and beyond, atmospheres are thought to be clear and continue to cool down. T dwarfs have effective temperatures between 1\,500 and $\sim$ 600 K. The transition to the Y-class is defined by the appearance of ammonia and the disappearance of alkali lines produced by the condensation of sodium and potassium.

Interestingly, transiting planets, most often hot Jupiters, have dayside magnitudes, brightness temperatures and colours that overlap with the entire ultra-cool dwarf range. For instance, WASP-12Ab, the intrinsically brightest planet in the current sample, is as hot as an M6 dwarf. Its inferred size is as large as a 0.16~$M_\odot$ star \citep{Baraffe:1998ly}. This would allow in principle to draw parallels between planets and ultra-cool dwarfs, especially so{ , since mass regimes of field brown dwarfs and extrasolar planets are overlapping} \citep{Latham:1989yq,Chauvin:2004lr,Caballero:2007mz,Deleuil:2008lr,Marois:2008fk,Hellier:2009fj,Sahlmann:2011qy,Siverd:2012ef,Diaz:2013fr,Delorme:2013cr,Naud:2014kk}.

\subsection{Near-infrared}

The J, H and K$_{\rm S}$ bands colour-magnitude diagrams contain a large number of field brown dwarfs (see \citet{Dupuy:2012lr} and references therein) but very few planets. Each of Figure \ref{fig:NIR}'s panels contain WASP-12Ab, the only planet with firm detections of its emission in each of those near-infrared bands { (M$_{\rm J} = 9.42$, M$_{\rm H} = 8.83$, M$_{\rm K_s} = 8.16$)}. A few more measurements were obtained on individual systems, but often in only one band (depicted as dotted lines). WASP-12Ab's location seems to agree well with the top of the ultra-cool dwarf distribution especially in the J$-$H colour. The two colours involving the K$_{\rm S}$-band would imply that the object is redder than most late M-dwarfs. However, a recent work by \citet{Rogers:2013zr} showed that eclipse depth measurements, notably in the K$_{\rm S}$ bandpass are likely to be biased towards deeper values. This in turns would make authors infer brighter planets, leading to a smaller magnitude and a redder colour index. \citet{Bean:2013pd} observed WASP-19b at low spectral resolution and consistently found shallower occultation depths than broad band measurements would imply.

It remains unclear whether irradiated planetary atmosphere should follow the same general behaviour that very low-mass stars and field brown dwarfs have, whether they would constitute their own sequence or agree with a blackbody (see Sec.~\ref{sec:bb} for a discussion on the matter). If indeed, irradiated planets and ultra-cool dwarfs were to coincide, then positioning a new measurement in a colour-magnitude diagram will become an efficient method to verify anyone's results. For instance, it can immediately be noticed how most K$_{\rm S}$ bands results imply redder colours than would otherwise be anticipated. 

By extension, obtaining a detection in one band would offer straight-forward predictions for the other two bands.  As an example, WASP-19b has an absolute magnitude in the H-band, $M_{\rm H} = 9.80 \pm 0.21$ (Tab. \ref{tab:planets}). Reading on the M$_{\rm H}$ vs J$-$H plot, we notice its magnitude intersects with the M \& L-dwarf sequence at J$-$H$=0.6\pm0.1$. This leads to $M_{\rm J} = 10.40 \pm 0.23$, that we can convert into an apparent magnitude. WASP-19b can be predicted to have $m_{\rm J} = 17.60 \pm 0.21$, on a par with WASP-12Ab's  measurement (Tab.~\ref{tab:app_pl}).

\begin{figure*}  
\begin{center}  
	\begin{subfigure}[b]{0.24\textwidth}
		\includegraphics[width=\textwidth]{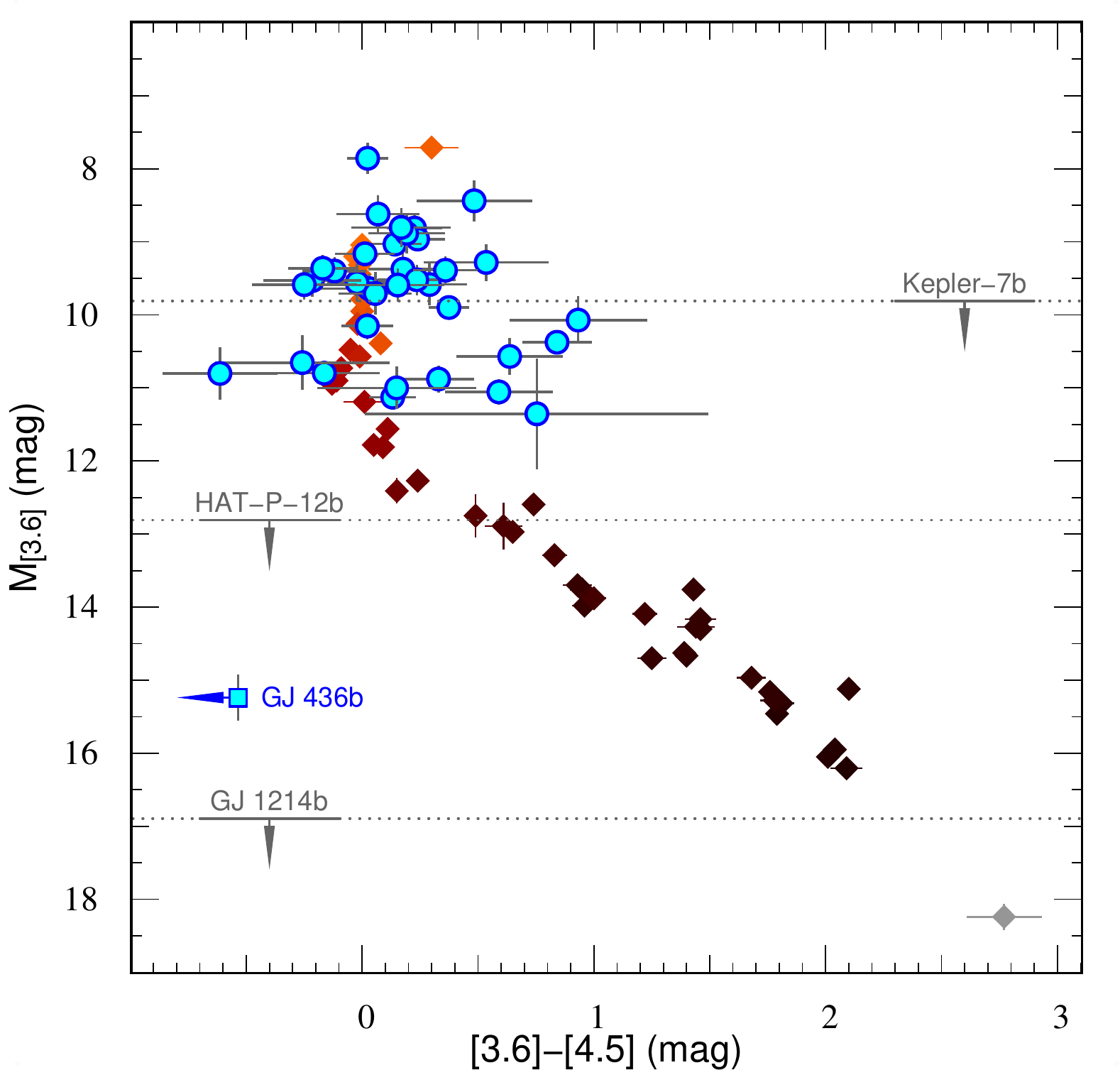}  
		\label{fig:M_3p6_a}  
	\end{subfigure}
	\begin{subfigure}[b]{0.24\textwidth}
		\includegraphics[width=\textwidth]{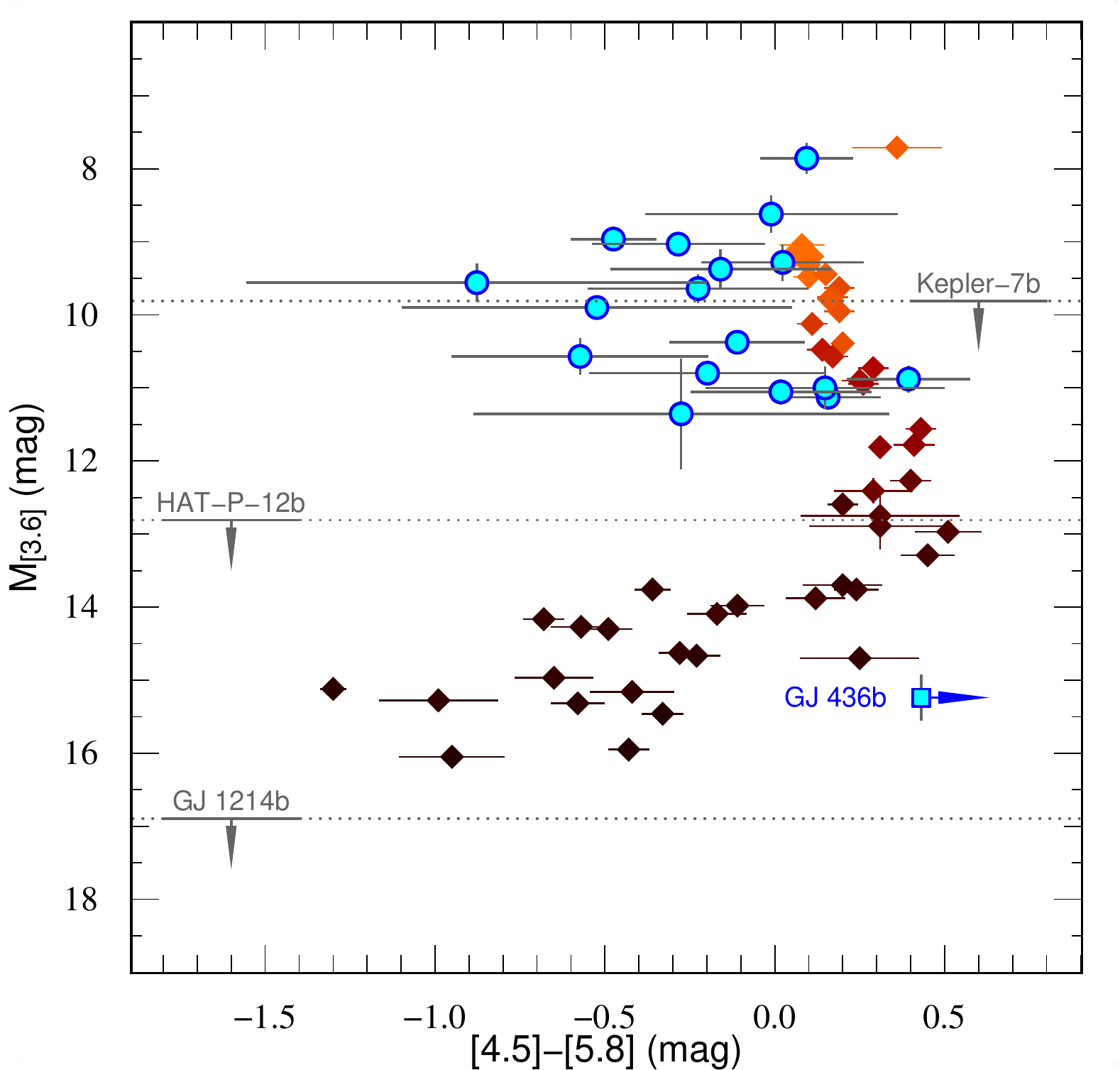}  
		\label{fig:M_3p6_b}  
	\end{subfigure}
	\begin{subfigure}[b]{0.24\textwidth}
		\includegraphics[width=\textwidth]{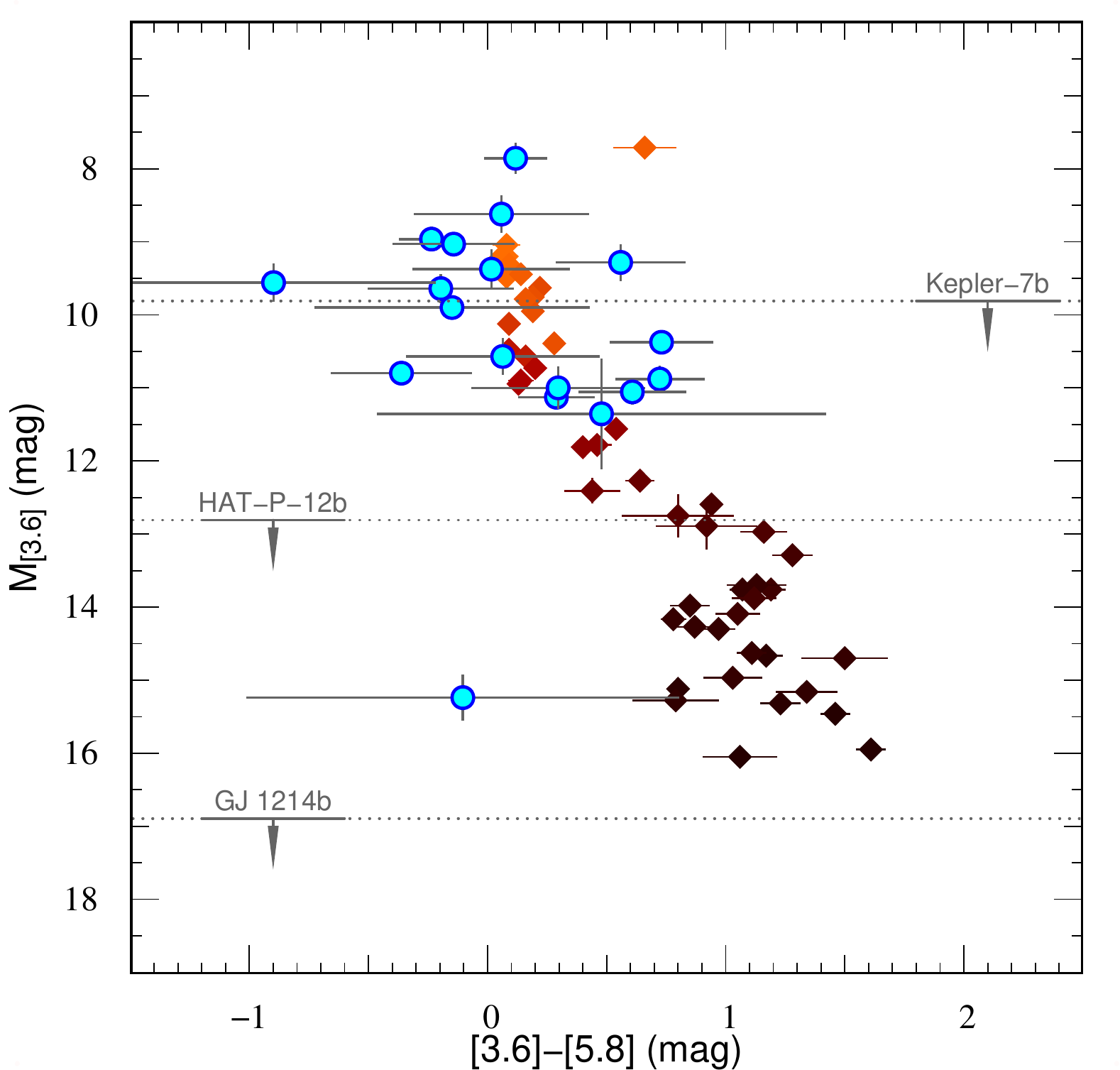}  
		\label{fig:M_3p6_c}  
	\end{subfigure}
	\begin{subfigure}[b]{0.24\textwidth}
		\includegraphics[width=\textwidth]{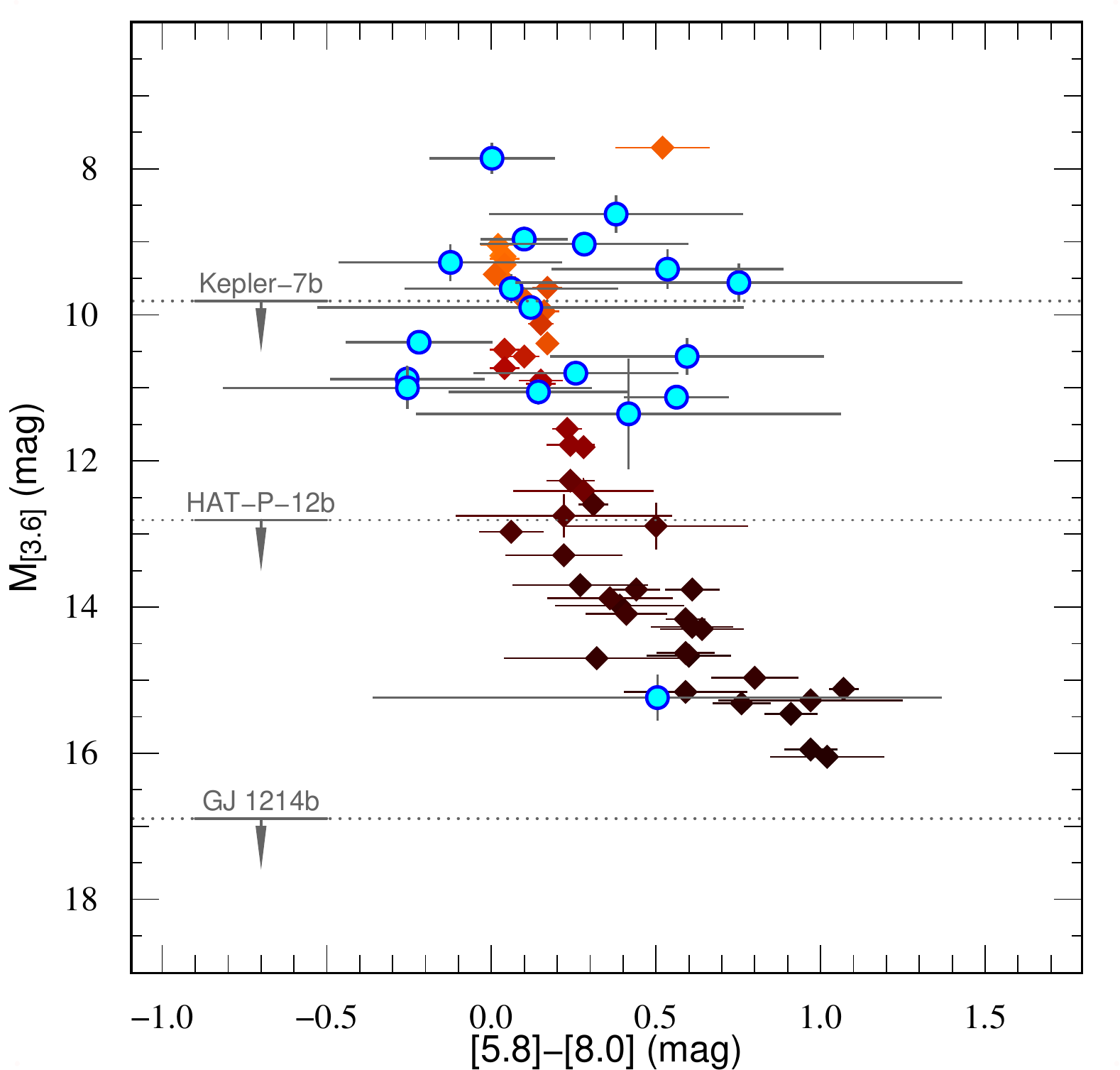}  
		\label{fig:M_3p6_d}  
	\end{subfigure}

	\begin{subfigure}[b]{0.24\textwidth}
		\includegraphics[width=\textwidth]{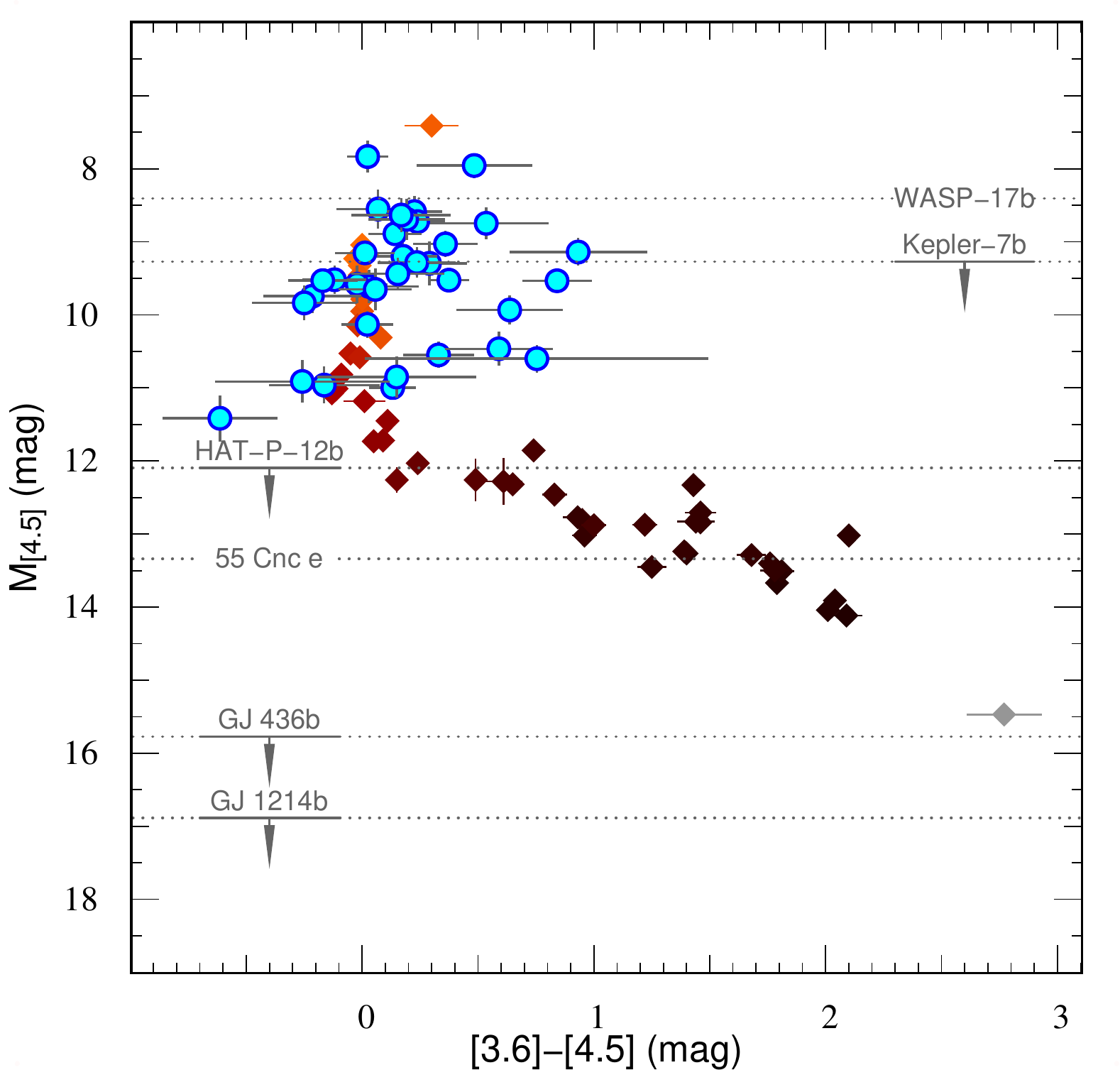}  
		\label{fig:M_4p5_a}  
	\end{subfigure}
	\begin{subfigure}[b]{0.24\textwidth}
		\includegraphics[width=\textwidth]{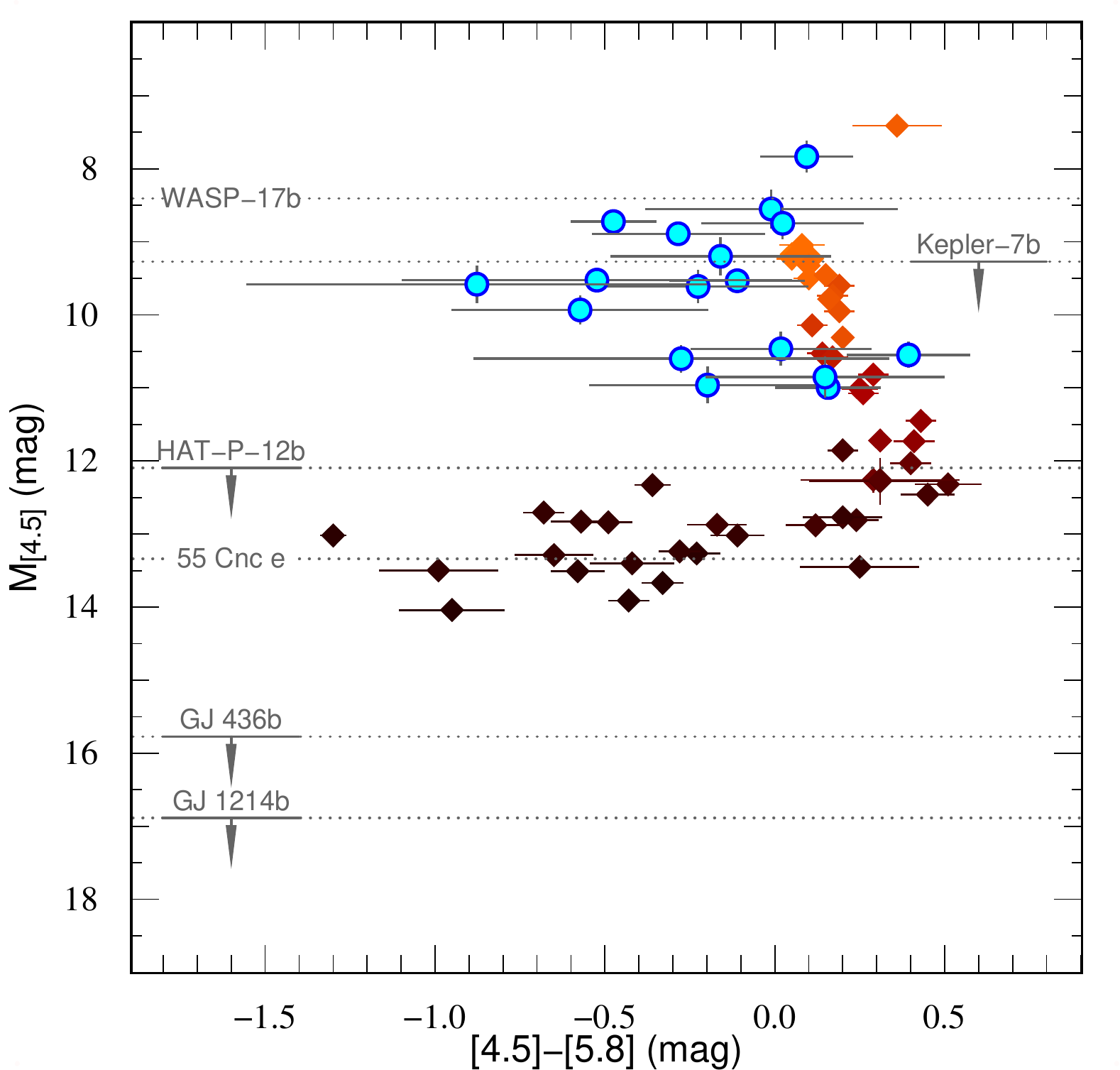}  
		\label{fig:M_4p5_b}  
	\end{subfigure}
	\begin{subfigure}[b]{0.24\textwidth}
		\includegraphics[width=\textwidth]{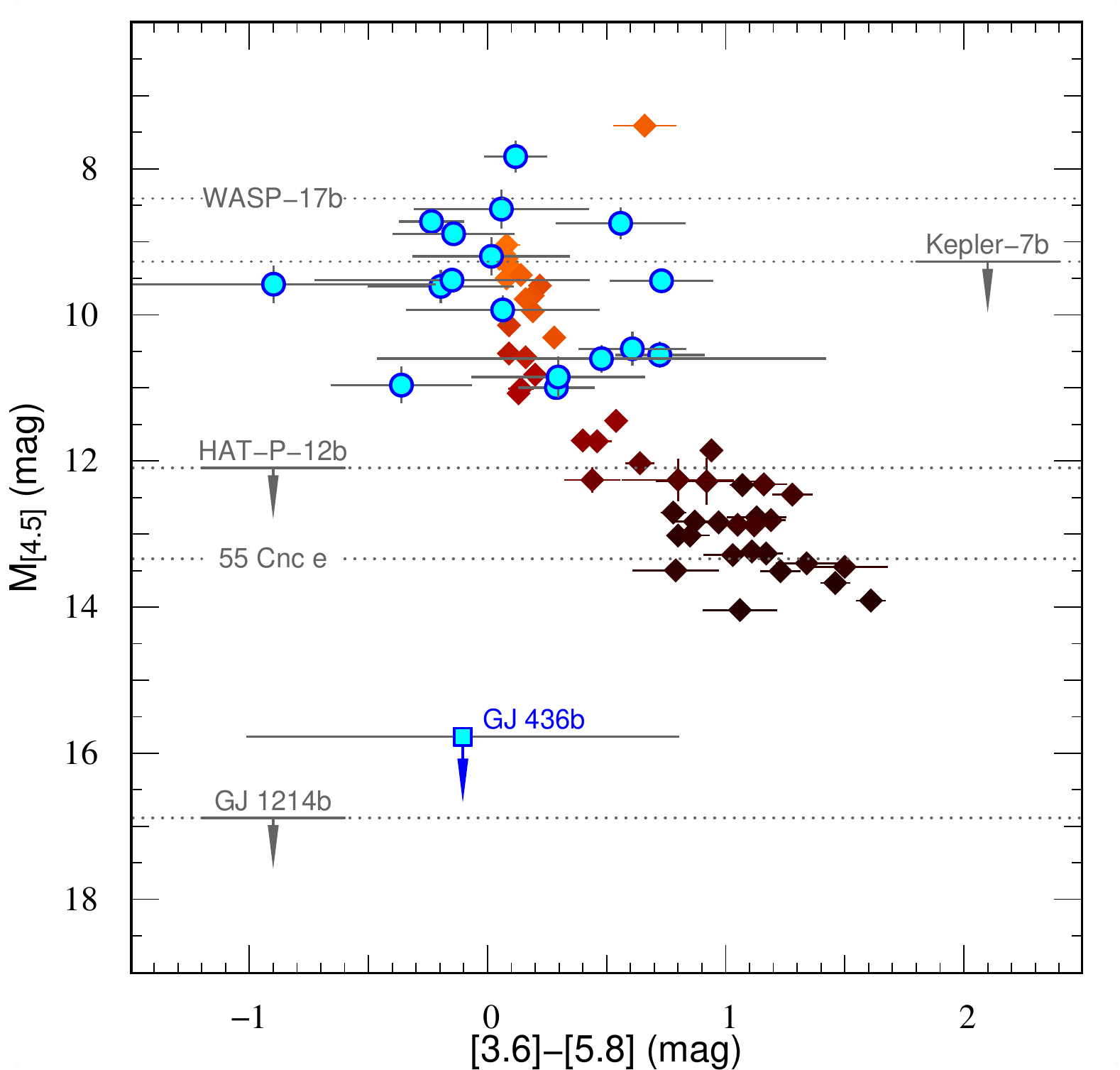}  
		\label{fig:M_4p5_c}  
	\end{subfigure}
	\begin{subfigure}[b]{0.24\textwidth}
		\includegraphics[width=\textwidth]{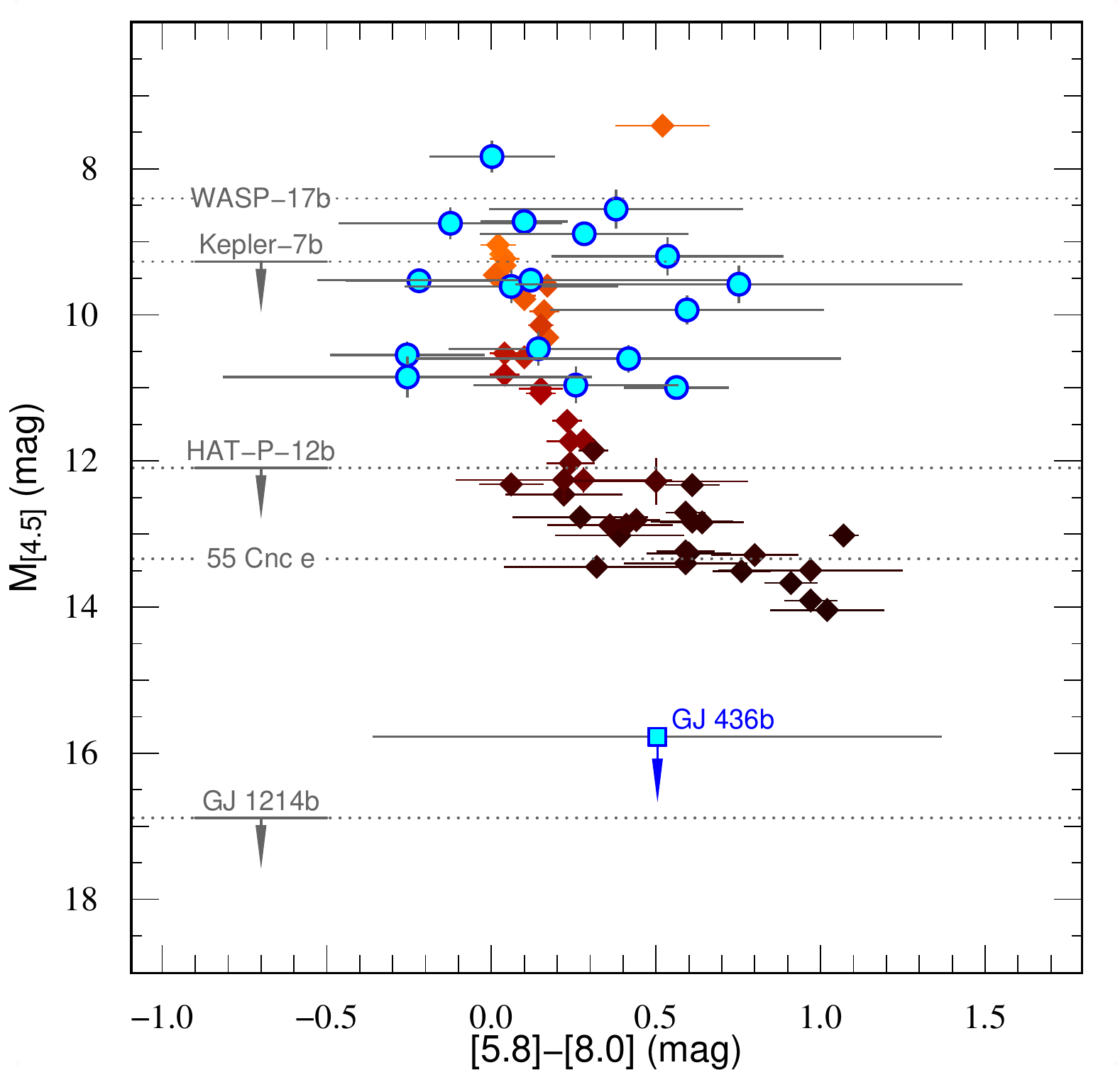}  
		\label{fig:M_4p5_d}  
	\end{subfigure}

	\begin{subfigure}[b]{0.24\textwidth}
		\includegraphics[width=\textwidth]{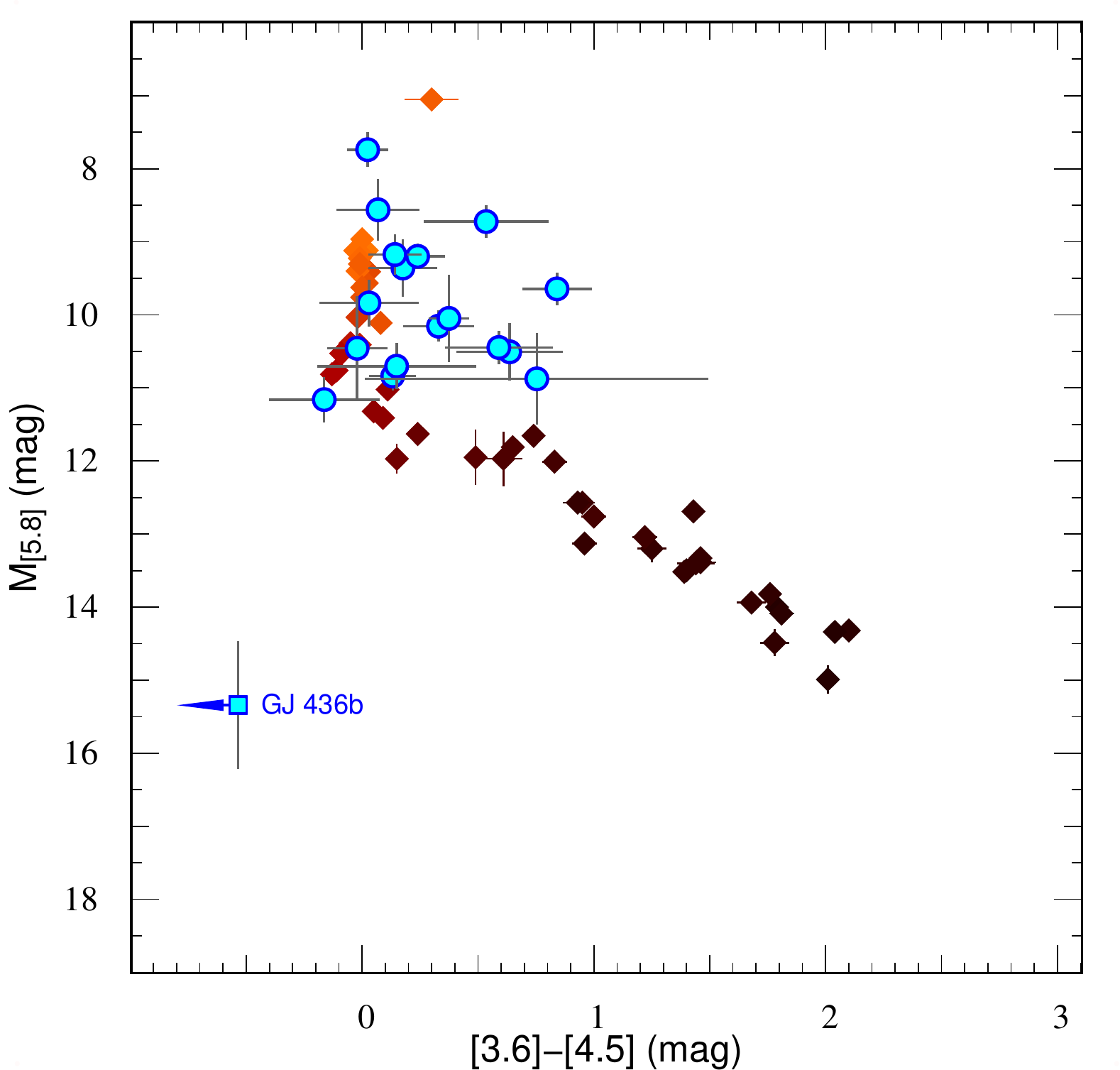}  
		\label{fig:M_5p8_a}  
	\end{subfigure}
	\begin{subfigure}[b]{0.24\textwidth}
		\includegraphics[width=\textwidth]{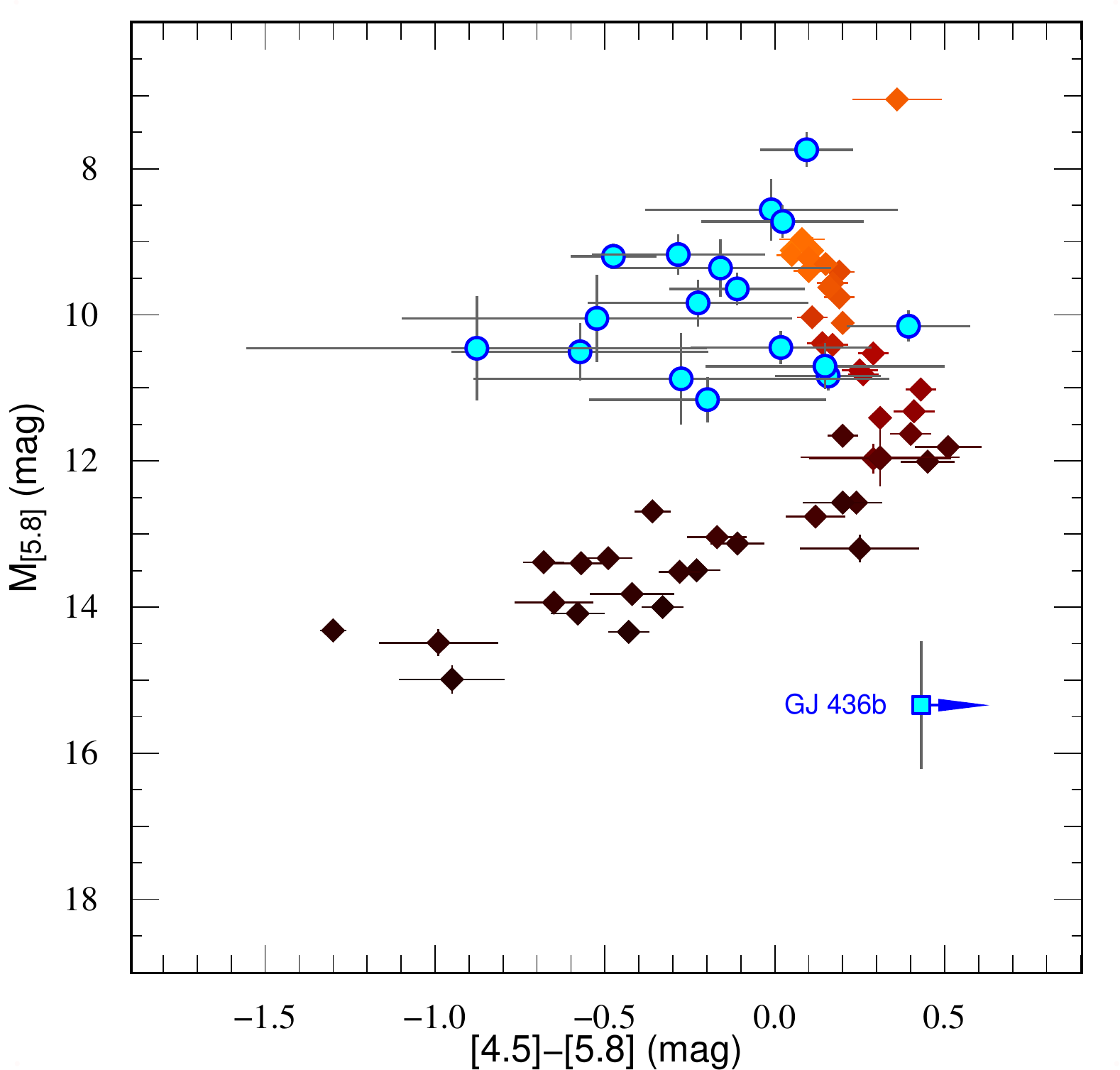}  
		\label{fig:M_5p8_b}  
	\end{subfigure}
	\begin{subfigure}[b]{0.24\textwidth}
		\includegraphics[width=\textwidth]{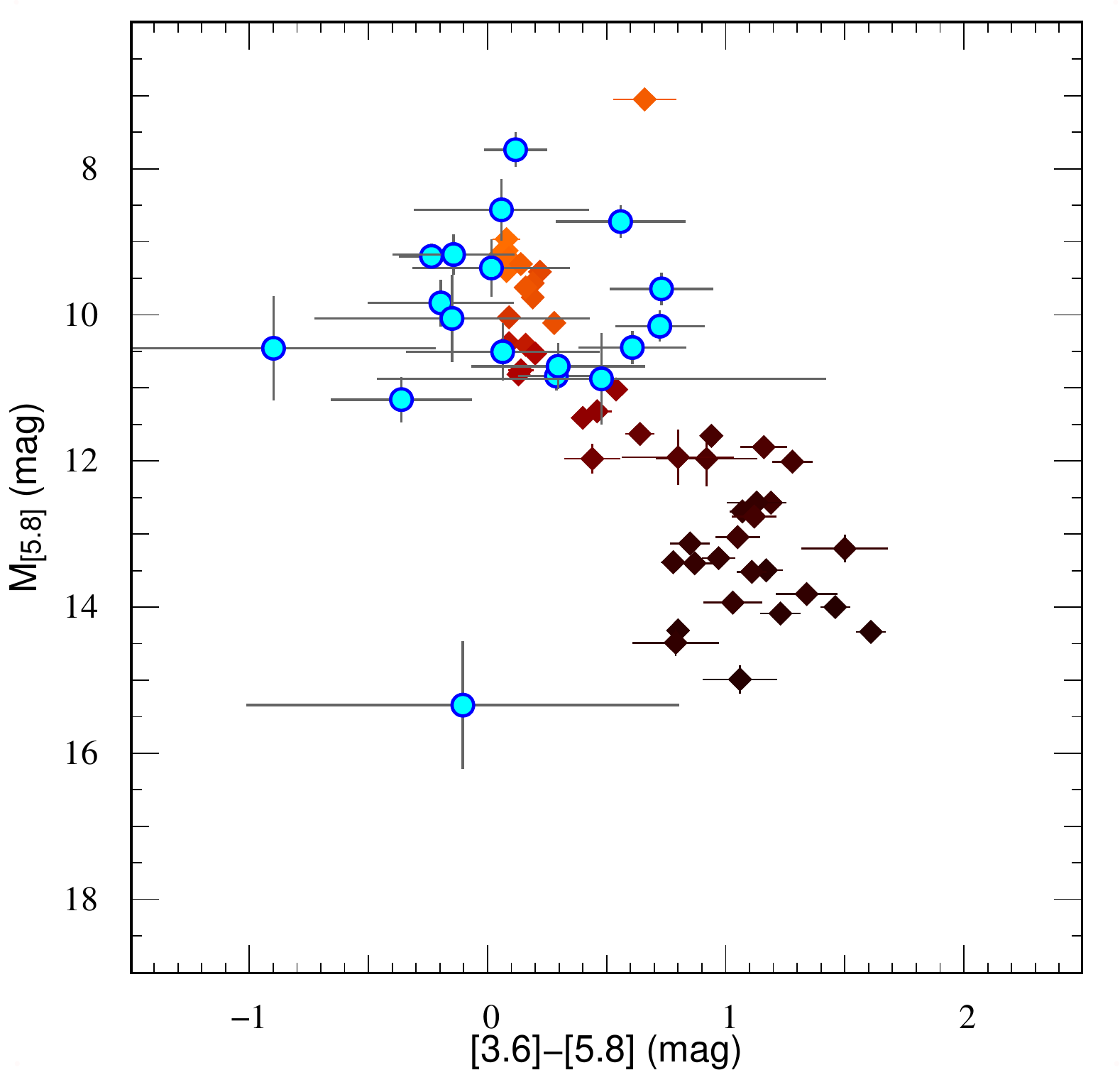}  
		\label{fig:M_5p8_c}  
	\end{subfigure}
	\begin{subfigure}[b]{0.24\textwidth}
		\includegraphics[width=\textwidth]{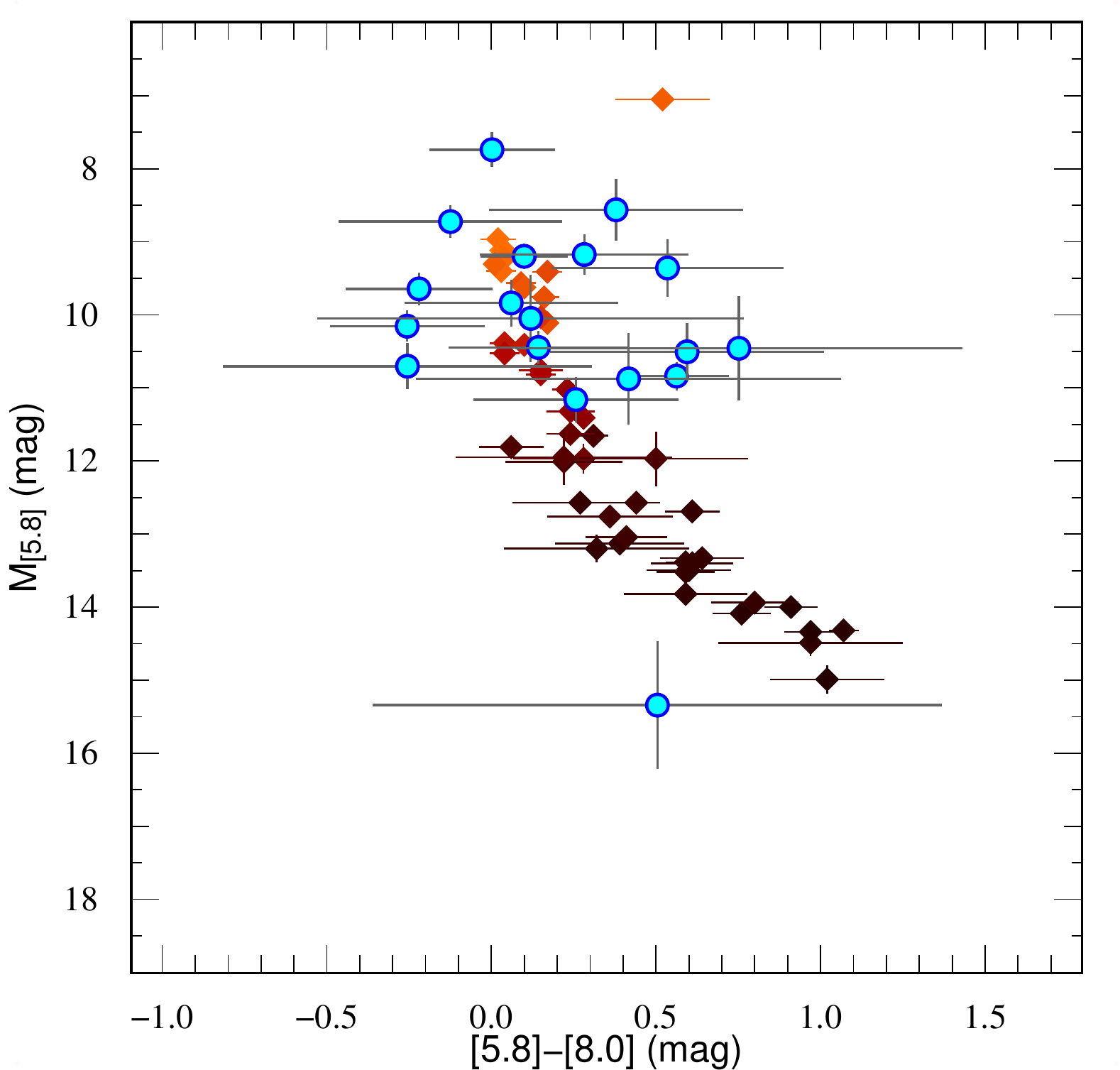}  
		\label{fig:M_5p8_d}  
	\end{subfigure}

	\begin{subfigure}[b]{0.24\textwidth}
		\includegraphics[width=\textwidth]{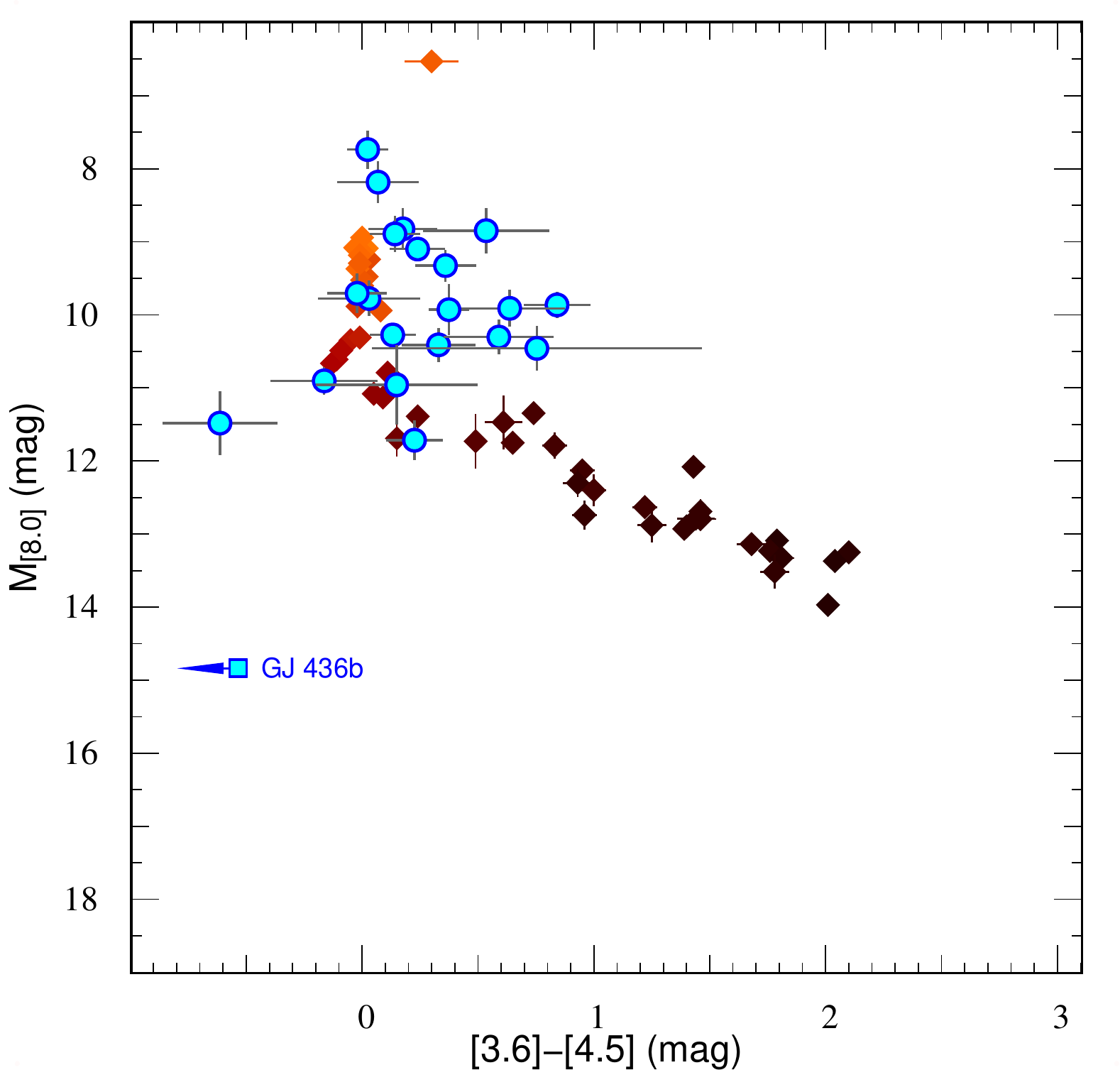}  
		\label{fig:M_8p0_a}  
	\end{subfigure}
	\begin{subfigure}[b]{0.24\textwidth}
		\includegraphics[width=\textwidth]{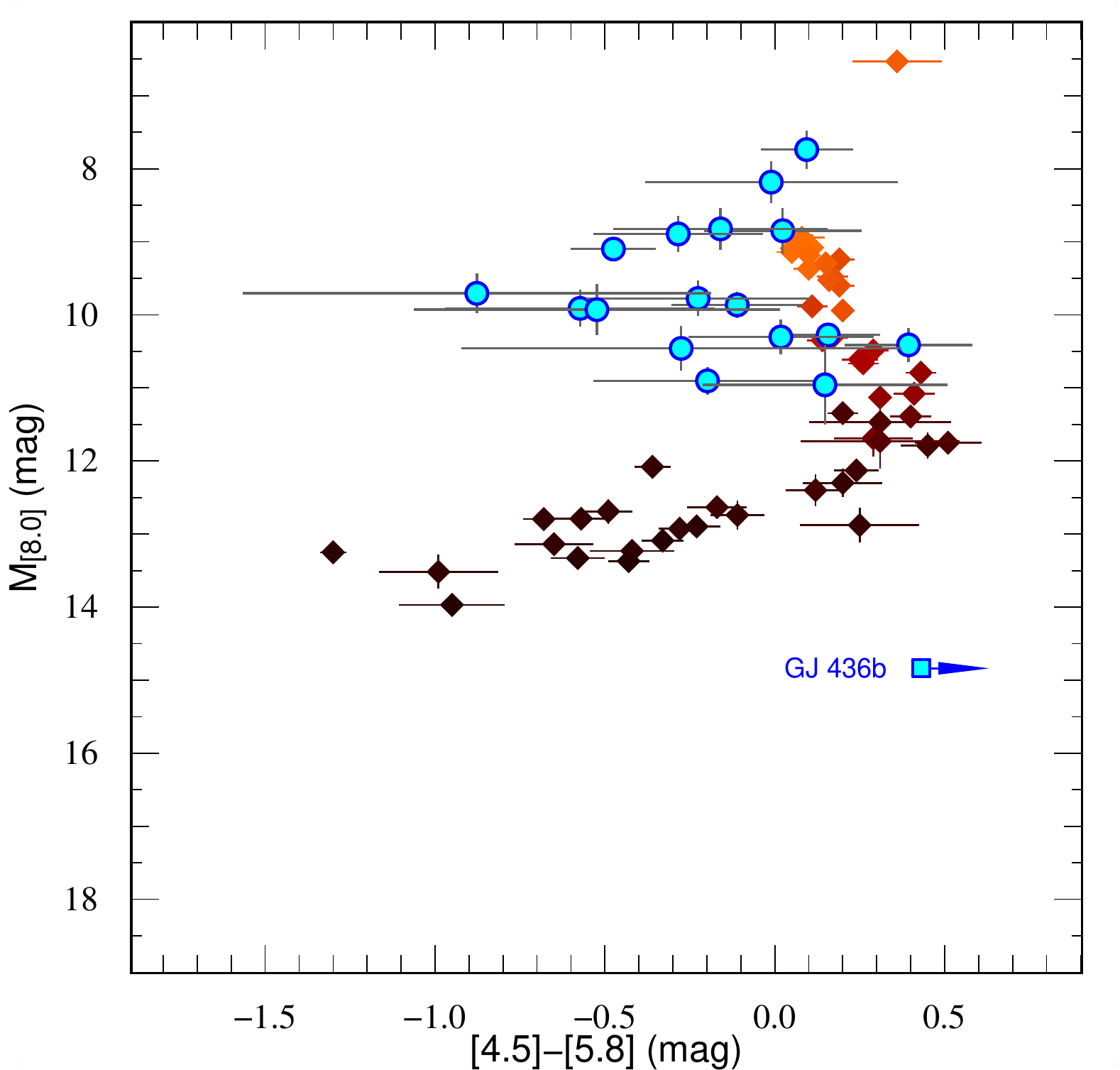}  
		\label{fig:M_8p0_b}  
	\end{subfigure}
	\begin{subfigure}[b]{0.24\textwidth}
		\includegraphics[width=\textwidth]{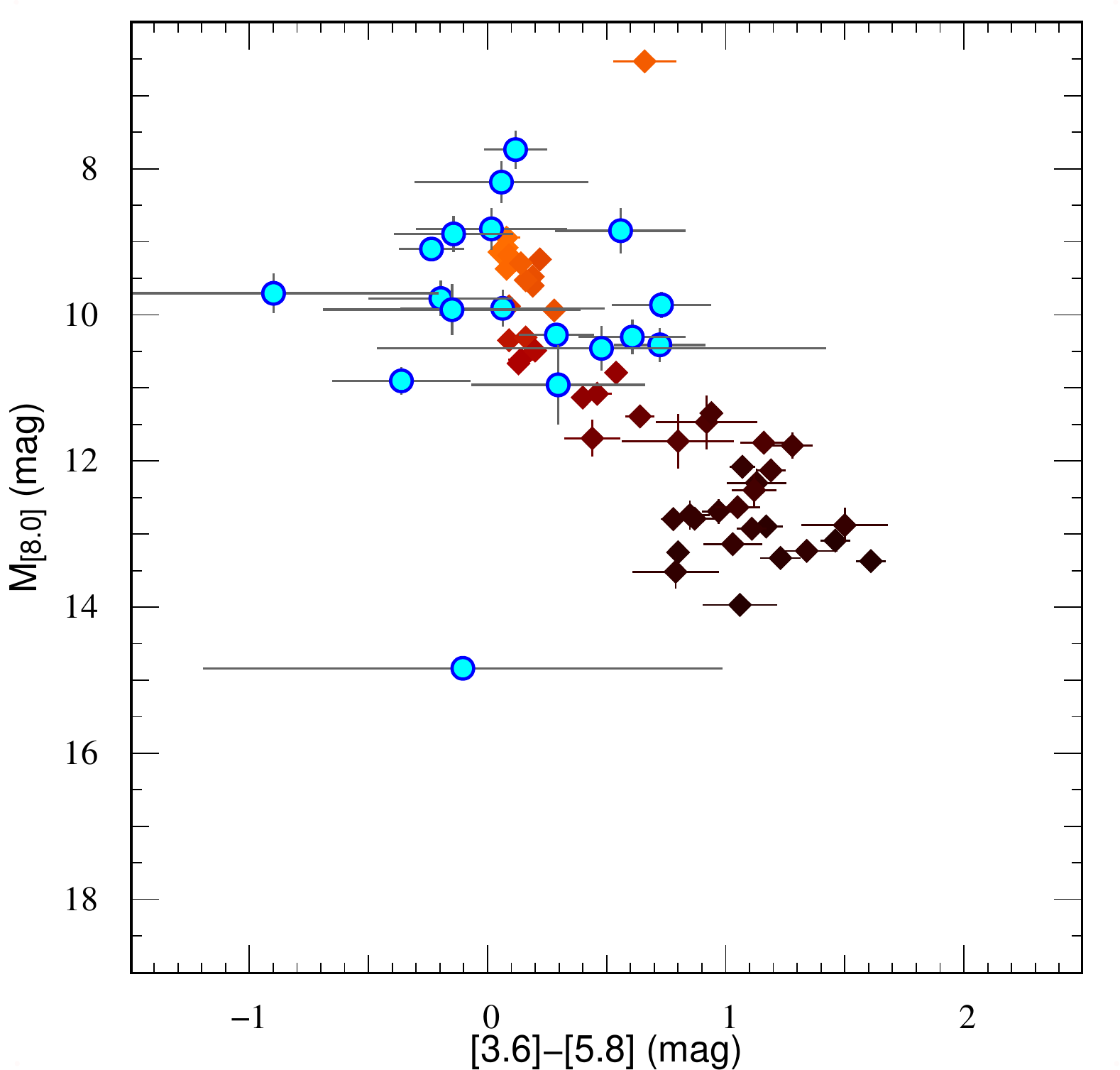}  
		\label{fig:M_8p0_c}  
	\end{subfigure}
	\begin{subfigure}[b]{0.24\textwidth}
		\includegraphics[width=\textwidth]{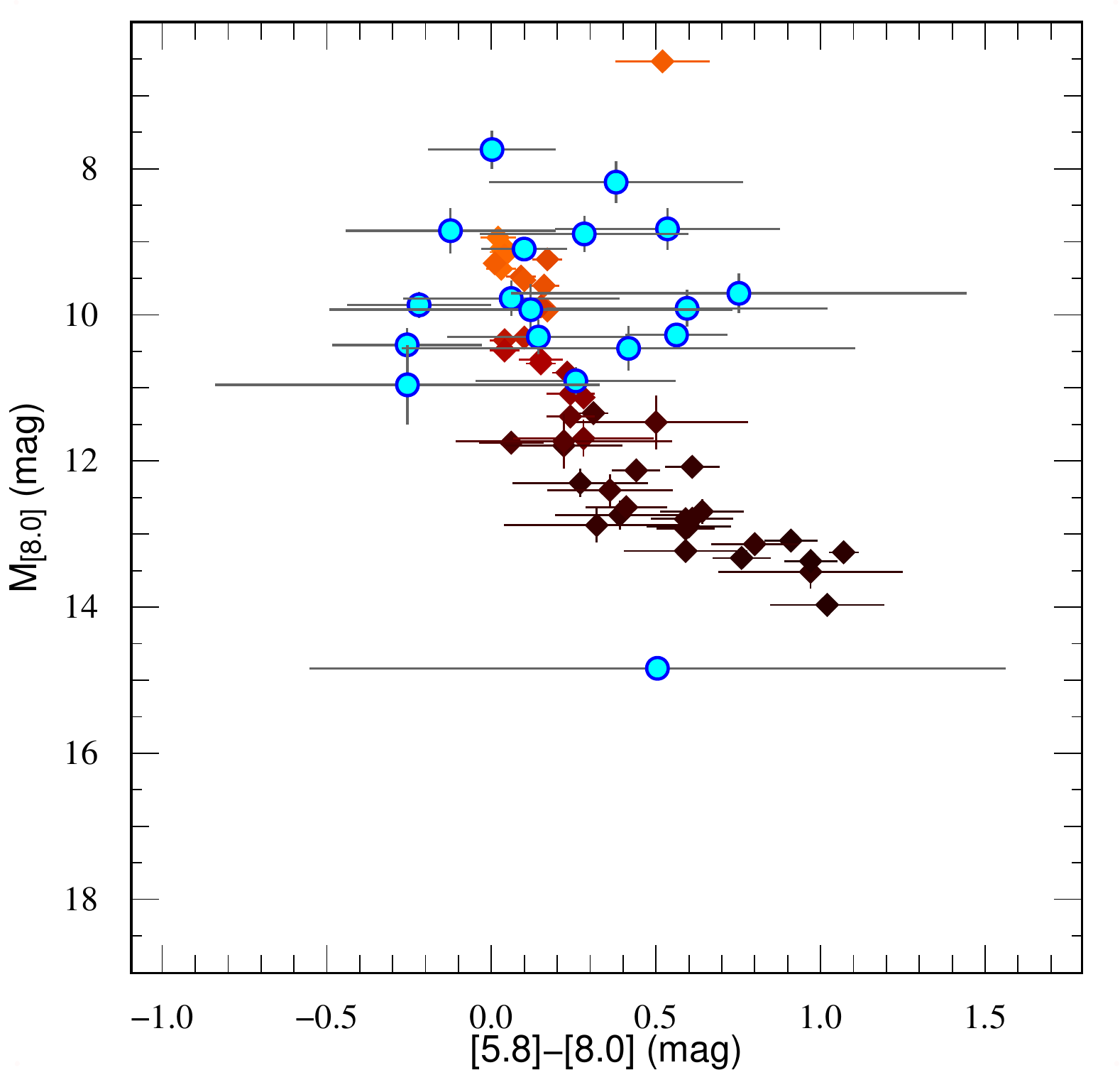}  
		\label{fig:M_8p0_d}  
	\end{subfigure}
	\begin{subfigure}[b]{0.9\textwidth}
		\includegraphics[width=\textwidth]{bar2S.pdf}  
		\label{fig:bar}  
	\end{subfigure}

	\caption{Mid-infrared colour-magnitude diagrams, using {\it Spitzer}'s IRAC photometric system. The blue dots show the dayside emission of transiting planets observed during occultation. Squares and arrows represent upper limits. Lines labelled with the name of a planet show the position of systems { where colour or absolute magnitude is missing} (not all cases are represented, for clarity). The coloured diamonds underlying the plots are ultra-cool dwarfs and directly imaged planets, whose magnitudes are listed in \citet{Dupuy:2012lr}. Colours represent the spectral class of the object, spanning from M5 (orange) to Y1 (black). { The only unclassified object here, in grey, is WD 0806-661B \citep{Luhman:2012lr}}.
}\label{fig:MidIR}  
\end{center}  
\end{figure*}

\subsection{Mid-infrared}

In the mid-infrared, all the bands that were considered are the {\it Spitzer}'s IRAC { channels}. Both ultra-cool dwarfs and exoplanets have been observed extensively, especially so in the IRAC 1 \& 2 centred at 3.6 and 4.5 $\mu$m. Compared to the seven systems presented in \citet{Triaud:2014kq}, { the first diagrams in the top two rows} in Fig.~\ref{fig:MidIR} show a marked increase in the number of objects. 

\subsubsection{[3.6]$-$[4.5]}

The M \& L-dwarf sequence is colour-less in those bands. Objects get fainter for decreasing temperatures. As brown dwarfs transition towards the T sequence, a sharp turn occurs, caused by the widening and deepening methane absorption band at 3.3$\mu$m, revealed by the recession of dust clouds in brown dwarfs' atmospheres (e.g. \citealt{Patten:2006uq}, and references therein). This leads to increasingly redder colours with increasing magnitudes. The clarity of this pattern is handy to compare planets and brown dwarfs together. { So far no planet that has had its emission detected clearly falls} within the T-range. Good contenders can be found in HAT-P-12b \citep{Hartman:2009rt} whose upper limit places it beyond the methane {\it kink} and in WASP-80b that has a reported effective temperature around 800\,K \citep{Triaud:2013sk,Mancini:2014lr}. All currently measured hot Jupiters can therefore be compared to the M \& L sequence (GJ\,436b, a Neptune, is kept aside for now).

Despite significant scatter, one can notice that objects are not located completely at random. No object redder than [3.6]$-$[4.5]  = 1 for example exists.  All planets but two have colours compatible or redder than brown dwarfs. This gets clearer for absolute magnitudes { in the redder channels}. Only GJ\,436b { ([3.6]$-$[4.5]  $<$ 0.6)} and WASP-8Ab { ([3.6]$-$[4.5]  = 0.6)} are significantly bluer, two eccentric planets (a third eccentric planet, HAT-P-2b { ([3.6]$-$[4.5]  = 0)}, is compatible with the { colourless} L-sequence). 

{ The scatter in colour increases} for increasing magnitudes: objects brighter than the median magnitude (GJ\,436b removed) consistently have an RMS in colour lower than objects fainter than the median magnitude. { This is not because intrinsically fainter planets produce weaker (and harder to measure) occultations. Some of the most significant detections (for instance HD\,189733Ab (M$_{\rm [3.6]}  = 11.1$, [3.6]$-$[4.5]  = 0.1), HD\,209458b (M$_{\rm [3.6]}  = 10.4$, [3.6]$-$[4.5]  = 0.8))  are amongst the fainter planets. The graphs shuffled borderline and significant measurements by using absolute magnitudes. A clear detection arises because the host star and the planet are bright in apparent magnitudes, for instance thanks to their proximity to the Solar system.} 

The known hot Jupiters' diversity in radius (0.8 to 2 R$_{\rm jup}$), which does not exist for field brown dwarfs, cannot be held responsible for the scatter either. A change in radius translates with a decrease in absolute magnitude, but no change in colour as shown in Fig.~\ref{fig:bb_MidIR} when we compare with blackbodies, the current effects are much larger. This forces us to turn to other processes such as an increased diversity (in atmospheric  structure or in absorbents) at colder temperatures, or to some intrinsic variability (with an amplitude $\sim 1.5$ mag). If such is the case, repeated measurements should be attempted. 



\subsubsection{[4.5]$-$[5.8]}\label{subsec:4p5-5p8}

Brown dwarfs face a similar pattern than in the previous subsection, but orientated in the opposite direction. It also marks the transition between the L and T spectral classes. With decreasing temperatures, CO (that has absorption in the IRAC 2 bandpass) reacts with H$_2$ to produce CH$_4$, it also produces H$_2$O that has several important absorption features around 5.8$\mu$m. This makes the atmosphere become increasingly bluer with decreasing effective temperature.

The hot Jupiters, again, are all located in the absolute magnitude range of the M \& L-sequence. Apart from GJ\,436b, all are marginally bluer than their ultra-cool dwarf counterparts. Would we consider each planet individually, we would conclude that each is consistent with the M \& L-sequence when in fact the general population clearly is not. It is systematically biased towards the blue: They have a mean colour inferior to 0 when all brown dwarfs are above 0 in the same absolute magnitude range. Water absorption has been noticed in several transmitted spectra (e.g. \citealt{Deming:2013zm}), which would indicate that planets may depart from ultra-cool dwarfs' atmospheres in that water absorption appears at higher temperatures. 

{ Alternatively planetary atmospheres and ultra-cool dwarfs could be reconciled if ultra-cool dwarfs contain an absorbant around 4.5 $\mu$m that planets do not possess. If present, it would increase the planets' absolute magnitudes in the IRAC 2 channel {  at 4.5 $\mu$m}, moving each point closer to 0.}

\subsubsection{[3.6]$-$[5.8] \& [5.8]$-$[8.0]}

Those two colours show a redward trend with decreasing luminosity. At [3.6]$-$[5.8] and at [5.8]$-$[8.0] planets and ultra-cool dwarf overlap very well: as many objects are found on either side of the brown dwarf sequence showing statistical agreement. Planets may be slightly offset towards redder colours, in [5.8]$-$[8.0] but only marginally so at the moment.

This agreement between planets and ultra-cool dwarfs could in principle act as a sort of calibration, validating that measurements in those bands are well estimated (in value and error bar). However, we have to remember here that hot Jupiters are significantly larger than the typical brown dwarf ($\sim$ 1.3--1.6 R$_{\rm jup}$ vs 0.8--0.9 R$_{\rm jup}$). { Reducing the planets size to the brown dwarf level should normally lead the planets to be dimmer by 0.8 to 1.5 magnitudes} (see Fig. \ref{fig:bb_MidIR} and Sec. \ref{sec:bb}). At first { sights}, both classes of objects should not be compatible. { The fact that both groups have similar absolute magnitudes, indicates that hot Jupiters have lower surface emissivity than ultra-cool dwarfs.}

\begin{figure*}  
\begin{center}  
	\begin{subfigure}[b]{0.33\textwidth}
		\includegraphics[width=\textwidth]{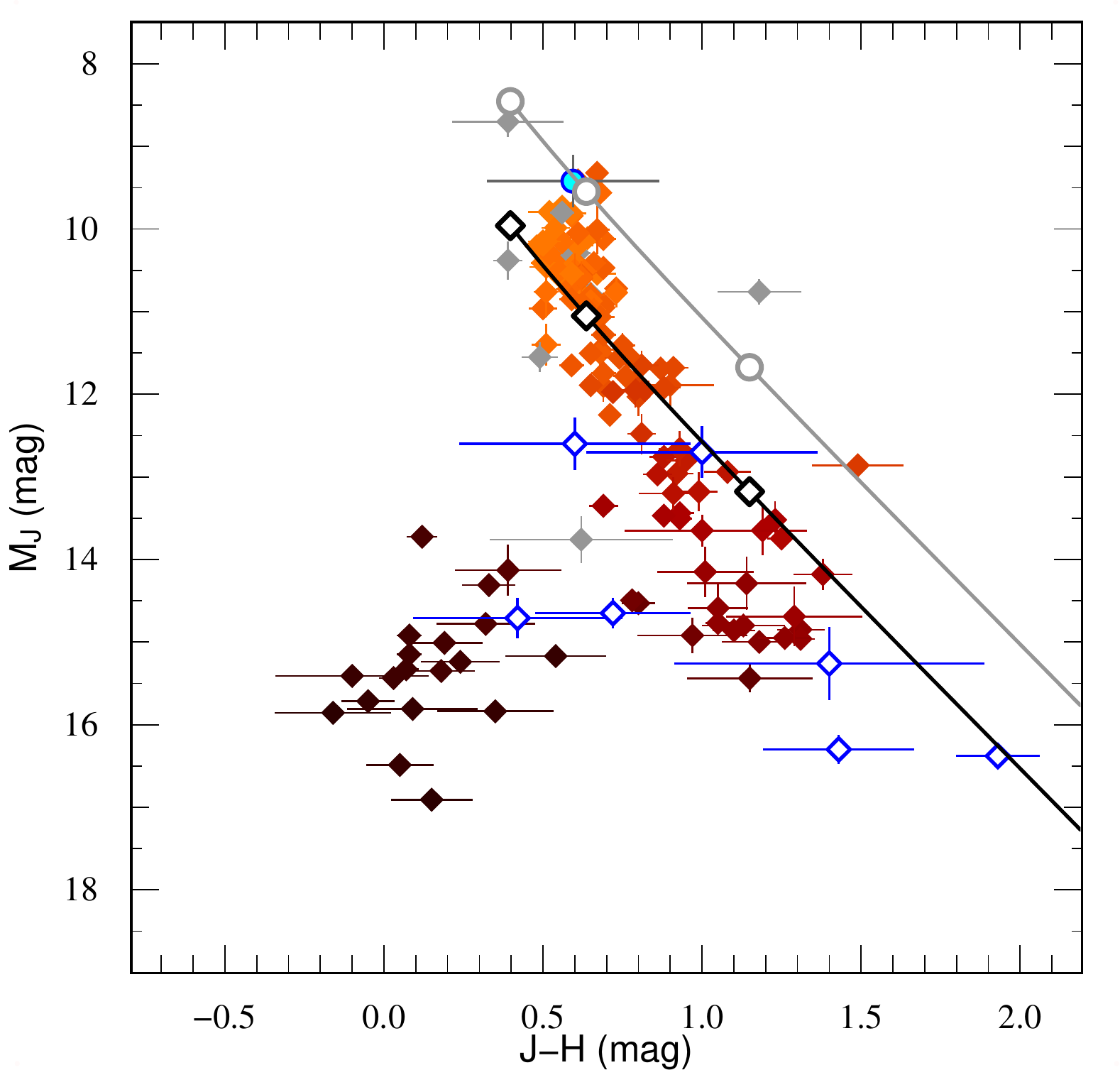}  
		\label{fig:M_J_a}  
	\end{subfigure}
	\begin{subfigure}[b]{0.33\textwidth}
		\includegraphics[width=\textwidth]{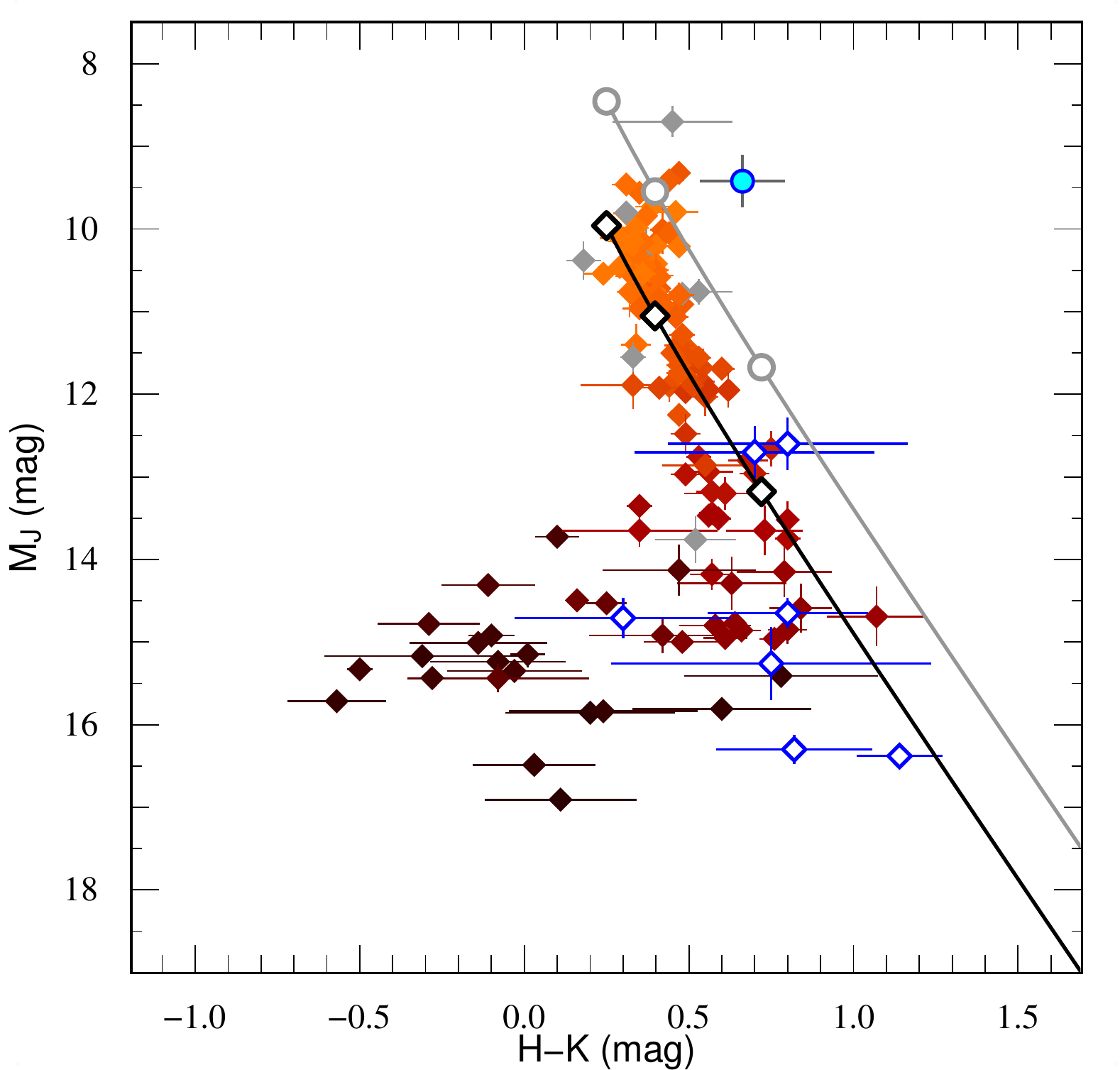}  
		\label{fig:M_J_b}  
	\end{subfigure}
	\begin{subfigure}[b]{0.33\textwidth}
		\includegraphics[width=\textwidth]{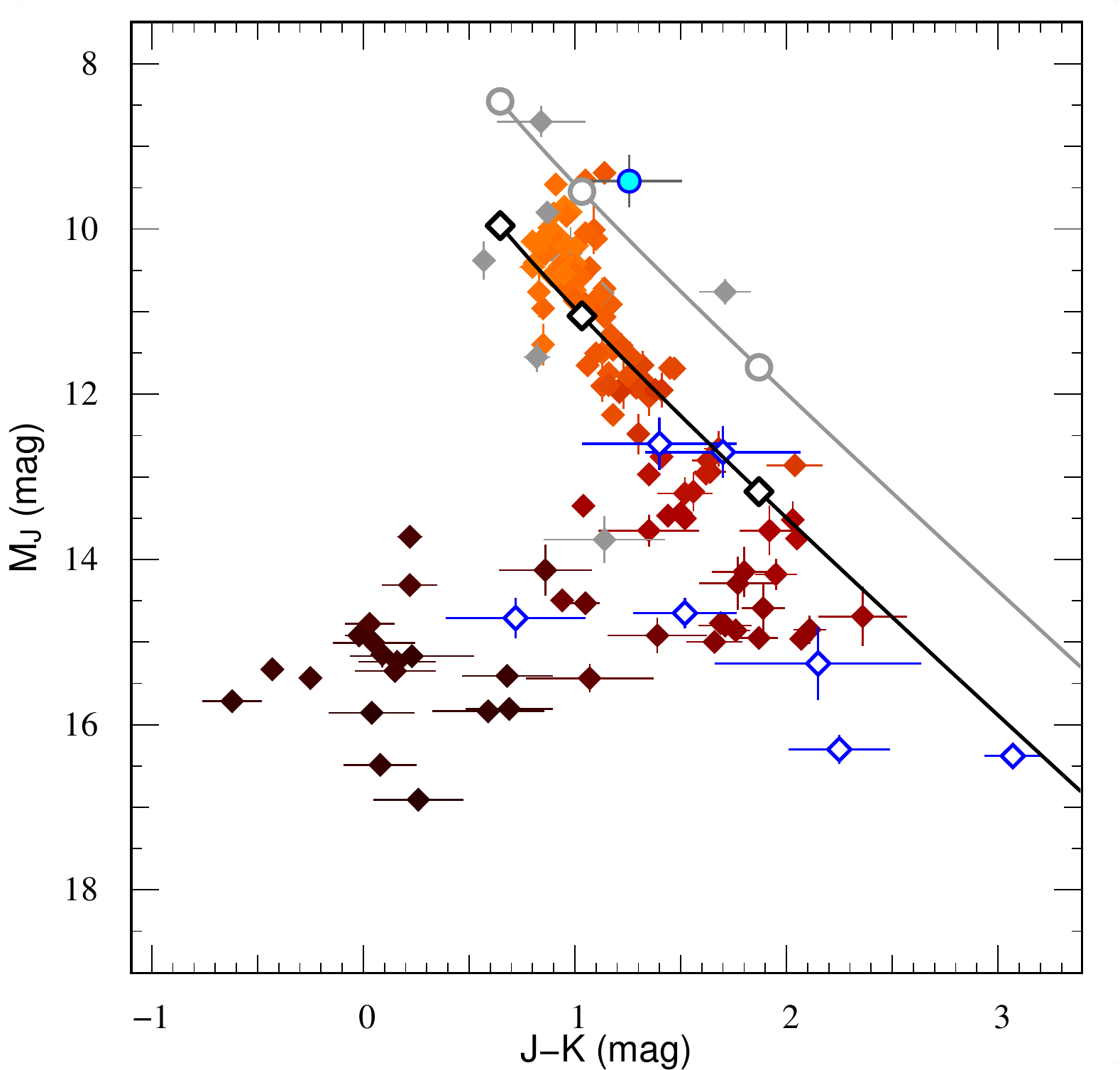}  
		\label{fig:M_J_c}  
	\end{subfigure}

	\caption{Same diagrams as the top line in Fig. \ref{fig:NIR} but showcasing the behaviour of blackbodies at 10 pc, whose effective temperature is changed while keeping its size constant. The plain grey line is for a 0.9~R$_{\rm Jup}$ object, similar to the radius of a brown dwarf, and the plain black line represents a 1.8 R$_{\rm Jup}$, the size of WASP-12Ab. The white-filled dots (0.9~R$_{\rm Jup}$) and diamonds (1.8 R$_{\rm Jup}$) along the blackbodies indicate the location of a 4\,000, 3\,000 and 2\,000~K object. For reference, the blue, empty diamonds highlight the position of young, directly detected exoplanets whose data is located in Table~\ref{tab:direct}.
	}\label{fig:bb_NIR}  
\end{center}  
\end{figure*}

\subsubsection{Summary from mid-IR colours}
 
If a reason is found to explain the apparent agreement at [3.6]$-$[5.8] \& [5.8]$-$[8.0] then we could conclude that the 4.5 $\mu$m band measurements are at the source of the observed divergence between irradiated gas giants and brown dwarfs in the [3.6]$-$[4.5] \& [4.5]$-$[5.8] colours. { Introducing some additional absorber within the planets' spectrum, around 4.5 $\mu$m,} would move planets closer to 0 in both diagrams while keeping the [3.6]$-$[5.8] \& [5.8]$-$[8.0] untouched. 
{ The fact that the intrinsically fainter planets display a greater divergence from the ultra-cool dwarfs in colours based on the 4.5 $\mu$m band, may imply that they have an increased atmospheric diversity, some of them with, and some without that absorbant.}
We prefer this interpretation over intrinsic variability whose otherwise required amplitude would seem too large to explain the data. The discrepant [4.5] band has been noticed by a number of authors, with \citet{Knutson:2009gf} proposing that a temperature-inversion in the temperature-pressure profile is responsible \citep{Fortney:2008lr}. However this interpretation has been disputed by \citet{Madhusudhan:2011hc}, who argue that disparities in relative abundances, notably the carbon to oxygen ratio, can reproduce the observations equally well.

A number of other measurements exists, notably observed in narrow bands \citep{Gillon:2009ai,Smith:2011cr,Crossfield:2012qy,Gillon:2012fj,Lendl:2013fk,Anderson:2013fk}, in the $z$' band \citep{Lopez-Morales:2010fr,Lendl:2013fk,Abe:2013kx} or observed by folding the {\it CoRoT} and {\it Kepler} lightcurves (e.g. \citet{Snellen:2009tg,Alonso:2009kl,Morris:2013eu,Demory:2013uq,Sanchis-Ojeda:2013if}). Because of a lack of measured brown dwarfs to compare them to and often, because of a lack of apparent magnitudes in those particular bands, it seemed futile to do this exercise at this time. It will however become something worth investigating.

\section{Comparison with blackbodies}\label{sec:bb}

\begin{figure*}  
\begin{center}  
	\begin{subfigure}[b]{0.4\textwidth}
		\includegraphics[width=\textwidth]{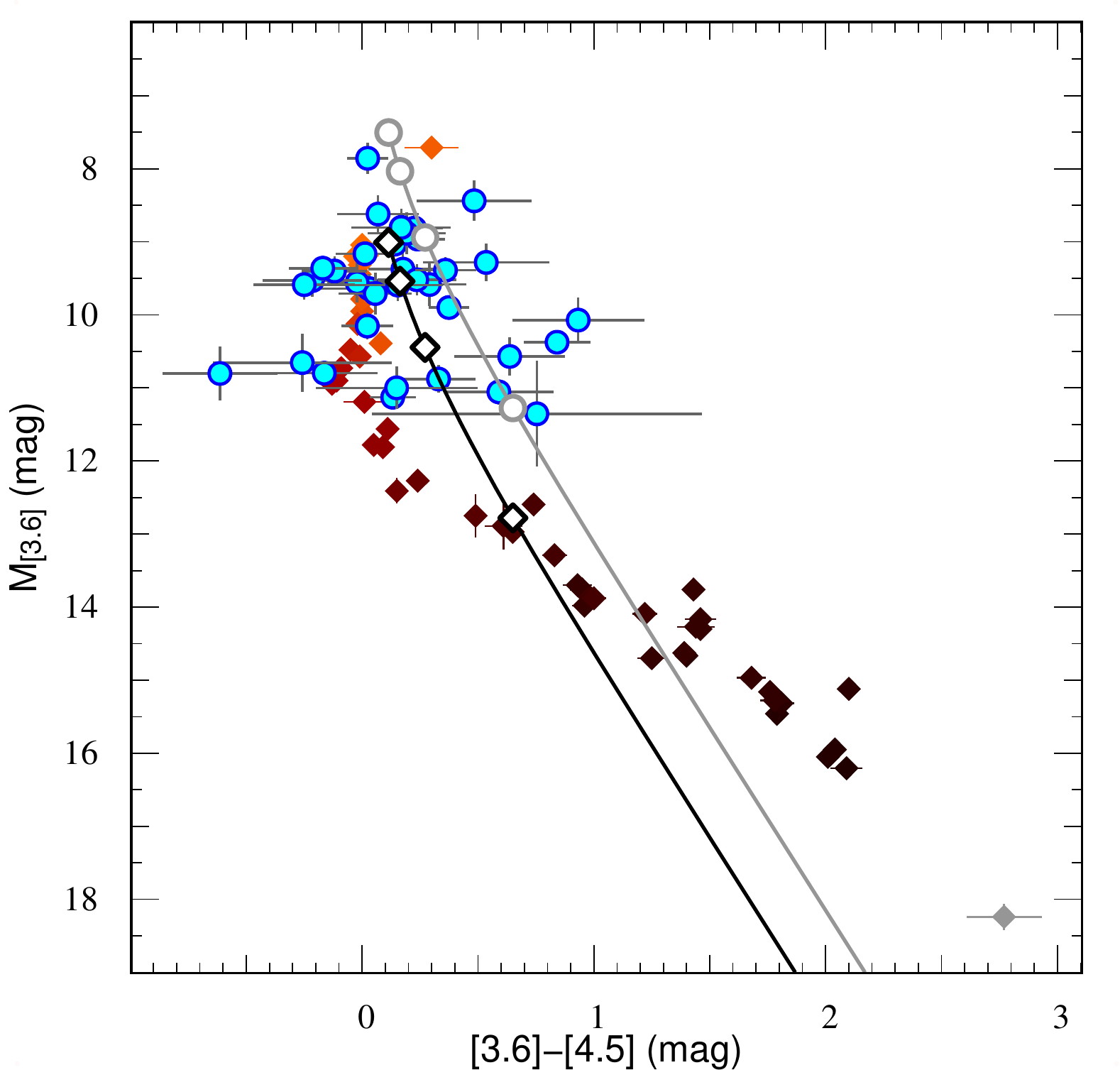}  
		\label{fig:M_3p6_a}  
	\end{subfigure}
	\begin{subfigure}[b]{0.4\textwidth}
		\includegraphics[width=\textwidth]{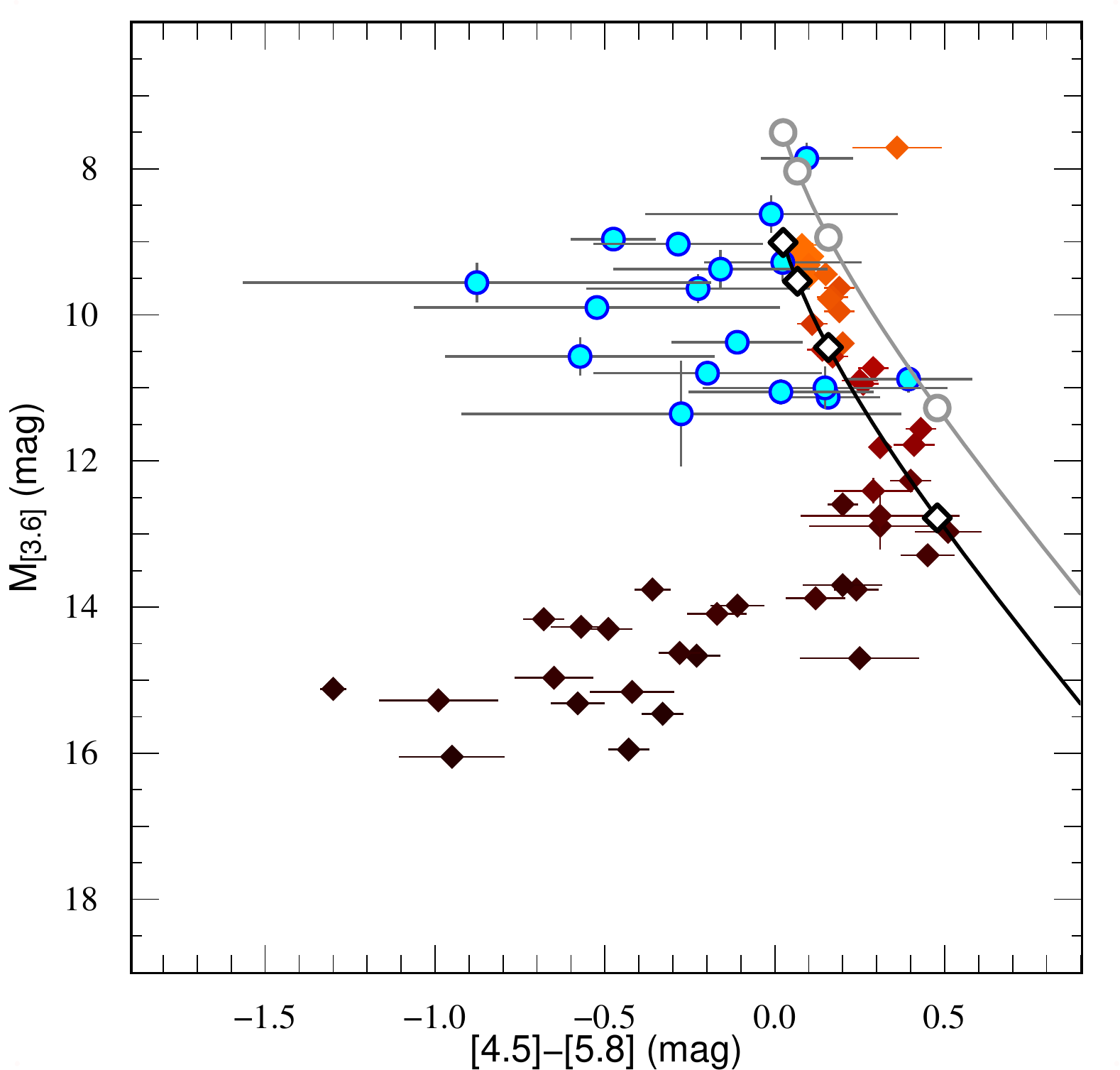}  
		\label{fig:M_3p6_b}  
	\end{subfigure}
	\begin{subfigure}[b]{0.4\textwidth}
		\includegraphics[width=\textwidth]{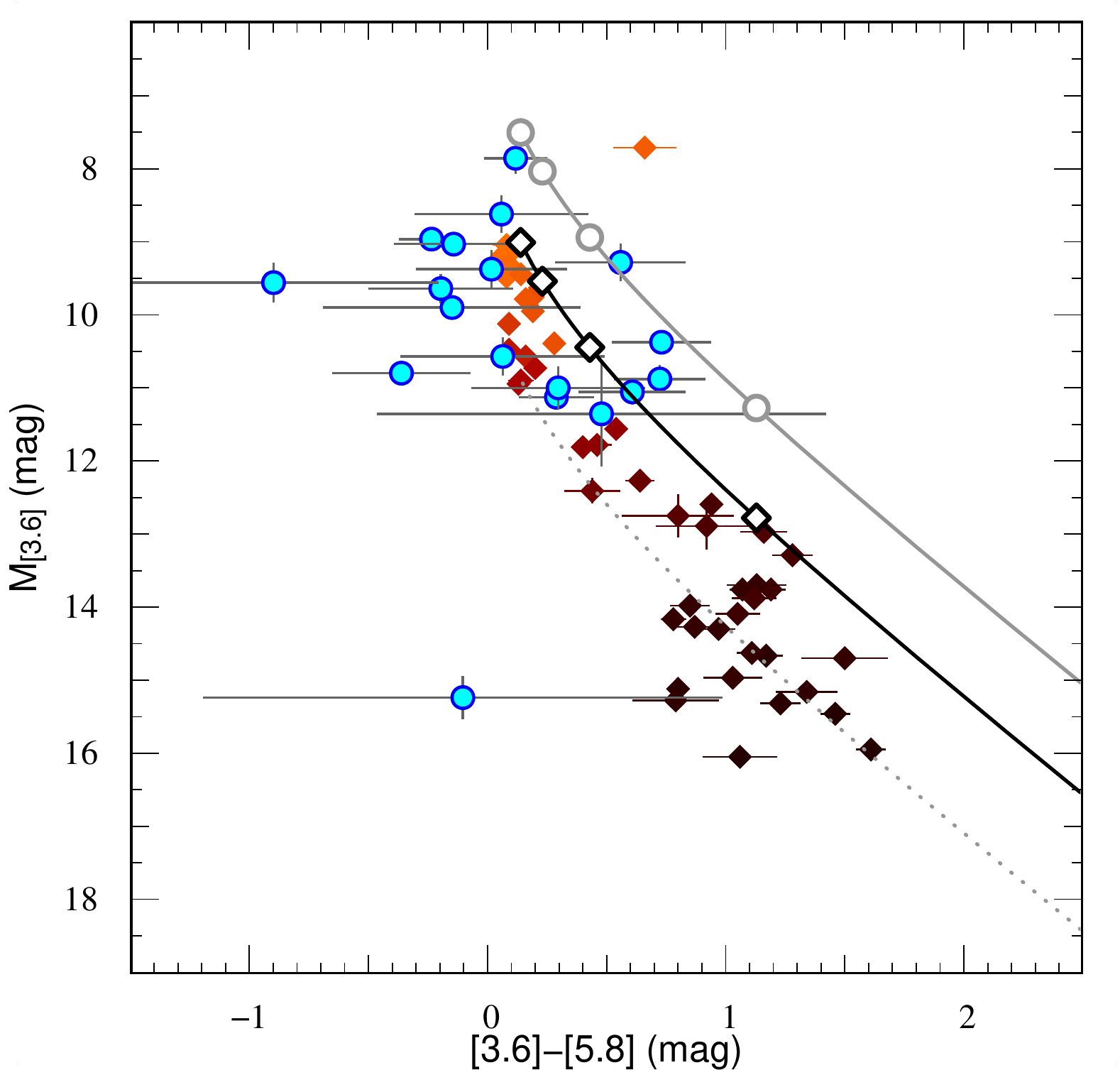}  
		\label{fig:M_3p6_c}  
	\end{subfigure}
	\begin{subfigure}[b]{0.4\textwidth}
		\includegraphics[width=\textwidth]{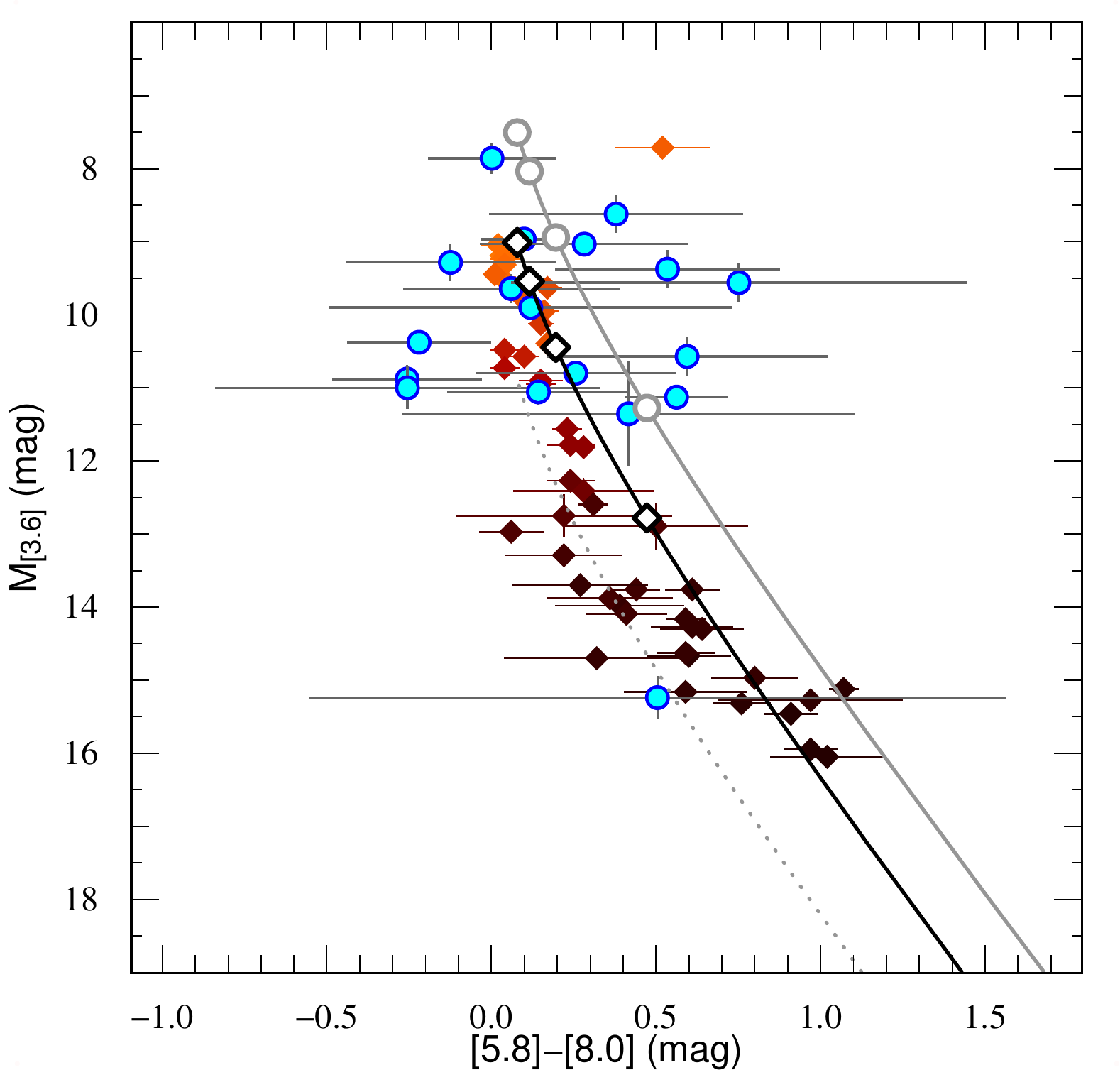}  
		\label{fig:M_3p6_d}  
	\end{subfigure}
	\caption{Same diagrams as the top line in Fig. \ref{fig:MidIR} but showcasing the behaviour of blackbodies at 10 pc whose effective temperature is changed while keeping its size constant. In plain grey, is drawn a 0.9 R$_{\rm Jup}$ object, similar to the radius of a brown dwarf, and in plain black a 1.8 R$_{\rm Jup}$, the size of WASP-12Ab. The two bottom panels have an added dotted grey line, which is a blackbody the size of GJ\,436b {(0.38 R$_{\rm Jup}$)}. The { marks} along the blackbodies indicate the expected location of a 4\,000, 3\,000, 2\,000 and 1\,000~K object.
}\label{fig:bb_MidIR}  
\end{center}  
\end{figure*}

Hot Jupiter emission measurements are often compared to complex models and to blackbodies, with frequent claims that { planet spectra} are compatible with the shape expected of a blackbody. WASP-12Ab is one of the most noticeable examples \citep{Crossfield:2012qy}. \citet{Hansen:2014jk} surveyed the literature for objects whose emission has been detected in several datasets at the same wavelength and, taking the variation in results as a systematic error bar, found that planets have featureless spectra resembling blackbodies. 

To answer this claim, and also because we should not expect  irradiated planets and ultra-cool dwarfs to be exactly the same, plotting the location of blackbodies within a colour-magnitude diagram seemed warranted. The blackbody loci can provide context by revealing how brown dwarfs depart from a blackbody and how irradiated gas giants compare to these departures. Figs.~\ref{fig:bb_NIR} and \ref{fig:bb_MidIR} have a 0.9~R$_{\rm Jup}$ and a 1.8~R$_{\rm Jup}$ { sized black-body} plotted for all temperatures between 4\,000 and 400~K. Those sizes where chosen as they represent the maximum size brown dwarfs are expected to have (with an age $>$ 1 Gyr; \citealt{Baraffe:2003gf}), and the approximate size of WASP-12Ab, one of the largest known exoplanet. 

If planets were blackbodies their measurements should be comprised strictly between the 0.9 and 1.8 R$_{\rm Jup}$ blackbodies. They cannot be above and cannot be below that strip (except for HD\,149026b and GJ\,436b). In the near-infrared (the only transiting planet in Fig.~\ref{fig:bb_NIR}) and mid-infrared, WASP-12Ab is lying near or on top of the expected blackbody line, in absolute magnitude and colour. Its location is also slightly above the 3\,000~K mark, which is compatible with its estimated equilibrium temperature of  $2\,990\pm110$ K as provided by \citet{Crossfield:2012qy}.

Whether WASP-12Ab follows the behaviour of a late M dwarf better than a blackbody is irrelevant in this case: in all colours, the M \& L sequence intersects with the expected { blackbody line} at WASP-12Ab's location { in the colour-magnitude diagram}\footnote{{ reflected light likely plays no part in placing WASP-12Ab at this special location. It is expected to be about three orders of magnitude fainter than thermal emission \citep{Seager:2010kx}}}. The planet is where it ought to be. Having only few examples to work with, { we added to Fig.~\ref{fig:bb_NIR}} the directly imaged planets (Table~\ref{tab:direct}). { Apart from the recently announced GU Psc b \citep{Naud:2014kk},} those young planets show good agreement with their M \& L-dwarf counter parts, but continue redder and fainter instead of turning into the blueward L-T transition, { not unlike grey atmospheres.} { Irradiated planets could follow blackbodies, the ultra-cool dwarf's sequence, the path of the young directly imaged planets, or their own sequence. To differentiate between these four solutions,} measurements of cooler transiting planets are required in near-infrared bands. HAT-P-12b and WASP-80b are good contenders. 

In the mid-infrared, the picture is more complex. In the M$_{\rm [3.6]}$ vs [3.6]$-$[4.5] diagram, there are seven planets redder or brighter than the 1.8 R$_{\rm Jup}$ blackbody. Thirteen systems are bluer or fainter than the 0.9 R$_{\rm Jup}$ blackbody. Due to the dispersion (increasing with increasing magnitude), neither the brown dwarfs, nor the blackbodies would seem to better explain all the measurements. We note that only two systems are more than 1$\sigma$ above the 1.8 R$_{\rm Jup}$ blackbody (HD\,209458b and XO-4b), and one (WASP-8Ab) is away from the brown dwarfs. All other gas giants lie in agreement with a triangular confinement bordered by the blackbody on one side and the ultra-cool dwarf atmospheres on the other two. Targeting planets at the cool junction between the T-dwarfs and the blackbody expectations will show if planets follow the T-sequence, a blackbody, or their own sequence (for example when reflected starlight starts producing a strong effect). This means studying gas giants cooler than 1\,000 K (whose size would presumably be closer to 0.9 than 1.8 R$_{\rm Jup}$).

The M$_{\rm [3.6]}$ vs [4.5]$-$[5.8] diagram shows that the L-sequence is slightly brighter than a 0.9 R$_{\rm Jup}$ blackbody would predict, but generally follows the same slope. Brown dwarfs clearly depart when they transition to the T spectral class. In section~\ref{subsec:4p5-5p8} we noted the blueward bias of hot Jupiters. This is strengthened when compared to a blackbody. Planets clearly depart. If each measurement is only 1 to $2\sigma$ away, what we lack in precision we gain in the number of systems measured. Hot Jupiters are not featureless. Again here, the departure between the brown dwarfs and the blackbody happens { below} 1\,000~K.

Gas-giants and ultra-cool dwarfs agree well in M$_{\rm [3.6]}$ vs [3.6]$-$[5.8]. { However planets do not match the expectations of a blackbody: All} but four planets are found bluer or fainter than the 0.9 R$_{\rm Jup}$ { blackbody line. The fact that planets follow the same { slope} as a blackbody suggests a behaviour similar to a grey atmosphere, implying that opacities in these bands are grey.}  Hot-Jupiters are not blackbodies and here behave more like dwarfs do. The final diagram, plotting  M$_{\rm [3.6]}$ vs [5.8]$-$[8.0] shows good agreement: brown dwarfs appear to follow the expected blackbody (being slightly below, maybe evidence they are slightly smaller than 0.9 R$_{\rm Jup}$), so do hot Jupiters but with a large scatter. This would indicate that the opacity is grey in these bands and approach Planck's law.

The location of a blackbody with the size of GJ\,436b (0.38 R$_{\rm Jup}$) was added and goes right through its measurement { at [5.8]-[8.0]}. A change in radius is only a translation in absolute magnitude. GJ\,436b sits right at the 1\,000 K marks, which would imply a similar temperature, much higher than its estimated equilibrium temperature of $\sim$700 K \citep{Deming:2007fp}. If { this is} not the indication of excess energy produced by its on-going tidal circularisation \citep{Maness:2007ys,Beust:2012rt}, this should be seen as a reminder { that effective temperature is different from equilibrium temperature and that} touching the blackbody sequence, does not mean a measurement agrees with it, as temperature too needs to be accounted for. Shape is not all.








\section{Discussion \& Conclusion}

We computed photometric distances that allowed us to obtain the absolute magnitude of occulting planets. They were used to compile colour-magnitude diagrams. Planets on their own would not offer much information. This is why we compared { their location in these diagrams}, to the location of very low-mass stars and field brown dwarfs, and to the behaviour expected of pure blackbodies. { By defining a blackbody sequence with a lower size of 0.9 R$_{\rm Jup}$ and an upper one of 1.8 R$_{\rm Jup}$}, we describe a locus in the form of a strip where all hot Jupiters should congregate would they follow Planck's law.

In the near-infrared, three clear conclusions can be drawn:
\begin{itemize}
\item { Planets are brighter in K$_{\rm S}$ band measurements}, and in average redder than the M \& L brown dwarf sequence (this probably has an instrumental origin).
\item WASP-12Ab is as much compatible with a blackbody as with the M \& L sequences, because that is the location where both intersect. 
\item A clear distinction between irradiated gas giants following a brown dwarf behaviour, the young directly-imaged planets, or a blackbody will emerge for equilibrium temperatures cooler than $\sim$ 2\,000~K.
\end{itemize}

In the mid-infrared we obtained the following general trends:
\begin{itemize}
\item Gas giants are only in agreement with the blackbody locus in the [5.8]$-$[8.0] colour. Deviations, made significant by the number of objects considered, in the other colours imply that planet are not pure blackbodies, although individual objects may appear to be.
\item Gas giants are bluer in the [4.5]$-$[5.8] colour than a blackbody or the M \& L brown dwarf sequence. This shows that hot Jupiters are not featureless.
\item Combining this with an increased scatter as magnitudes increase in the [3.6]$-$[4.5], provides support that some gas giants are missing an absorbant at 4.5 $\mu$m.
\item This affects only certain planets making us conclude that atmospheric diversity increases with decreasing absolute magnitude, presumably, with decreasing equilibrium temperature.
\item Clearly associating planets to the brown dwarf locus or to the blackbody strip can be made by obtaining { the emission (dayside or nightside)} of gas giants with effective temperatures below 1\,000 K at [3.6] and [4.5].
\end{itemize}

It is worth noting at this point that the observed increase in atmospheric diversity is found under the upper limits placed by \citet{Demory:2013uq} on Kepler-7b. This planet's detected occultation and phase curve in the {\it Kepler} bandpass have been interpreted as reflected light from an inhomogeneous, high albedo, cloud layer, mostly located on the dayside. From studying HD\,189733Ab's phase curve inside a colour-magnitude diagram, \citet{Triaud:2014kq} made a similar inference: the presence of clouds can hide the effect of some absorbing species, or can locally change the atmospheric chemistry. We can therefore wonder whether the existence of clouds can be linked to the presence or absence of an absorbing feature in {\it Spitzer}'s 4.5 $\mu$m channel that leads to the scatter present in the [3.6]$-$[4.5] and [4.5]$-$[5.8] colours.

If brown dwarf atmospheres and irradiated exoplanets are set to coincide, then it is perhaps not surprising that since most exoplanets fall in the range occupied by the M \& L types, they too would have an opaque cloud layer at least on the dayside. Clouds are likely to leak over the terminator covering transmitted features. This provides context to the frequently announced featureless transmission spectra on several exoplanets (e.g. \citealt{Bean:2010vn,Berta:2012fk,Sing:2013xe,Jordan:2013dn}). GJ\,436b is found on the continuation of the M \& L sequences, and too shows a featureless transmission spectrum \citep{Knutson:2014fk}. The scatter in colour of the emitted spectra for the colder of the transiting gas giants can give hope that some will possess an inhomogeneous cloud cover, revealing the deeper parts of their atmospheres through cloud holes. Using colour-magnitude diagrams would become a useful tool to select the right exoplanet sample before attempting an observing campaign aimed at producing transmission spectra. 

{  \citet{Burrows:2014qf} point out, in their supplementary materials, that for an equivalent emerging flux, the spectra of an irradiated and of an isolated planet are dissimilar, notably by possessing widely different temperature-pressure profiles. The widening range in colour could also originate from distinctions in the impacting irradiative stellar flux, or on how this energy affects different atmospheres. An irradiated planet, for instance emits more strongly at 4.5 $\mu$m than its isolated equivalent.}\\

An obvious extension of this work would be to explore other colours, notably in some narrow bands where successful occultations measurements have been obtained by a number of { investigators}. Ultra-cool dwarf magnitudes can be obtained from the many spectra that have been acquired of these objects and integrating over the correct bandpasses. It would be interesting to know whether those fall into regions sensitive to additional species, which could greatly help our understanding of exoplanetary atmospheres. For instance, \citet{Demory:2013uq} have shown how bright Kepler-7 is in { {\it Kepler}'s optical bandpass, Kmag,} compared to its mid-infrared magnitude. It is therefore likely that a { Kmag-[3.6] or a Kmag-[4.5]} would be a  tracer of cloudy structures on the dayside of exoplanets. We cannot but encourage authors to report apparent magnitudes in the bands that they report occultations in.

From studying those diagrams we can make judgements about the most interesting planets to obtain emission measurements on. 
Some objects are particular in deviating from the global trends we outlined above, with the clearest example found with GJ\,436b. Its small size is not sufficient to explain its discrepancy. The absence of a detection in the 4.5 $\mu$m band signifies it is the bluest object in the current sample in the [3.6]$-$[4.5] colour, and the reddest in [4.5]$-$[5.8]. While being broadly consistent with the shape of a blackbody, its inferred effective temperature ($\sim$ 1\,000 K) appears unreasonably high. The study of the other smaller planets, GJ\,1214\,b \citep{Charbonneau:2009fj}, GJ\,3470\,b \citep{Bonfils:2012yu} and HD\,97658b \citep{Dragomir:2013lr} can show if they manifest an atmospheric behaviour similar to GJ\,436b's.

Arguably there are now enough measurements over the M \& L sequences; it is scientifically interesting to reserve our ressources to extend beyond that range. Going further up along the M sequence would need hot Jupiters orbiting A stars (like WASP-33 \citep{Collier-Cameron:2010lr,Deming:2012rr}) that are hard to come about and hard to analyse: many A stars are within the instability strip and display oscillations (WASP-33 is a $\delta$ Scuti). Exploring further down, closer to  the T regime, especially for equilibrium temperatures below 1\,000 K can be achieved by targeting longer period planets (WASP-8Ab for example is close to the L-T transition \citep{Queloz:2010lr,Cubillos:2013kx}). { The main issue in observing colder planets are the weak signals that can be expected from them. This can be mitigated by selecting host} stars of late spectral classes such as WASP-80 \citep{Triaud:2013sk}.


So far very few transiting (or occulting) brown dwarfs have been detected \citep{Deleuil:2008lr,Anderson:2011fk,Bouchy:2011lr, Siverd:2012ef,Diaz:2013fr}. Orbiting hot, and large stars their occultation can be hard to obtain, but are doable \citep{Beatty:2014lr}. However, those brown dwarfs are mostly found on short orbits, like hot Jupiters. They have inferred temperatures similar to M or L objects but differ from usual brown dwarfs in that { they} are inflated. Because of their size, they fall on isochrones younger than the inferred age of the star they orbit \citep{Triaud:2013lr}. Proximity acts like a rejuvenation. Obtaining several brightness measurements over the M, L and T range, preferably on long period objects would in principle procure a radius calibration for field brown dwarfs.

\section*{Acknowledgments}

The authors would like to thank Franck Selsis, Mercedes L\'opez-Morales, Jacqueline Radigan, Mickael Bonnefoy, Josh Winn, Kevin Schlaufman and Jay Pasachoff for inspiring reflections, for explanations -- for reminders -- and for providing comments and reactions to the text. We would like to also thank and acknowledge the influence of our referee, Hans Deeg, whose suggestions improved the paper and helped clarify it.

A. H.\,M.\,J. Triaud is a Swiss National Science Foundation fellow under grant number P300P2-147773. 

This publication makes use of data products from the following projects, which were obtained through the  \href{http://simbad.u-strasbg.fr/simbad/}{Simbad} and \href{http://vizier.u-strasbg.fr/viz-bin/VizieR}{VizieR} services hosted at the \href{http://cds.u-strasbg.fr}{CDS-Strasbourg}:
\begin{itemize}
\item The Two Micron All Sky Survey (2MASS), which is a joint project of the University of Massachusetts and the Infrared Processing and Analysis Center/California Institute of Technology, funded by the National Aeronautics and Space Administration and the National Science Foundation.
\item The Wide-field Infrared Survey Explorer (WISE), which is a joint project of the University of California, Los Angeles, and the Jet Propulsion Laboratory/California Institute of Technology, funded by the National Aeronautics and Space Administration.
\item The Tycho2 catalog \citep{Hog:2000uq}.
\item The Amateur Sky Survey (TASS) \citep{Droege:2006yq}.
\item The Fourth U.S. Naval Observatory CCD Astrograph Catalogue (UCA4) \citep{Zacharias:2013fj}.
\item The AAVSO Photometric All-Sky Survey (APASS), funded by the Robert Martin Ayers Sciences Fund.
\end{itemize}

We gathered the {\it Spitzer} Space Telescope data from the \href{http://sha.ipac.caltech.edu/applications/Spitzer/SHA/}{Spitzer Heritage Archive}.
References to exoplanetary systems were obtained by an extensive use of the paper repositories, \href{http://adsabs.harvard.edu/abstract_service.html}{ADS} and \href{http://arxiv.org/archive/astro-ph}{arXiv}, but also through frequent visits to the \href{http://exoplanet.eu}{exoplanet.eu} \citep{Schneider:2011lr} and \href{http://exoplanets.org}{exoplanets.org} \citep{Wright:2011fj} websites.

\bibliographystyle{mn2e}
\bibliography{1Mybib.bib}

\appendix

\section{Obtaining calibrated apparent magnitudes with {\it Spitzer}}\label{app:phot}


Apparent magnitudes in all four IRAC bands are based on IRAC images calibrated by the standard {\it Spitzer}
pipeline (version S18.18 or S18.25 depending on their availability at the time of the data reduction). They are delivered to the community as Basic Calibrated Data (BCD) sets and can be easily found at the {\it Spitzer} Heritage Archive\footnote{\href{http://sha.ipac.caltech.edu/applications/Spitzer/SHA/}{http://sha.ipac.caltech.edu/applications/Spitzer/SHA/}}. According to the brightness of each targets, some sets were observed in the IRAC channels in sub-array mode, some in full-array mode and a number in both. This forced us to employ two different data reductions.  The sub-array mode offers a high temporal resolution for observing very bright sources (available exposure times : 0.02, 0.1 and 0.4 seconds) on a portion of the array detector (32$\times$32-pixel). The full-array mode provides 256$\times$256-pixel (5.22' $\times$ 5.22') frames for longer exposure times of 2, 12, 30 and 100 seconds.

\subsection{Aperture photometry}

Each BCD set provided by sub-array mode is composed of 64 sub-array images. These data are reduced according to the {\tt EXOPHOT} {\it pyraf} pipeline following Lanotte et al. (in prep) to get raw light curves. For each sub-array image, a 2-D elliptical Gaussian profile fit is performed on the point spread function (PSF) of the target to obtain its PSF centre coordinates. We operate aperture photometry thanks to the {\tt IRAF/DAOPHOT}\footnote{{\tt IRAF} is distributed by the National Optical Astronomy Observatory, which is operated by the Association of Universities for Research in Astronomy, Inc., under cooperative agreement with the National Science Foundation.}  software \citep{Stetson:1987kl}. For each sub-array image, the software measures the stellar flux on apertures centred on our estimated PSF locations, ranging from 2.5 to 5.9 pixels by increments of 0.1 pixel, and subtracts the background level evaluated in an annulus extending from 12 to 15 pixels from the centre of aperture. For each block of 64 sub-array images, the discrepant values for the measurements of the $x$- and $y$-position, and the stellar and background flux are rejected using a 3-$\sigma$ median clipping. The remaining measurements in each BCD set are averaged.

The full-array mode images are reduced in the same way, except that the PSF centres are determined by a flux-weighted centroid. This method is better adapted to lower signal-to-noise data. 

At this stage, the first measurements of each light curve are discarded if they correspond to deviant values for all or some of the the external parameters (detector or pointing stabilisation). Finally we perform for each light curve a moving median filtering to discard outlier measurements due, for instance,  to cosmic hits. We also reject the measurements during a planetary transit, if present, to always consider the total stellar flux. Ideally one should measure the flux coming from the stellar system only during the occultation of the planet to only consider the stellar flux. However the planetary emission is negligible in comparison to flux variations induced by instrumental effects such as the `pixel-phase' and the `ramp' effects. The first one lies in the dependence of the observed flux with the stellar centroid location on the pixel of the IRAC InSb (3.6 and 4.5 $\mu$m) arrays. It is due to the inhomogeneous intra-pixel sensitivity combined to the jitter of the telescope and to the poor sampling of the PSF.  The second effect is the increase of the detector response at the start of AORs and is attributed to a charge-trapping mechanism resulting in a dependence of the gain of the pixels to their illumination history. We refer the reader to \citet{Knutson:2008qy} and references therein for more informations about these instrumental systematics.  

The pixel phase response changed at the beginning of the {\it Warm} mission, with the consequence that the correction map of the cryogenic phase of {\it Spitzer} could not be used for all the data. Since no complete correction map is available for the {\it Warm} phase of {\it Spitzer} at the time of our analysis, we do not correct the flux measurements for the intra-pixel sensitivity.
In practice, those intra-pixel flux variations are partially averaged out thanks to variations in the location of the PSF during an observational run. We do not model the `ramp' effect but simply remove the more affected sequence of measurements.


For each dataset (called AOR = Astronomical Observation Request in {\it Spitzer} terminology), we average all remaining measured stellar fluxes computed for each radius separately. We then apply the appropriate aperture correction to determine the stellar flux as it would be falling into a circular aperture radius of 10 pixels. This is carried out in order to remain consistent with the magnitude calibrations present in \citet{Reach:2005lr}.
The IRAC instrument handbook provides aperture corrections for different aperture radii and background annuli. However only three aperture corrections can be applied for the sub-array mode data, so that we generate other aperture correction factors to coincide with all our photometric apertures. Indeed the accuracy of the flux measurement resides in the choice of the photometric aperture radius. While small aperture radii are dominated by imprecisions due to under-sampling the PSF and pixel to pixel response, larger radii are affected by larger background contributions.  
We thus perform aperture photometry on deconvolved images reconvolved by the best-fitting partial PSF model to derive the aperture corrections required for deriving the observed flux of the star. The deconvolution photometry is made using {\tt DECPHOT} following a procedure described in \citet{Gillon:2006yg} and optimised for {\it Spitzer} data by Lanotte et al. (in prep). {\tt DECPHOT} is based on the image-deconvolution method of \citet{Magain:1998fv} that, contrarily to traditional deconvolution methods, respects the sampling theorem of \citet{Shannon:1949af} and preserves the photometric flux.
The aperture corrections are normalised to the flux falling into a circular aperture radius of 10 pixels subtracted to the background level measured in an annulus from 12-20 pixels.

Then we average all flux corrected for aperture and take the resulting value as the observed flux measurement for the dataset. The mean of the errors on each corrected flux is taken as our error bar on the measured stellar flux. We convert the measured flux in Jansky and apply the colour and inter-pixel corrections\footnote{see \S 4.4 and 4.5 of the {\it{Spitzer Observer's Manual}} and http://irsa.ipac.caltech.edu/data/SPITZER/docs/irac/warmfeatures/}.
Finally the flux densities are converted into Vega apparent magnitudes using the zero-magnitude flux densities computed by \citet{Reach:2005lr}. 
The associated error bars are dominated by the uncertainty in the absolute calibration.

\subsection{Deconvolution of blended stars}

Two systems in our sample (CoRoT-2 and WASP-8) are blended by a visual companion. \citet{Gillon:2010zl} and \citet{Deming:2011dp} have evaluated { the dilution factor:} the correction { to the measured flux} needed to remove the dilution caused by CoRotT-2A's visual companion. Their correction factors at 4.5 $\mu$m return a magnitude disparity of $\sim$0.3 mag using our measured fluxes using the method described above.  No similar work has been done for WASP-8. In order to  measure the dilution factor induced in the flux measurement with a higher precision, we performed once again a deconvolution of the data for those two stars. We used {\tt DECPHOT} to operate aperture photometry on model images considering two stars or the target only. We compute the dilution factor for both systems using all our aperture radii to reduce the errors of the inferred factors. The standard deviations of CoRoT-2 and WASP-8 fluxes due to the change of aperture radius are  { 0.11 and 0.07 \%, respectively, at} 3.6 $\mu$m, and 0.04 and 0.08 \% at 4.5 $\mu$m. For comparison, the standard deviations of isolated target fluxes due to the change of aperture radius are encompassed between 0.01 and 0.06 \%. Table~\ref{tab:dilution} gives dilution factors according to some aperture radius, the target, and the instrument. With these factors, fluxes for each aperture are corrected and the same procedure as described { in the} previous section is carried out to yield corrected apparent magnitudes.

\begin{table}
\centering
\caption{\label{tab:dilution} Dilution factors in the stellar flux from CoRoT-2A and WASP-8A caused by their visual companion. These factors are estimated for a range of aperture radii. }
\begin{tabular}{c|cccc}
\hline
 Aperture radius	 & \multicolumn{4}{c}{Dilution (\%)} \\ \noalign{\smallskip}
       				 & \multicolumn{2}{c}{CoRoT-2A} & \multicolumn{2}{c}{WASP-8A}\\ \noalign{\smallskip}
 	(pixels)	& [3.6] & [4.5] & [3.6] & [4.5]\\ \noalign{\smallskip}
\hline  \noalign{\smallskip}
2.5 & 4.08 & 2.72 & 0.85 & 1.23 \\
3.0 & 7.51 & 5.80 & 2.54 & 3.56 \\ 
3.5 & 12.98 & 11.31 & 6.34 & 7.87 \\ 
4.0 & 17.44 & 15.60 & 9.90 & 11.83 \\ 
4.5 & 18.91 & 17.19 & 11.43 & 13.76 \\ 
\hline  \noalign{\smallskip}
\end{tabular}
\end{table}

\section{Tables}
\begin{table} 
  \caption{Absolute magnitudes reported for some directly imaged planets.}\label{tab:direct}
  \begin{tabular}{lrrrl}
  \hline
  Name & M$_{\rm J}$ & M$_{\rm H}$ & M$_{\rm K_s}$ & Refs\\
  \hline
$\kappa$\,And\,b	&$12.7\pm0.3$		&$11.7\pm0.2$		&$11.0\pm0.4$		&1\\
HR\,8799\,b		&$16.30\pm0.16$	&$14.87\pm0.17$	&$14.05\pm0.08$	&2\\
HR\,8799\,c		&$14.65\pm0.17$	&$13.93\pm0.17$	&$13.13\pm0.08$	&2\\
HR\,8799\,d		&$15.26\pm0.43$	&$13.86\pm0.22$	&$13.11\pm0.12$	&2\\
2M\,1207\,b		&$16.38\pm0.09$	&$14.45\pm0.09$	&$13.31\pm0.08$	&3,4\\
$\beta$\,Pic\,b		&$12.6\pm0.3$		&$12.0\pm0.2$		&$11.2\pm0.1$		&5,6,7\\
GU Psc\,b			&$14.71\pm0.23$	&$14.29\pm0.23$	&$13.99\pm0.23$	&8\\
\hline
\end{tabular}\\
{ References:} (1) \citet{Carson:2013eu}; (2) \citet{Marois:2008fk};  (3) \citet{Chauvin:2004lr}; (4) \citet{Mohanty:2007rw}; (5) \citet{Lagrange:2009it}; (6) \citet{Bonnefoy:2011la}; (7) \citet{Bonnefoy:2013yu}; (8) \citet{Naud:2014kk}\\
\end{table}

\landscape

\newpage

\begin{table}
\scriptsize 
  \caption{Parallactic distances and apparent magnitudes from the literature, presented with photometric distances, obtained from absolute visual magnitudes $M_{\rm V}$ computed as described in the text, using the Vega-based magnitude convention.}\label{tab:stars}
  \begin{tabular}{lcccrrrrrrrl}
  \hline
   Name     &   parallactic     & photometric     & distance modulus          & \multicolumn{7}{c}{apparent magnitudes} & Refs\\
                   & distance (pc)    & distance (pc) &  ($m_{\rm V}-M_{\rm V}$)   &m$_{\rm J}$ & m$_{\rm H}$ & m$_{\rm K_s}$ & m$_{\rm [3.6]}$ & m$_{\rm [4.5]}$ & m$_{\rm [5.8]}$ & m$_{\rm [8.0]}$ & \\
 \hline
HD\,189733A&$19.45\pm0.26$	&$19.1\pm1.0$		&$1.41\pm0.12$&$ 6.073\pm0.032$		&$ 5.587\pm0.031$		&$ 5.541\pm0.021$		&$5.450	\pm0.065	$&$5.530\pm0.061	$&$5.971\pm0.063	$	&$5.968\pm0.059$		& 	1,2,3,5\\ 
HD\,209458	& $49.6\pm2.0$ 	&$49.0\pm2.2$		&$3.45\pm0.10$&$ 6.591\pm0.020$		&$ 6.366\pm0.038$		&$ 6.308\pm0.026$		&$6.258	\pm0.043	$&$6.305\pm0.035	$&$6.791\pm0.064	$	&$6.765\pm0.057$		& 	1,2,3,5\\ 
HD\,80606  	&--				&$65.8\pm3.9$		&$4.09\pm0.13$&$ 7.702\pm0.030$		&$ 7.400\pm0.034$		&$ 7.316\pm0.020$		&$7.257	\pm0.063	$&$7.348\pm0.037	$&--					&$7.742\pm0.061$		& 	1,3,5\\ 
HD\,149026	&$79.4\pm4.4$		&$80.8\pm4.0$		&$4.54\pm0.11$&$ 7.118\pm0.024$		&$ 6.899\pm0.018$		&$ 6.819\pm0.017$		&$6.840	\pm0.062	$&$6.827\pm0.045	$&$7.305\pm0.062	$	&$7.230\pm0.060$		& 	1,2,3,5\\ 
GJ\,436  		&$10.14\pm0.24$	&$*6.1\pm0.9$		&$-1.07\pm0.35$&$ 6.900\pm0.024$		&$ 6.319\pm0.023$		&$ 6.073\pm0.016$		&$5.889	\pm0.031	$&$5.836\pm0.023	$&$6.277\pm0.062	$	&$6.265\pm0.023$		& 	1,2,3,7\\ 
GJ\,1214 		&$14.55\pm0.13$	&$*9.1\pm4.4$		&$-0.2\pm1.4    $&$ 9.750\pm0.024$		&$ 9.094\pm0.024$		&$ 8.782\pm0.020$		&$8.488	\pm0.064	$&$8.397\pm0.060	$&--					&--					& 	1,3,4\\ 
55\,Cnc  		&$12.34\pm0.11$	&$12.2\pm0.9$		&$0.44\pm0.17$&$ 4.768\pm0.244$		&$ 4.265\pm0.234$		&$ *4.015\pm0.036$		&$ 4.09	\pm 0.11	$&$4.065\pm0.062	$&--					&--					& 	1,2,3,5\\ 
TReS-1  		&--				&$129.7\pm8.7$	&$5.56\pm0.15$&$10.294\pm0.022$		&$ 9.887\pm0.021$		&$ 9.819\pm0.019$		&$9.779	\pm0.047	$&$9.779\pm0.044	$&$10.232\pm0.049	$	&$10.241\pm0.048$		& 	1,3,5\\ 
TReS-2  		&--				&$195.3\pm12.0$	&$6.45\pm0.14$&$10.232\pm0.020$		&$ 9.920\pm0.026$		&$ 9.846\pm0.022$		&$9.782	\pm0.045	$&$9.790\pm0.043	$&$10.206\pm0.069	$	&$10.252\pm0.070$		& 	1,3,5\\ 
TReS-3  		&--				&$258.5\pm16.1$	&$7.06\pm0.14$&$11.015\pm0.022$		&$10.655\pm0.020$		&$10.608\pm0.017$		&$10.550	\pm0.064	$&$10.599\pm0.064	$&$11.029\pm0.076	$	&$11.03\pm 0.14$		& 	1,3,6\\ 
TReS-4  		&--				&$576.0\pm56.7$	&$8.80\pm0.22$&$10.583\pm0.018$		&$10.350\pm0.015$		&$10.330\pm0.019$		&$10.264	\pm0.064	$&$10.279\pm0.063	$&$10.700\pm0.073	$	&$10.741\pm0.074$		& 	1,3,5\\ 
XO-1    		&--				&$177.9\pm10.7$	&$6.25\pm0.13$&$ 9.939\pm0.022$		&$ 9.601\pm0.017$		&$ 9.527\pm0.015$		&$9.465	\pm0.061	$&$9.515\pm0.061	$&$9.947\pm0.069	$	&$9.967\pm0.075$		& 	1,3,5\\ 
XO-2    		&--				&$156.0\pm8.8$	&$5.97\pm0.13$&$ 9.744\pm0.022$		&$ 9.340\pm0.026$		&$ 9.308\pm0.021$		&$9.236	\pm0.064	$&$9.294\pm0.060	$&$9.725\pm0.078	$	&$9.733\pm0.072$		& 	1,3,5\\ 
XO-3    		&--				&$185.7\pm11.8$	&$6.34\pm0.14$&$ 9.013\pm0.029$		&$ 8.845\pm0.018$		&$ 8.791\pm0.019$		&$8.754	\pm0.039	$&$8.757\pm0.037	$&$9.210\pm0.067	$	&$9.213\pm0.067$		& 	1,3,5\\ 
XO-4    		&--				&$308.2\pm19.6$	&$7.44\pm0.14$&$ 9.667\pm0.021$		&$ 9.476\pm0.022$		&$ 9.406\pm0.023$		&$9.386	\pm0.066	$&$9.409\pm0.061	$&--					&--					& 	1,3,5\\ 
HAT-P-1B		&--				&$129.6\pm5.9$	&$5.56\pm0.10$&$ 9.156\pm0.026$		&$ 8.923\pm0.030$		&$ 8.858\pm0.018$		&$8.875	\pm0.064	$&$8.853\pm0.065	$&$9.278\pm0.069	$	&$9.308\pm0.072$		& 	1,3,7\\ 
HAT-P-2 		&$114.3\pm9.8$	&$125.3\pm13.1$	&$5.49\pm0.24$&$ 7.796\pm0.027$		&$ 7.652\pm0.038$		&$ 7.603\pm0.020$		&$7.544	\pm0.063	$&$7.603\pm0.043	$&$8.075\pm0.063	$	&$8.054\pm0.062$		& 	1,2,3,5\\ 
HAT-P-3 		&--				&$166.4\pm14.4$	&$6.11\pm0.20$&$ 9.936\pm0.022$		&$ 9.542\pm0.028$		&$ 9.448\pm0.025$		&$9.382	\pm0.065	$&$9.450\pm0.062	$&--					&--					& 	1,3,5\\ 
HAT-P-4 		&--				&$293.5\pm19.4$	&$7.34\pm0.15$&$10.100\pm0.022$		&$ 9.837\pm0.020$		&$ 9.770\pm0.020$		&$9.749	\pm0.072	$&$9.799\pm0.068	$&--					&--					& 	1,3,5\\ 
HAT-P-6 		&--				&$277.8\pm19.1$	&$7.22\pm0.15$&$ 9.558\pm0.023$		&$ 9.440\pm0.018$		&$ 9.313\pm0.019$		&$9.289	\pm0.067	$&$9.301\pm0.063	$&--					&--					& 	1,3,5\\ 
HAT-P-7 		&--				&$320.8\pm17.4$	&$7.53\pm0.12$&$ 9.555\pm0.030$		&$ 9.344\pm0.029$		&$ 9.334\pm0.018$		&$9.291	\pm0.068	$&$9.281\pm0.043	$&$9.727\pm0.081	$	&$9.759\pm0.068$		& 	1,3,5\\ 
HAT-P-8 		&--				&$227.8\pm12.7$	&$6.79\pm0.12$&$ 9.214\pm0.022$		&$ 9.004\pm0.018$		&$ 8.953\pm0.013$		&$8.942	\pm0.064	$&$8.932\pm0.060	$&--					&--					& 	1,3,5\\ 
HAT-P-12		&--				&$139.1\pm9.6$	&$5.72\pm0.16$&$10.794\pm0.023$		&$10.236\pm0.022$		&$10.108\pm0.016$		&$10.084	\pm0.048	$&$10.135\pm0.047	$&--					&--					& 	1,3,7\\ 
HAT-P-23		&--				&$355.0\pm40.8$	&$7.75\pm0.27$&$11.103\pm0.022$		&$10.846\pm0.022$		&$10.791\pm0.020$		&$10.822	\pm0.066	$&$10.770\pm0.068	$&--					&--					& 	1,3,5\\ 
WASP-1  		&--				&$346.4\pm34.8$	&$7.70\pm0.23$&$10.586\pm0.019$		&$10.364\pm0.016$		&$10.276\pm0.018$		&$10.234	\pm0.047	$&$10.237\pm0.044	$&$ 10.65\pm 0.19	$	&$ 10.71\pm 0.13$		& 	1,3,5\\ 
WASP-2  		&--				&$153.9\pm8.3$	&$5.94\pm0.12$&$10.166\pm0.027$		&$ 9.752\pm0.026$		&$ 9.632\pm0.024$		&$9.588	\pm0.066	$&$9.606\pm0.065	$&$ 10.02\pm0.11	$	&$10.032\pm0.071$		& 	1,3,7\\ 
WASP-3  		&--				&$251.4\pm18.8$	&$7.00\pm0.17$&$ 9.603\pm0.020$		&$ 9.407\pm0.014$		&$ 9.361\pm0.015$		&$9.366	\pm0.045	$&$9.356\pm0.063	$&$ 9.773\pm0.067	$	&$9.758\pm0.072$		& 	1,3,5\\ 
WASP-4  		&--				&$280.9\pm31.1$	&$7.24\pm0.25$&$11.179\pm0.025$		&$10.842\pm0.026$		&$10.746\pm0.021$		&$10.710	\pm0.051	$&$10.732\pm0.046	$&--					&--					& 	1,3,5\\ 
WASP-5  		&--				&$318.6\pm19.9$	&$7.52\pm0.14$&$10.949\pm0.022$		&$10.650\pm0.025$		&$10.598\pm0.023$		&$10.539	\pm0.069	$&$10.590\pm0.072	$&--					&--					& 	1,3,7\\ 
WASP-8A 	&--				&$85.1\pm10.7$	&$4.65\pm0.29$&$ 8.501\pm0.027$		&$ 8.218\pm0.049$		&$ 8.086\pm0.023$		&$8.084	\pm0.085	$&$8.162\pm0.077	$&--					&$8.552\pm0.063$		& 	1,3,5\\ 
WASP-12A 	&--				&$436.3\pm37.3$	&$8.20\pm0.19$&$10.477\pm0.021$		&$10.228\pm0.022$		&$10.188\pm0.020$		&$10.111	\pm0.042	$&$10.100\pm0.038	$&$10.541\pm0.074	$	&$ 10.55\pm 0.10$		& 	1,3,5\\ 
WASP-14 		&--				&$208.4\pm15.9$	&$6.60\pm0.17$&$ 8.869\pm0.021$		&$ 8.650\pm0.019$		&$ 8.621\pm0.019$		&$8.586	\pm0.046	$&$8.576\pm0.043	$&--					&$ 9.034\pm 0.064$		& 	1,3,5\\ 
WASP-17 		&--				&$476.0\pm36.0$	&$8.39\pm0.17$&$10.509\pm0.027$		&$10.319\pm0.024$		&$10.224\pm0.027$		&$10.196	\pm0.064	$&$10.193\pm0.065	$&--					&$10.66\pm 0.11$		& 	1,3,5\\ 
WASP-18 		&$99\pm11$		&$122.6\pm6.7$	&$5.44\pm0.12$&$ 8.409\pm0.018$		&$ 8.231\pm0.055$		&$ 8.131\pm0.027$		&$8.098	\pm0.046	$&$8.115\pm0.044	$&$8.561\pm0.047	$	&$8.573\pm0.063$		& 	1,2,3,5\\ 
WASP-19 		&--				&$275.9\pm13.4$	&$7.20\pm0.11$&$10.911\pm0.026$		&$10.602\pm0.022$		&$10.481\pm0.023$		&$10.445	\pm0.047	$&$10.486\pm0.047	$&$10.91\pm 0.13	$	&$10.754\pm0.085$		& 	1,3,7\\ 
WASP-24 		&--				&$332.5\pm23.8$	&$7.61\pm0.16$&$10.457\pm0.022$		&$10.219\pm0.026$		&$10.148\pm0.023$		&$10.132	\pm0.067	$&$10.156\pm0.066	$&--					&--					& 	1,3,5\\ 
WASP-26A 	&--				&$293.4\pm20.5$	&$7.34\pm0.16$&$10.021\pm0.022$		&$ 9.775\pm0.023$		&$ 9.690\pm0.023$		&$9.680	\pm0.075	$&$9.708\pm0.067	$&--					&--					& 	1,4,5\\ 
WASP-33 		&$116\pm11$		&$123.1\pm7.2$	&$5.45\pm0.13$&$ 7.581\pm0.021$		&$ 7.516\pm0.024$		&$ 7.468\pm0.024$		&$7.427	\pm0.066	$&$7.438\pm0.060	$&--					&--					& 	1,2,3,5\\ 
WASP-43 		&--				&$106.1\pm7.2$	&$5.31\pm0.24$&$ 9.995\pm0.024$		&$ 9.397\pm0.025$		&$ 9.267\pm0.026$		&$9.129	\pm0.063	$&$9.214\pm0.061	$&--					&--					& 	1,3,7\\ 
WASP-48 		&--				&$466.0\pm49.0$	&$8.34\pm0.24$&$10.627\pm0.025$		&$10.441\pm0.032$		&$10.372\pm0.022$		&$10.340	\pm0.065	$&$10.360\pm0.069	$&--					&--					& 	1,3,5\\ 
CoRoT-1 		&--				&$715.2\pm58.2$	&$9.27\pm0.18$&$12.462\pm0.029$		&$12.218\pm0.026$		&$12.149\pm0.027$		&$12.12	\pm0.12	$&$12.114\pm0.095	$&--					&--					& 	1,3,7\\ 
CoRoT-2A	&--				&$255.0\pm16.2$	&$7.03\pm0.14$&$10.783\pm0.028$		&$10.438\pm0.037$		&$10.310\pm0.031$		&$10.297	\pm0.071	$&$10.309\pm0.078	$&--					&$10.627\pm0.072$		& 	1,3,7\\ 
Kepler-7		&--				&$893.3\pm50.5$	&$9.76\pm0.13$&$11.833\pm0.020$		&$11.601\pm0.022$		&$11.535\pm0.020$		&$11.545	\pm0.077	$&$11.536\pm0.074	$&--					&--					& 	1,3,7\\ 
KELT-1		&--				&$251.3\pm13.0$	&$7.00\pm0.12$&$9.682\pm0.022$			&$9.534\pm0.030$		&$9.437\pm0.019$		&$9.390	\pm0.063	$&$9.405\pm0.061	$&--					&--					&	1,3,5\\

\hline
\end{tabular}\\
{ References:} (1) this paper; (2) \citet{van-Leeuwen:2007qy};  (3) \citet{Cutri:2003rt}; (4) \citet{Anglada-Escude:2013lr}; (5) \citet{Hog:2000uq}; (6) \citet{Droege:2006yq}; (7) \citet{Zacharias:2013fj}\\
$\star$ flagged in the 2MASS catalog
\end{table}
\clearpage
\newpage

\begin{table}
\scriptsize 
  \caption{Adopted distance and estimated stellar absolute magnitudes using data presented in table~\ref{tab:stars}, in the Vega-based magnitude convention.}\label{tab:abs}
  \begin{tabular}{lccrrrrrrr}
  \hline
   Name     &   adopted         & Absolute      		     	  	& \multicolumn{7}{c}{absolute magnitudes} \\
                   & distance (pc)   & M$_{\rm V}$			& M$_{\rm J}$ & M$_{\rm H}$ & M$_{\rm K_s}$ & M$_{\rm [3.6]}$ & M$_{\rm [4.5]}$ & M$_{\rm [5.8]}$ & M$_{\rm [8.0]}$ \\
 \hline
HD\,189733A	&$19.1\pm1.0$   	&$6.27\pm0.12$	&$ 4.67\pm0.12$	&$ 4.18\pm0.12$	&$ 4.14\pm0.11$	&$ 4.04\pm0.13$	&$ 4.12\pm0.13$	&$ 4.57\pm0.13$	&$ 4.56\pm0.13$	 \\ 
HD\,209458	&$49.0\pm2.2$   	&$4.18\pm0.10$	&$ 3.140\pm0.099$	&$ 2.915\pm0.099$	&$ 2.86\pm0.10$	&$ 2.80\pm0.11$	&$ 2.85\pm0.10$	&$ 3.34\pm0.11$	&$ 3.31\pm0.11$ 	\\ 
HD\,80606  	&$65.8\pm3.9$   	&$4.91\pm0.13$	&$ 3.61\pm0.13$	&$ 3.31\pm0.13$	&$ 3.22\pm0.13$	&$ 3.17\pm0.14$	&$ 3.26\pm0.13$	& --				&$ 3.65\pm0.14$	 \\ 
HD\,149026	&$80.8\pm4.0$   	&$3.60\pm0.11$	&$ 2.58\pm0.11$	&$ 2.36\pm0.11$	&$ 2.28\pm0.11$	&$ 2.30\pm0.12$	&$ 2.29\pm0.12$	&$ 2.77\pm0.13$	&$ 2.69\pm0.13$	 \\ 
GJ\,436  		&$10.14\pm0.24$ 	&$11.68\pm0.35$	&$ 6.870\pm0.057$	&$ 6.289\pm0.056$	&$ 6.043\pm0.052$	&$ 5.858\pm0.063$	&$ 5.806\pm0.054$	&$ 6.25\pm0.082$	&$ 6.235\pm0.056$ \\ 
GJ\,1214 		&$14.55\pm0.13$ 	&$14.9\pm1.4   $	&$ 8.936\pm0.030$	&$ 8.280\pm0.031$	&$ 7.968\pm0.028$	&$ 7.674\pm0.065$	&$ 7.58\pm0.062$	& --				& -- 				\\ 
55\,Cnc  		&$12.2\pm0.9$   	&$5.50\pm0.17$	&$ 4.33\pm0.29$	&$ 3.83\pm0.29$	&$ 3.58\pm0.17$	&$ 3.66\pm0.19$	&$ 3.63\pm0.17$	& --				& -- 				\\ 
TReS-1  		&$129.7\pm8.7$  	&$5.86\pm0.11$	&$ 4.73\pm0.14$	&$ 4.32\pm0.16$	&$ 4.25\pm0.15$	&$ 4.21\pm0.16$	&$ 4.21\pm0.16$	&$ 4.67\pm0.16$	&$ 4.68\pm0.15$ 	\\ 
TReS-2  		&$195.3\pm12.0$ 	&$4.80\pm0.11$	&$ 3.78\pm0.13$	&$ 3.47\pm0.14$	&$ 3.39\pm0.19$	&$ 3.33\pm0.14$	&$ 3.34\pm0.14$	&$ 3.75\pm0.15$	&$ 3.80\pm0.15$	 \\ 
TReS-3  		&$258.5\pm16.1$ 	&$5.33\pm0.12$	&$ 3.95\pm0.14$	&$ 3.59\pm0.18$	&$ 3.55\pm0.14$	&$ 3.49\pm0.16$	&$ 3.54\pm0.15$	&$ 3.97\pm0.16$	&$ 3.97\pm0.19$ 	\\ 
TReS-4  		&$576.0\pm56.7$ 	&$3.13\pm0.16$	&$ 1.78\pm0.21$	&$ 1.55\pm0.22$	&$ 1.53\pm0.22$	&$ 1.46\pm0.23$	&$ 1.48\pm0.22$	&$ 1.90\pm0.23$	&$ 1.94\pm0.23$ 	\\ 
XO-1    		&$177.9\pm10.7$ 	&$5.00\pm0.12$	&$ 3.69\pm0.13$	&$ 3.35\pm0.13$	&$ 3.28\pm0.13$	&$ 3.21\pm0.15$	&$ 3.26\pm0.15$	&$ 3.70\pm0.15$	&$ 3.72\pm0.16$ 	\\ 
XO-2    		&$156.0\pm8.8$  	&$5.28\pm0.10$	&$ 3.78\pm0.13$	&$ 3.37\pm0.14$	&$ 3.34\pm0.13$	&$ 3.27\pm0.14$	&$ 3.33\pm0.14$	&$ 3.76\pm0.14$	&$ 3.77\pm0.14$ 	\\ 
XO-3    		&$185.7\pm11.8$ 	&$3.51\pm0.14$	&$ 2.67\pm0.14$	&$ 2.50\pm0.14$	&$ 2.45\pm0.14$	&$ 2.41\pm0.15$	&$ 2.41\pm0.15$	&$ 2.87\pm0.15$	&$ 2.87\pm0.15$ 	\\ 
XO-4    		&$308.2\pm19.6$ 	&$3.37\pm0.13$	&$ 2.22\pm0.13$	&$ 2.03\pm0.14$	&$ 1.96\pm0.14$	&$ 1.94\pm0.15$	&$ 1.96\pm0.16$	& --				& --				 \\ 
HAT-P-1B	 	&$129.6\pm5.9$  	&$4.31\pm0.10$	&$ 3.59\pm0.10$	&$ 3.36\pm0.10$	&$ 3.29\pm0.10$	&$ 3.31\pm0.12$	&$ 3.29\pm0.12$	&$ 3.71\pm0.12$	&$ 3.74\pm0.12$	 \\ 
HAT-P-2 		&$125.3\pm13.1$ 	&$3.20\pm0.24$	&$ 2.30\pm0.23$	&$ 2.16\pm0.23$	&$ 2.11\pm0.24$	&$ 2.05\pm0.24$	&$ 2.11\pm0.24$	&$ 2.59\pm0.24$	&$ 2.56\pm0.24$	 \\ 
HAT-P-3 		&$166.4\pm14.4$ 	&$5.75\pm0.13$	&$ 3.83\pm0.20$	&$ 3.44\pm0.18$	&$ 3.34\pm0.18$	&$ 3.28\pm0.20$	&$ 3.34\pm0.20$	& --				& -- 				\\ 
HAT-P-4 		&$293.5\pm19.4$ 	&$3.78\pm0.13$	&$ 2.76\pm0.15$	&$ 2.50\pm0.14$	&$ 2.43\pm0.15$	&$ 2.41\pm0.16$	&$ 2.46\pm0.16$	& --				& -- 				\\ 
HAT-P-6 		&$277.8\pm19.1$ 	&$3.25\pm0.15$	&$ 2.34\pm0.15$	&$ 2.22\pm0.15$	&$ 2.09\pm0.15$	&$ 2.07\pm0.16$	&$ 2.08\pm0.16$	& --				& -- 				\\ 
HAT-P-7	 	&$320.8\pm17.4$ 	&$2.95\pm0.11$	&$ 2.02\pm0.12$	&$ 1.81\pm0.12$	&$ 1.80\pm0.12$	&$ 1.76\pm0.13$	&$ 1.75\pm0.13$	&$ 2.20\pm0.14$	&$ 2.23\pm0.14$ 	\\ 
HAT-P-8 		&$227.8\pm12.7$ 	&$3.57\pm0.12$	&$ 2.43\pm0.12$	&$ 2.22\pm0.12$	&$ 2.17\pm0.13$	&$ 2.15\pm0.14$	&$ 2.14\pm0.14$	& --				& -- 				\\ 
HAT-P-12		&$139.1\pm9.6$  	&$7.05\pm0.15$	&$ 5.08\pm0.15$	&$ 4.52\pm0.15$	&$ 4.39\pm0.15$	&$ 4.37\pm0.16$	&$ 4.42\pm0.16$	& --				& -- 				\\ 
HAT-P-23		&$355.0\pm40.8$ 	&$4.31\pm0.18$	&$ 3.35\pm0.26$	&$ 3.09\pm0.26$	&$ 3.04\pm0.26$	&$ 3.07\pm0.27$	&$ 3.02\pm0.25$	& --				& -- 				\\ 
WASP-1  		&$346.4\pm34.8$ 	&$3.61\pm0.19$	&$ 2.89\pm0.22$	&$ 2.67\pm0.22$	&$ 2.58\pm0.23$	&$ 2.54\pm0.21$	&$ 2.54\pm0.23$	&$ 2.95\pm0.30$	&$ 3.01\pm0.26$ 	\\ 
WASP-2  		&$153.9\pm8.3$  	&$5.88\pm0.12$	&$ 4.23\pm0.18$	&$ 3.82\pm0.12$	&$ 3.70\pm0.12$	&$ 3.65\pm0.13$	&$ 3.67\pm0.14$	&$ 4.08\pm0.18$	&$ 4.10\pm0.13$ 	\\ 
WASP-3  		&$251.4\pm18.8$ 	&$3.63\pm0.16$	&$ 2.60\pm0.17$	&$ 2.40\pm0.16$	&$ 2.36\pm0.16$	&$ 2.36\pm0.17$	&$ 2.35\pm0.17$	&$ 2.77\pm0.18$	&$ 2.76\pm0.18$ 	\\ 
WASP-4  		&$280.9\pm31.1$ 	&$5.24\pm0.11$	&$ 3.94\pm0.24$	&$ 3.60\pm0.25$	&$ 3.50\pm0.24$	&$ 3.47\pm0.26$	&$ 3.49\pm0.25$	& --				& -- 				\\ 
WASP-5  		&$318.6\pm19.9$ 	&$4.63\pm0.14$	&$ 3.43\pm0.14$	&$ 3.13\pm0.14$	&$ 3.08\pm0.14$	&$ 3.02\pm0.15$	&$ 3.07\pm0.15$	& --				& -- 				\\ 
WASP-8A 	&$85.1\pm10.7$  	&$5.14\pm0.29$	&$ 3.85\pm0.28$	&$ 3.57\pm0.28$	&$ 3.44\pm0.28$	&$ 3.43\pm0.30$	&$ 3.51\pm0.29$	& --				&$ 3.90\pm0.28$ 	\\ 
WASP-12A 	&$436.3\pm37.3$ 	&$3.37\pm0.11$	&$ 2.28\pm0.19$	&$ 2.03\pm0.19$	&$ 1.99\pm0.20$	&$ 1.91\pm0.20$	&$ 1.90\pm0.19$	&$ 2.34\pm0.20$	&$ 2.35\pm0.21$ 	\\ 
WASP-14 		&$208.4\pm15.9$ 	&$3.15\pm0.17$	&$ 2.27\pm0.17$	&$ 2.06\pm0.17$	&$ 2.03\pm0.17$	&$ 1.99\pm0.17$	&$ 1.96\pm0.17$	& --				&$ 2.44\pm0.18$ 	\\ 
WASP-17 		&$476.0\pm36.0$ 	&$3.20\pm0.13$	&$ 2.12\pm0.17$	&$ 1.93\pm0.17$	&$ 1.84\pm0.16$	&$ 1.81\pm0.18$	&$ 1.80\pm0.17$	& --				&$ 2.27\pm0.20$ 	\\ 
WASP-18 		&$122.6\pm6.7$  	&$3.86\pm0.12$	&$ 2.97\pm0.12$	&$ 2.79\pm0.13$	&$ 2.69\pm0.13$	&$ 2.66\pm0.13$	&$ 2.67\pm0.13$	&$ 3.12\pm0.14$	&$ 3.13\pm0.13$ 	\\ 
WASP-19 		&$275.9\pm13.4$ 	&$5.11\pm0.10$	&$ 3.70\pm0.11$	&$ 3.40\pm0.10$	&$ 3.28\pm0.11$	&$ 3.24\pm0.12$	&$ 3.28\pm0.12$	&$ 3.71\pm0.17$	&$ 3.55\pm0.13$ 	\\ 
WASP-24 		&$332.5\pm23.8$ 	&$3.61\pm0.12$	&$ 2.85\pm0.16$	&$ 2.61\pm0.16$	&$ 2.54\pm0.16$	&$ 2.52\pm0.17$	&$ 2.55\pm0.16$	& --				& -- 				\\ 
WASP-26A	&$293.4\pm20.5$ 	&$3.96\pm0.12$	&$ 2.68\pm0.15$	&$ 2.44\pm0.15$	&$ 2.35\pm0.15$	&$ 2.34\pm0.17$	&$ 2.37\pm0.17$	& --				& -- 				\\ 
WASP-33 		&$123.1\pm7.2$  	&$2.69\pm0.13$	&$ 2.13\pm0.13$	&$ 2.06\pm0.13$	&$ 2.02\pm0.13$	&$ 1.98\pm0.14$	&$ 1.99\pm0.14$	& --				& -- 				\\ 
WASP-43 		&$106.1\pm7.2$  	&$7.36\pm0.23$	&$ 4.87\pm0.15$	&$ 4.27\pm0.15$	&$ 4.14\pm0.15$	&$ 4.00\pm0.15$	&$ 4.08\pm0.16$	& --				& -- 				\\ 
WASP-48 		&$466.0\pm49.0$ 	&$3.38\pm0.20$	&$ 2.29\pm0.23$	&$ 2.10\pm0.24$	&$ 2.03\pm0.23$	&$ 2.00\pm0.24$	&$ 2.02\pm0.24$	& --				& -- 				\\ 
CoRoT-1 		&$715.2\pm58.2$ 	&$4.29\pm0.18$	&$ 3.19\pm0.18$	&$ 2.95\pm0.18$	&$ 2.88\pm0.17$	&$ 2.85\pm0.21$	&$ 2.84\pm0.21$	& --				& -- 				\\ 
CoRoT-2A	&$255.0\pm16.2$ 	&$5.25\pm0.11$	&$ 3.75\pm0.14$	&$ 3.41\pm0.14$	&$ 3.28\pm0.14$	&$ 3.26\pm0.16$	&$ 3.28\pm0.16$	& --				&$ 3.59\pm0.16$	 \\ 
Kepler-7		&$893.3\pm50.5$ 	&$3.25\pm0.12$	&$ 2.08\pm0.12$	&$ 1.85\pm0.12$	&$ 1.78\pm0.12$	&$ 1.79\pm0.15$	&$ 1.78\pm0.14$	& --				& -- 				\\ 
KELT-1		&$251.3\pm13.0$ 	&$3.70\pm0.10$	&$ 2.68\pm0.12$	&$ 2.53\pm0.12$	&$ 2.44\pm0.11$	&$ 2.39\pm0.13$	&$ 2.40\pm0.13$	& --				& -- 				\\ 

\hline
\end{tabular}
\end{table}

\clearpage
\newpage

\begin{table}
\scriptsize 
  \caption{Apparent magnitudes for the dayside of occulting extrasolar planets in the Vega-based magnitude convention. References in the last column include the discovery paper and the published flux drop at occultation. List compiled with the help
  of \href{http://exoplanets.org}{exoplanets.org} \citep{Wright:2011fj} and of the SAO/NASA ADS paper repository.
  }\label{tab:app_pl}
  \begin{tabular}{lcrrrrrrrl}
  \hline
   Name   & distance modulus & \multicolumn{7}{c}{apparent magnitudes} & Refs\\
                 &  ($m-M$)   &    m$_{\rm J}$ & m$_{\rm H}$ & m$_{\rm K_s}$ & m$_{\rm [3.6]}$ & m$_{\rm [4.5]}$ & m$_{\rm [5.8]}$ & m$_{\rm [8.0]}$ & \\
 \hline
HD\,189733A\,b&$1.41\pm0.12$	& --  				& --  				& --  				&$12.532\pm0.070$	&$12.400\pm0.064$	&$12.24\pm0.13$	&$11.680\pm0.071$	&1,2,3\\
HD\,209458\,b	&$3.45\pm0.10$	& --  				& --  				& --  				&$13.83\pm0.11$	&$12.984\pm0.083$	&$13.09\pm0.17$	&$13.31\pm0.13$	&1,4,5,6,7\\
HD\,80606\,b	&$4.09\pm0.13$	& --  				& --  				& --  				& --   				& --   				& --  				&$15.24\pm0.23$	&1,8,9\\
HD\,149026\,b	&$4.54\pm0.11$	& --  				& --  				& --  				&$15.33\pm0.10$	&$15.50\pm0.20$	&$15.70\pm0.27$	&$15.44\pm0.14$	&1,10,11\\
GJ\,436\,b		&$-1.07\pm0.35$	& --  				& --  				& --  				&$15.27\pm0.28$	&$> 15.8$			&$15.4\pm1.1$		&$14.868\pm0.093$	&1,12,13,14\\
GJ\,1214\,b	&$-0.2\pm1.4$		& --  				& --  				& --  				&$> 17.7$			&$> 17.7	$		& --  				& --   				&1,15,16\\
55\,Cnc\,e		&$0.44\pm0.17$	& --  				& --  				& --  				& --   				&$13.77\pm0.27$	& --  				& --   				&1,17,18,19,20,21\\
TReS-1\,b		&$5.56\pm0.15$	& --  				& --  				& -- 				& --   				&$17.73\pm0.23$	& --   				&$16.86\pm0.18$	&1,22,23\\
TReS-2\,b		&$6.45\pm0.14$	& --  				& --  				&$17.86\pm0.23$	&$17.02\pm0.20$	&$16.39\pm0.12$	&$16.96\pm0.37$	&$16.36\pm0.20$	&1,24,25,26\\
TReS-3\,b		&$7.06\pm0.14$	& --  				&$> 18.9$			&$17.80\pm0.15$	&$16.70\pm0.12$	&$16.67\pm0.18$	&$16.90\pm0.27$	&$16.84\pm0.18$	&1,27,28,29\\
TReS-4\,b		&$8.80\pm0.22$	& --  				& --   				& --  				&$17.42\pm0.11$	&$17.35\pm0.13$	&$17.36\pm0.34$	&$16.98\pm0.17$	&1,30,31\\
XO-1\,b		&$6.25\pm0.13$	& --  				& --   				& --  				&$17.13\pm0.11$	&$16.80\pm0.11$	&$16.41\pm0.15$	&$16.66\pm0.17$	&1,32.33\\
XO-2\,b		&$5.97\pm0.13$	& --  				& --   				& --  				&$16.96\pm0.24$	&$16.82\pm0.24$	&$16.67\pm0.26$	&$16.92\pm0.52$	&1,34,35\\
XO-3\,b		&$6.34\pm0.14$	& --  				& --   				& --  				&$16.24\pm0.059$	&$15.869\pm0.058$	&$16.39\pm0.53$	&$16.27\pm0.30$	&1,36,37\\
XO-4\,b		&$7.44\pm0.14$	& --  				& --   				& --  				&$17.52\pm0.26$	&$16.58\pm0.10$	& --   				& --   				&1,38,39\\
HAT-P-1B\,b	&$5.56\pm0.10$	& --  				& --   				&$16.26\pm0.27$	&$16.62\pm0.12$	&$16.03\pm0.20$	&$16.01\pm0.18$	&$15.87\pm0.20$	&1,40,41,42\\
HAT-P-2\,b	&$5.49\pm0.24$	& --  				& --   				& --  				&$15.05\pm0.096$	&$15.07\pm0.079$	&$15.95\pm0.68$	&$15.19\pm0.10$	&1,43,44\\
HAT-P-3\,b	&$6.11\pm0.20$	& --  				& --   				& --  				&$16.76\pm0.32$	&$17.02\pm0.20$	& --  				& --  				&1,45,46\\
HAT-P-4\,b	&$7.34\pm0.15$	& --  				& --   				& --  				&$16.87\pm0.15$	&$17.08\pm0.15$	& --  				& --  				&1,47,46\\
HAT-P-6\,b	&$7.22\pm0.15$	& --  				& --   				& --  				&$16.62\pm0.10$	&$16.738\pm0.087$	& --  				& --  				&1,48,39\\
HAT-P-7\,b	&$7.53\pm0.12$	& --  				& --   				& --  				&$16.81\pm0.21$	&$16.28\pm0.16$	&$16.25\pm0.16$	&$16.38\pm0.27$	&1,49,50\\
HAT-P-8\,b	&$6.79\pm0.12$	& --  				& --   				& --  				&$16.15\pm0.10$	&$16.319\pm0.097$	& --  				& --  				&1,51,39\\
HAT-P-12\,b	&$5.72\pm0.16$	& --  				& --   				& --  				&$> 18.5$			&$17.8$			& --  				& --  				&1,52,46\\
HAT-P-23\,b	&$7.75\pm0.27$	& --  				& --   				& --  				&$17.34\pm0.10$	&$17.05\pm0.11$	& --  				& --  				&1,53,54\\
WASP-1\,b	&$7.70\pm0.23$	& --  				& --   				& --  				&$17.07\pm0.10$	&$16.896\pm0.099$	&$17.06\pm0.30$	&$16.52\pm0.17$	&1,55,56\\
WASP-2\,b	&$5.94\pm0.12$	& --  				& --   				& --  				&$17.29\pm0.69$	&$16.54\pm0.13$	&$16.81\pm0.63$	&$16.39\pm0.27$	&1,55,56\\
WASP-3\,b	&$7.00\pm0.17$	& --  				& --   				&$16.27\pm0.18$	& --   				& --   				& --  				& --  				&1,57,58\\
WASP-4\,b	&$7.24\pm0.25$	& --  				& --   				&$17.58\pm0.13$	&$16.95\pm0.12$	&$16.894\pm0.095$	& --  				& --  				&1,59,60,61\\
WASP-5\,b	&$7.52\pm0.14$	& --  				& --   				& --  				&$17.10\pm0.13$	&$17.35\pm0.17$	& --  				& --  				&1,62,63\\
WASP-8A\,b	&$4.65\pm0.29$	& -- 				& --   				& --  				&$15.45\pm0.20$	&$16.06\pm0.14$	& --  				&$16.13\pm0.31$	&1,64,65\\
WASP-12A\,b	&$8.20\pm0.19$	&$17.62\pm0.24$	&$17.03\pm0.17$	&$16.363\pm0.050$	&$16.056\pm0.053$&$16.032\pm0.065$	&$15.94\pm0.12$	&$15.94\pm0.15$	&1,66,67\\
WASP-14\,b	&$6.60\pm0.17$	& --  				& --   				& --  				&$15.406\pm0.060$&$15.18\pm0.10$	& --  				&$15.89\pm0.14$	&1,68,69\\
WASP-17\,b	&$8.39\pm0.17$	& --  				& --   				& --  				& -- 				&$16.793\pm0.089$	& --  				&$17.22\pm0.21$	&1,70,71\\
WASP-18\,b	&$5.44\pm0.12$	& --  				& --   				& --  				&$14.405\pm0.088$	&$14.166\pm0.072$	&$14.641\pm0.099$	&$14.541\pm0.081$	&1,72,73,74\\
WASP-19\,b	&$7.20\pm0.11$	& --  				&$17.00\pm0.19$	& --  				&$16.235\pm0.072$	&$16.094\pm0.075$	&$16.38\pm0.23$	&$16.10\pm0.21$	&1,75,76\\
WASP-24\,b	&$7.61\pm0.16$	& --  				& --   				& --  				&$17.13\pm0.11$	&$16.89\pm0.12$	& --  				& --  				&1,77,78\\
WASP-26A\,b	&$7.34\pm0.16$	& --  				& --   				& --  				&$16.93\pm0.14$	&$16.78\pm0.13$	& --  				& --  				&1,79,80\\
WASP-33\,b	&$5.45\pm0.13$	& --  				& --   				&$14.00\pm0.12$	&$13.89\pm0.23$	&$13.406\pm0.082$	& --  				& --  				&1,81,82,83\\
WASP-43\,b	&$5.31\pm0.24$	& --  				&$16.80\pm0.17$	&$16.05\pm0.17$	&$15.281\pm0.077$	&$15.26\pm0.074$	& --  				& --  				&1,84,85,86\\
WASP-48\,b	&$8.34\pm0.24$	& --  				& --   				& --  				&$17.22\pm0.11$	&$17.03\pm0.12$	& --  				& --  				&1,87,54\\
CoRoT-1\,b	&$9.27\pm0.18$	& --  				&$19.31\pm0.44$	&$18.33\pm0.15$	&$18.07\pm0.16$	&$17.91\pm0.13$	& --  				& --  				&1,88,89,90,91\\
CoRoT-2A\,b	&$7.03\pm0.14$	& --  				& --   				&$17.3\pm1.2$		&$16.421\pm0.092$	&$16.062\pm0.087$& --   				&$16.36\pm0.14$	&1,92,89\\
Kepler-7\,b	&$9.76\pm0.13$	& --  				& --   				& --  				&$> 19.6$			&$> 19.0$			& --   				& --   				&1,94,95\\
KELT-1\,b		&$7.00\pm0.12$	& --  				& --   				& --  				&$16.165\pm0.085$	&$16.152\pm0.089$	& --   				& --   				&1,96,97\\

\hline
\end{tabular}\\
{ References:} (1) this paper; (2) \citet{Bouchy:2005lr};  (3) \citet{Knutson:2012ys}; (4) \citet{Charbonneau:2000dp}; (5) \citet{Henry:2000fk}; (6) \citet{Mazeh:2000qy}; (7) \citet{Knutson:2008qy}; (8) \citet{Naef:2001uq}; (9) \citet{Laughlin:2009uq}; (10) \citet{Sato:2005fj}; 
(11) \citet{Stevenson:2012vn}; (12) \citet{Butler:2004vn}; (13) \citet{Gillon:2007yq}; (14) Lanotte et al. (in prep); (15) \citet{Charbonneau:2009fj}; (16) \citet{Gillon:2014ix}; (17) \citet{McArthur:2004qy}; (18) \citet{Dawson:2010fk}; (19) \citet{Winn:2011qf}; (20) \citet{Demory:2011kx};
(21) \citet{Demory:2012uq}; (22) \citet{Alonso:2004rm}; (23) \citet{Charbonneau:2005fj}; (24) \citet{ODonovan:2006gf}; (25) \citet{ODonovan:2010ys}; (26) \citet{Croll:2010rt}; (27) \citet{ODonovan:2007ly}; (28) \citet{Fressin:2010ly}; (29) \citet{Croll:2010zr}; (30) \citet{Mandushev:2007ul}; 
(31) \citet{Knutson:2009gf}; (32) \citet{McCullough:2006qf}; (33) \citet{Machalek:2008ve}; (34) \citet{Burke:2007db}; (35) \citet{Machalek:2009ul}; (36) \citet{Johns-Krull:2008qf}; (37) \citet{Machalek:2010pd}; (38) \citet{McCullough:2008bh}; (39) \citet{Todorov:2012lq}; (40) \citet{Bakos:2007fj}; 
(41) \citet{Todorov:2010rr}; (42) \citet{de-Mooij:2011cr}; (43) \citet{Bakos:2007yq}; (44) \citet{Lewis:2013qy}; (45) \citet{Torres:2007eu}; (46) \citet{Todorov:2013wd}; (47) \citet{Kovacs:2007vn}; (48) \citet{Noyes:2008ys}; (49) \citet{Pal:2008fr}; (50) \citet{Christiansen:2010nx}; 
(51) \citet{Latham:2009qe}; (52) \citet{Hartman:2009rt}; (53) \citet{Bakos:2011lr}; (54) \citet{ORourke:2014rp}; (55) \citet{Collier-Cameron:2007fj}; (56) \citet{Wheatley:2010lr}; (57) \citet{Pollacco:2008hc}; (58) \citet{Zhao:2012fk}; (59) \citet{Wilson:2008lr}; (60) \citet{Beerer:2011uq}; 
(61) \citet{Caceres:2011qy}; (62) \citet{Anderson:2008sp}; (63) \citet{Baskin:2013yg}; (64) \citet{Queloz:2010lr}; (65) \citet{Cubillos:2013kx}; (66) \citet{Hebb:2009lr}; (67) \citet{Crossfield:2012qy}; (68) \citet{Joshi:2009jt}; (69) \citet{Blecic:2013wq}; (70) \citet{Anderson:2010fj};
(71) \citet{Anderson:2011kx}; (72) \citet{Hellier:2009fj}; (73) \citet{Nymeyer:2011ve}; (74) \citet{Maxted:2013qf}; (75) \citet{Hebb:2010fk}; (76) \citet{Anderson:2013fk}; (77) \citet{Street:2010xy}; (78) \citet{Smith:2012lq}; (79) \citet{Smalley:2010qy}; (80) \citet{Mahtani:2013dq};
(81) \citet{Collier-Cameron:2010lr}; (82) \citet{Deming:2012rr}; (83) \citet{de-Mooij:2013wd}; (84) \citet{Hellier:2011fk}; (85) \citet{Blecic:2014ao}; (86) \citet{Wang:2013eu}; (87) \citet{Enoch:2011yq}; (88) \citet{Barge:2008ys}; (89) \citet{Deming:2011dp}; (90) \citet{Rogers:2009qe};
(91) \citet{Zhao:2012mz}; (92) \citet{Alonso:2008fk};  (93) \citet{Alonso:2010hc}; (94) \citet{Latham:2010lr}; (95) \citet{Demory:2013uq}; (96) \citet{Siverd:2012ef}; (97) \citet{Beatty:2014lr}.

\end{table}

\clearpage
\newpage

\begin{table}
\scriptsize 
  \caption{Absolute magnitudes for the dayside of occulting extrasolar planets in the Vega-based magnitude convention. 
  }\label{tab:planets}
  \begin{tabular}{lrrrrrrr}
  \hline
   Name     & \multicolumn{7}{c}{absolute magnitudes} \\
                   &    M$_{\rm J}$ & M$_{\rm H}$ & M$_{\rm K_s}$ & M$_{\rm [3.6]}$ & M$_{\rm [4.5]}$ & M$_{\rm [5.8]}$ & M$_{\rm [8.0]}$  \\
 \hline
HD\,189733A\,b&--				&--				&--				&$ 11.13\pm0.14$	&$ 11.00\pm0.13$	&$ 10.83\pm0.18$	&$ 10.27\pm0.14$	\\
HD\,209458\,b	&--				&--				&--				&$ 10.37\pm0.15$	&$ 9.53\pm0.13$	&$ 9.64\pm0.20$	&$ 9.86\pm0.17$	\\
HD\,80606\,b  	&--				&--				&--				&--				&--				&--				&$ 11.15\pm0.28$	\\
HD\,149026\,b	&--				&--				&--				&$ 10.80\pm0.15$	&$ 10.96\pm0.23$	&$ 11.16\pm0.31$	&$ 10.90\pm0.18$	\\
GJ\,436\,b		&--				&--				&--				&$ 15.24\pm0.28$	&$ > 15.8$		&$ 15.3\pm1.1$	&$ 14.84\pm0.10$	\\
GJ\,1214\,b	&--				&--				&--				&$ > 16.9$		&$ > 16.9$		&--				&--				\\
55\,Cnc\,e 	&--				&--				&--				&--				&$ 13.34\pm 0.31$	&--				&--				\\
TReS-1\,b  	&--				&--				&--				&--				&$ 12.17\pm0.29$	&--				&$ 11.30\pm0.23$	\\
TReS-2\,b  	&--				&--				&$ 11.41\pm0.25$	&$ 10.57\pm0.24$	&$ 9.93\pm0.18$	&$ 10.51\pm0.36$	&$ 9.91\pm0.24$	\\
TReS-3\,b  	&--				&$ >11.8$			&$ 10.74\pm0.20$	&$ 9.64\pm0.20$	&$ 9.61\pm0.22$	&$ 9.84\pm0.31$	&$ 9.78\pm0.23$	\\
TReS-4\,b  	&--				&--				&--				&$ 8.62\pm0.25$	&$ 8.55\pm0.25$	&$ 8.56\pm0.42$	&$ 8.18\pm0.27$	\\
XO-1\,b    		&--				&--				&--				&$ 10.88\pm0.17$	&$ 10.55\pm0.17$	&$ 10.15\pm0.19$	&$ 10.41\pm0.22$	\\
XO-2\,b    		&--				&--				&--				&$ 11.00\pm0.29$	&$ 10.85\pm0.27$	&$ 10.70\pm0.30$	&$ 10.96\pm0.52$	\\
XO-3\,b    		&--				&--				&--				&$ 9.90\pm0.15$	&$ 9.52\pm0.16$	&$ 10.05\pm0.54$	&$ 9.93\pm0.34$	\\
XO-4\,b    		&--				&--				&--				&$ 10.07\pm0.29$	&$ 9.14\pm0.18$	&--				&--				\\
HAT-P-1B\,b 	&--				&--				&$ 10.70\pm0.31$	&$ 11.05\pm0.16$	&$ 10.46\pm0.22$	&$ 10.44\pm0.20$	&$ 10.30\pm0.22$	\\
HAT-P-2\,b 	&--				&--				&--				&$ 9.56\pm0.26$	&$ 9.58\pm0.25$	&$ 10.46\pm0.72$	&$ 9.71\pm0.25$	\\
HAT-P-3\,b 	&--				&--				&--				&$ 10.65\pm0.38$	&$ 10.91\pm0.27$	&--				&--				\\
HAT-P-4\,b 	&--				&--				&--				&$ 9.53\pm0.20$	&$ 9.75\pm0.20$	&--				&--				\\
HAT-P-6\,b 	&--				&--				&--				&$ 9.40\pm0.19$	&$ 9.52\pm0.17$	&--				&--				\\
HAT-P-7\,b 	&--				&--				&--				&$ 9.28\pm0.23$	&$ 8.75\pm0.20$	&$ 8.72\pm0.20$	&$ 8.85\pm0.34$	\\
HAT-P-8\,b 	&--				&--				&--				&$ 9.36\pm0.16$	&$ 9.53\pm0.16$	&--				&--				\\
HAT-P-12\,b	&--				&--				&--				&$> 12.8$			&$ >12.1	$		&--				&--				\\
HAT-P-23\,b	&--				&--				&--				&$ 9.58\pm0.28$	&$ 9.29\pm0.27$	&--				&--				\\
WASP-1\,b  	&--				&--				&--				&$ 9.37\pm0.24$	&$ 9.20\pm0.24$	&$ 9.36\pm0.38$	&$ 8.82\pm0.28$	\\
WASP-2\,b  	&--				&--				&--				&$ 11.35\pm0.72$	&$ 10.60\pm0.18$	&$ 10.88\pm0.58$	&$ 10.46\pm0.28$	\\
WASP-3\,b  	&--				&--				&$ 9.21\pm0.20$	&--				&--				&--				&--				\\
WASP-4\,b  	&--				&--				&$ 10.34\pm0.27$	&$ 9.71\pm0.28$	&$ 9.65\pm0.27$	&--				&--				\\
WASP-5\,b  	&--				&--				&--				&$ 9.59\pm0.19$	&$ 9.84\pm0.22$	&--				&--				\\
WASP-8A\,b  	&--				&--				&--				&$ 10.80\pm0.35$	&$ 11.41\pm0.31$	&--				&$ 11.48\pm0.40$	\\
WASP-12A\,b 	&$ 9.42\pm0.32$	&$ 8.83\pm 0.22$	&$ 8.16\pm0.20$	&$ 7.86\pm0.20$	&$ 7.83\pm0.20$	&$ 7.74\pm0.22$	&$ 7.74\pm0.24$	\\
WASP-14\,b 	&--				&--				&--				&$ 8.81\pm0.17$	&$ 8.59\pm0.19$	&--				&$ 9.30\pm0.22$	\\
WASP-17\,b 	&--				&--				&--				&--				&$ 8.40\pm0.18$	&--				&$ 8.84\pm0.27$	\\
WASP-18\,b 	&--				&--				&--				&$ 8.96\pm0.15$	&$ 8.72\pm0.14$	&$ 9.20\pm0.17$	&$ 9.10\pm0.14$	\\
WASP-19\,b 	&--				&$ 9.80\pm0.21$	&--				&$ 9.03\pm0.13$	&$ 8.89\pm0.13$	&$ 9.17\pm0.25$	&$ 8.89\pm0.24$	\\
WASP-24\,b 	&--				&--				&--				&$ 9.52\pm0.19$	&$ 9.28\pm0.18$	&--				&--				\\
WASP-26A\,b 	&--				&--				&--				&$ 9.59\pm0.21$	&$ 9.43\pm0.20$	&--				&--				\\
WASP-33\,b 	&--				&--				&$ 8.55\pm0.18$	&$ 8.44\pm0.25$	&$ 7.95\pm0.15$	&--				&--				\\
WASP-43\,b 	&--				&$ 11.67\pm0.23$	&$ 10.92\pm0.23$	&$ 10.15\pm0.16$	&$ 10.13\pm0.17$	&--				&--				\\
WASP-48\,b 	&--				&--				&--				&$ 8.88\pm0.25$	&$ 8.69\pm0.26$	&--				&--				\\
CoRoT-1\,b 	&--				&$ 10.04\pm0.52$	&$ 9.06\pm0.23$	&$ 8.80\pm0.24$	&$ 8.63\pm0.23$	&--				&--				\\
CoRoT-2A\,b 	&--				&--				&$ 10.3\pm 1.1$	&$ 9.39\pm0.17$	&$ 9.02\pm0.17$	&--				&$ 9.33\pm0.21$	\\
Kepler-7\,b	&--				&--				&--				&$ >9.8$			&$ >9.3$			&--				&--				\\
KELT-1\,b		&--				&--				&--				&$ 9.16\pm0.14$	&$ 9.15\pm0.14$	&--				&--				\\

\hline
\end{tabular}
\end{table}

\bsp

\label{lastpage}

\end{document}